\documentclass[reprint,prd,nofootinbib,superscriptaddress,11pt]{revtex4}
\usepackage{amssymb,wrapfig,amsmath,graphicx,epsfig,bm,color,verbatim,slashed,mathtools,xparse,adjustbox,dcolumn,float,relsize,tikz,cancel}
\usepackage[hidelinks]{hyperref}
\usepackage{caption,booktabs,array}
\usepackage{subcaption,multirow}
\usepackage{float}
\usepackage[english]{babel}
\usepackage[utf8]{inputenc}
\usepackage[mathletters]{ucs}
\usepackage{soul}

\usepackage[ntheorem]{empheq}

\newcolumntype{P}[1]{>{\centering\arraybackslash}p{#1}}
\newcolumntype{C}{>{$}c<{$}}
\AtBeginDocument{
\heavyrulewidth=.08em
\lightrulewidth=.05em
\cmidrulewidth=.03em
\belowrulesep=.65ex
\belowbottomsep=0pt
\aboverulesep=.4ex
\abovetopsep=0pt
\cmidrulesep=\doublerulesep
\cmidrulekern=.5em
\defaultaddspace=.5em
}

\pdfoutput=1
\newcommand{\func}{\operatorname}
\DeclareUnicodeCharacter{2212}{-}

\input{tcilatex}

\begin{document}

\title{Global analysis and LHC study of a vector-like extension of the Standard Model with extra scalars}

\author{A. E. C\'{a}rcamo Hern\'{a}ndez}
\email{antonio.carcamo@usm.cl}
\affiliation{Universidad T\'{e}cnica Federico Santa Mar\'{\i}a,\\
 Casilla 110-V, Valpara\'{\i}so, Chile }
\affiliation{{Centro Cient\'{\i}fico-Tecnol\'ogico de Valpara\'{\i}so, Casilla 110-V,
Valpara\'{\i}so, Chile}}
\affiliation{{Millennium Institute for Subatomic Physics at High-Energy Frontier
(SAPHIR), Fern\'andez Concha 700, Santiago, Chile}}

\author{Kamila Kowalska}
\email{kamila.kowalska@ncbj.gov.pl}
\affiliation{National Centre for Nuclear Research,\\
Pasteura 7, 02-093 Warsaw, Poland}

\author{Huchan Lee}
\email{huchan.lee@ncbj.gov.pl}
\affiliation{National Centre for Nuclear Research,\\
Pasteura 7, 02-093 Warsaw, Poland}

\author{Daniele Rizzo}
\email{daniele.rizzo@ncbj.gov.pl}
\affiliation{National Centre for Nuclear Research,\\
Pasteura 7, 02-093 Warsaw, Poland}


\begin{abstract}
We perform a global analysis of a vector-like extension of the Standard Model, which also features additional doublet and singlet scalars. The usual Yukawa interactions are forbidden in this setup by an extra U(1) global symmetry and the masses of the second and third family quarks and leptons are generated via the mixing with the vector-like sector.
We identify three best-fit benchmark scenarios which satisfy the constraints imposed by the stability of the scalar potential, the perturbativity of the coupling constants, the measurement of the muon anomalous magnetic moment and the non-observation of the flavor violating tau decays.  
We show that dominant contributions to the muon $(g-2)$ originate in this model from the charged Higgs/neutral lepton one-loop diagrams, 
thus correcting an inaccurate statement than can be found in the literature.
We also perform a detailed LHC analysis of the benchmark scenarios. We investigate
the experimental constraints stemming from  direct searches for vector-like quarks, vector-like leptons and exotic scalars. While we show that the model is not currently tested by any collider experiment, we point out that decays of a heavy Higgs boson into
two tau leptons may offer a smoking gun signature for the model verification in upcoming runs
at the LHC.

\end{abstract}

\maketitle

\newpage

\tableofcontents

\newpage
\section{Introduction}\label{sec:introduction}

The origin of the flavor structure of the Standard Model (SM), i.e.~the observed hierarchy between fermion masses and mixing angles of the Cabibbo-Kobayashi-Maskawa (CKM) matrix, is one of the greatest mysteries of particle physics that still lacks a convincing and commonly accepted explanation. A number of New Physics (NP) ideas have been put forward in recent decades to address the flavor puzzle, among which the Froggatt-Nielsen (FN) mechanism~\cite{Froggatt:1978nt} and extra dimensions~\cite{Arkani-Hamed:1999ylh,Gherghetta:2000qt,Huber:2000ie} are those that admittedly received the most attention and applications. The underlying concept is to introduce a new quantity that in some sense would be ``larger'' than the electroweak symmetry breaking (EWSB) scale. This could be the vacuum expectation value (vev) of a flavon field, or a distance of a fermion field from the infra-red brane. Such a hierarchy of scales can be then translated into a hierarchy of masses and mixing angles of the SM quarks and leptons. A similar idea gave rise to the famous seesaw mechanism of neutrino mass generation~\cite{Minkowski:1977sc,Gell-Mann:1979vob,Yanagida:1979as,Glashow:1979nm,Mohapatra:1980yp,Schechter:1981cv,Schechter:1980gr}, where tiny values of the SM neutrino masses arise as a result of suppression of the EWSB scale by a very large Majorana mass. 

In Ref.~\cite{King:2018fcg} a FN-inspired model was proposed to explain the observed masses and mixing patterns of the SM fermions. The SM Yukawa interactions are forbidden in this setup by an extra abelian symmetry U$(1)_X$, which could be either global or local. 
The particle content of the model corresponds to the two-Higgs-doublet model (2HDM) extended by a full family of vector-like (VL) fermions charged under U$(1)_X$, and one U$(1)_X$-breaking singlet scalar which plays the role of a FN flavon. The large third-family Yukawa couplings are then effectively generated via mixing of the SM quarks and leptons with the SU$(2)_L$ doublet VL fermions, while the Yukawa couplings of the second family emerge from a seesaw-like construction, mediated by the heavy VL SU$(2)_L$ singlets.  

The rich structure of the model introduced in Ref.~\cite{King:2018fcg} makes it a perfect framework for providing a combined explanation both for the flavor pattern of the SM and for the miscellaneous anomalies which emerged  in recent years in collider experiments. 
In this context, lepton-flavor violating anomalies in the rare semi-leptonic decays of the $B$ mesons were analyzed in Refs.~\cite{King:2018fcg,Lee:2022sic}, $Z$-mediated Flavour Changing Neutral Currents in Ref.~\cite{CarcamoHernandez:2021yev}, and a deviation  from the SM prediction in the measured value of the anomalous magnetic moment of muon in Refs.~\cite{CarcamoHernandez:2019ydc,Hernandez:2021tii,Lee:2022sic}. In the latter study, in which the extra U$(1)_X$ symmetry was assumed to be global, five benchmark points were identified that could account for the muon $(g-2)$ anomaly and, at the same time, give rise to the mass and mixing patterns of the SM fermions. The scenarios pinpointed in Ref.~\cite{Hernandez:2021tii} were characterized by
relatively low ($\sim 200$ GeV) masses of the VL lepton doublets and large ($\sim 10$) quartic couplings of the scalar potential, which may 
indicate a loss of perturbativity at scales very close to the typical scale set by the masses of the NP particles in the analyzed model. 

In this study, we reassess the findings of Ref.~\cite{Hernandez:2021tii} improving and extending its analysis in several different directions. Firstly, we thoroughly discuss the impact of the most recent bounds from direct NP searches at the Large Hadron Collider~(LHC) on the allowed parameter space of the model, a topic which was not addressed in detail in Refs.~\cite{CarcamoHernandez:2019ydc,CarcamoHernandez:2021yev,Hernandez:2021tii,Lee:2022sic}. 
While we show that the model is not currently tested by any collider experiment, we point out that decays of a heavy Higgs boson into two tau leptons may offer a smoking gun signature for the detection of the model in the upcoming runs of the LHC.

Secondly, we demonstrate that the quartic and Yukawa couplings of the model are subject to strong constraints from their renormalization group (RG) running. In Ref.~\cite{Hernandez:2021tii} it was required that all the dimensionless parameters of the lagrangian remain perturbative (in a loose sense of being smaller than $\sqrt{4\pi}$ for the gauge/Yukawa and smaller than $4\pi$ for the scalar potential couplings) at the characteristic energy scale of the model.  We argue that such a simplistic implementation of the perturbativity bounds should be taken with a grain of salt. The breakdown of perturbativity usually calls for an extension of the  theoretical setup by extra degrees of freedom in order to cure pathological behavior of the running couplings, or/and for an inclusion of non-perturbative effects (like bound-state formation). If any of those arose at the scale specific to the original NP model, they would most likely affect its phenomenological predictions. Therefore, it is more correct to apply the perturbativity bounds to the running couplings evaluated at an energy scale which is high enough that the phenomenology of the specific NP model can be trusted. Once this improvement had been implemented in our study, we discovered that all the benchmark points found previously in Ref.~\cite{Hernandez:2021tii} were disfavored. 

Last but not least, we refine the derivation of the stability conditions for the scalar potential which in Refs.~\cite{Hernandez:2021tii,Lee:2022sic} was simplified to the 2HDM case by integrating out the singlet flavon field. In the current work we  derive all the relevant stability conditions in the full three-scalar setup, obtaining additional constraints on the quartic couplings. 

With all the improvements in place, we identify three benchmark scenarios that satisfy our theoretical and experimental requirements. While these best-fit points emerge from a random numerical scan, they present features that are generic for the model in study. Most importantly, we point out that a charged Higgs/heavy neutrino loop is a dominant contribution to the muon $(g-2)$ anomaly. This results from the fact that the competing neutral scalar/heavy charged lepton contributions are governed by the same Yukawa coupling that determines the tree-level muon mass and is thus required to be small. Once more, this finding is qualitatively different from the conclusions obtained in Refs.~\cite{Hernandez:2021tii,Lee:2022sic}, where only the charged lepton loops were considered. 

The structure of the paper is the following. In Sec.~\ref{sec:setup} we briefly review the field content of the model. We also show how the SM fermion masses and the CKM matrix are generated in this framework. Sec.~\ref{sec:scalar} is dedicated to the scalar sector of the theory. Tree-level scalar masses in the alignment limit are presented, as well as three-field potential stability conditions. 
Experimental constraints from the  flavor physics observables (muon $(g-2)$, rare $\tau$ decays, CKM anomaly) are examined in Sec.~\ref{sec:lepton}. 
In Sec.~\ref{sec:RG} we discuss the RG flow of the model couplings and we derive the corresponding  perturbativity bounds. Sec.~\ref{sec:num} comprises the numerical analysis of the model. We discuss the setup of our numerical scan and we identify three benchmark scenarios that satisfy all the theoretical and phenomenological constraints. In Sec.~\ref{sec:checkGBP} we present a detailed analysis of the LHC searches that may test the parameter space of the model. We summarize our findings in Sec.~\ref{sec:conclusion}. Appendices feature, respectively, explicit forms of the fermion (Appendix~\ref{sec:App_diag}) and scalar  (Appendix~\ref{sec:App_scal}) mass matrices, derivation of the bounded-from-below constraints (Appendix~\ref{sec:App_vs}), and the RG equations (Appendix~\ref{sec:App_RG}).

\section{Generation of fermion masses and mixing} \label{sec:setup}

We begin our study by reviewing the structure and the main properties of the model introduced in Ref.~\cite{King:2018fcg}. In the following, we focus mostly on these features of the model which play a pivotal role in the subsequent phenomenological analysis. Technical details of the model, including the analytical diagonalization of the fermion mass matrices and the derivation of the interaction vertices in the mass basis, can be found in Refs.~\cite{King:2018fcg,CarcamoHernandez:2021yev,Hernandez:2021tii,Lee:2022sic}. 

The particle content of the model is summarized in Table~\ref{tab:BSM_model}. The SM fermion sector, collectively denoted as $\psi_{i}$ ($\psi_i=Q_{iL},\,u_{iR},\,d_{iR},\,L_{iL},\,e_{iR}$ and $i=1,2,3$ stands for a generation index) is extended by one full family of VL fermions, indicated collectively as 
$(\psi_{4},\,\widetilde{\psi}_{4})$. We adopt the convention of using the left-chiral two-component Weyl spinors, therefore the subscripts $L,R$ indicate the names of the fermions, not the chiralities. The scalar sector contains, besides the usual SU(2)$_L$ Higgs doublet dubbed as $H_u$, an extra scalar doublet $H_d$ and a scalar singlet $\phi$. Note that all the NP particles and the Higgs doublet $H_u$ are charged under an extra global gauge symmetry U$(1)_X$, while the SM fermions are U$(1)_X$ singlets.  As a result, the ordinary SM Yukawa interactions are forbidden.

\begin{table}[t]
\centering\renewcommand{\arraystretch}{1.3} 
\begin{tabular}{c|ccccc|cccccc|cccccc|ccc}
\toprule
\toprule
Field & $Q_{iL}$ & $u_{iR}$ & $d_{iR}$ & $L_{iL}$ & $e_{iR}$ & $Q_{4L}$ & $u_{4R}$ & $d_{4R}$ & $L_{4L}$ & $e_{4R} $ & $\nu_{4R}$ & $\widetilde{Q}_{4R}$ & $\widetilde{u}_{4L}$ & $\widetilde{d}_{4L}$ & $\widetilde{L}_{4R}$ & $\widetilde{e}_{4L}$ & $\widetilde{\nu}_{4L}$ & $\phi$ & $H_u$ & $H_d$ \\ 
\midrule
SU$(3)_C$ & $\mathbf{3}$ & $\mathbf{\bar{3}}$ & $\mathbf{\bar{3}}$ & $\mathbf{1}$ & $\mathbf{1}$ & $\mathbf{3}$ & $\mathbf{\bar{3}}$ & $\mathbf{\bar{3}}$ & $\mathbf{1}$ & $\mathbf{1}$ & $\mathbf{1}$ & $\mathbf{\bar{3}}$ & $\mathbf{3}$ & $\mathbf{3}$ & $\mathbf{1}$ & $\mathbf{1}$ & $\mathbf{1}$ & $\mathbf{1}$ & $\mathbf{1}$ & $\mathbf{1}$ \\ 
SU$(2)_L$ & $\mathbf{2}$ & $\mathbf{1}$ & $\mathbf{1}$ & $\mathbf{2}$ & $\mathbf{1}$ & $\mathbf{2}$ & $\mathbf{1}$ & $\mathbf{1}$ & $\mathbf{2}$ & $\mathbf{1}$ & $\mathbf{1}$ & $\mathbf{2}$ & $\mathbf{1}$ & $\mathbf{1}$ & $\mathbf{2}$ & $\mathbf{1}$ & $\mathbf{1}$ & $\mathbf{1}$ & $\mathbf{2}$ & $\mathbf{2}$ \\ 
U$(1)_Y$ & $\frac{1}{6}$ & $-\frac{2}{3}$ & $\frac{1}{3}$ & $-\frac{1}{2}$ & $1$ & $\frac{1}{6}$ & $-\frac{2}{3}$ & $\frac{1}{3}$ & $-\frac{1}{2}$ & $1$ & $0$ & $-\frac{1}{6}$ & $\frac{2}{3}$ & $-\frac{1}{3}$ & $\frac{1}{2} $ & $-1$ & $0$ & $0$ & $\frac{1}{2}$ & $-\frac{1}{2}$ \\ 
U$(1)_X$ & $0$ & $0$ & $0$ & $0$ & $0$ & $1$ & $1$ & $1$ & $1$ & $1$ & $1$ & $-1$ & $-1$ & $-1$ & $-1$ & $-1$ & $-1$ & $1$ & $-1$ & $-1$ \\ 
\bottomrule
\bottomrule
\end{tabular}
\caption{The particle content of the NP model considered in this study.} 
\label{tab:BSM_model}
\end{table}

All the renormalizable Yukawa interactions between the SM and NP fermions which are allowed by the extended gauge symmetry can be schematically written as:
\begin{multline}\label{eqn:Lag_ren}
\mathcal{L}_{\func{ren}}^{\func{Yukawa}} = y_{i4}^{\psi} \psi_{iL} H \psi_{4R} + y_{4j}^{\psi} \psi_{4L} H \psi_{jR} + x_{i4}^{\psi} \psi_{iL} \phi\, \widetilde{\psi}_{4R} + x_{4j}^{\psi} \widetilde{\psi}_{4L} \phi\, \psi_{jR}\\ 
+ M_{4}^{\psi}\, \psi_{4L} \widetilde{\psi}_{4R} + M_{4}^{\widetilde{\psi}}\, \widetilde{\psi}_{4L} \psi_{4R} + \func{h.c.},
\end{multline}
where $H$ is either $H_{u}$ or $H_{d}$ and $M_{4}^{\psi}$ ($M_{4}^{\widetilde{\psi}}$) denotes the VL doublet (singlet) mass parameter. Note that with the U$(1)_X$ charges given in Table~\ref{tab:BSM_model} the scalar $H_u$ only couples to the up-type quarks, while $H_d$ to the down-type quarks and charged leptons, reminiscent of the 2HDM Type-II model. 

\subsection{Hierarchy of masses}\label{sec:hirer}

Once the neutral components of the scalar fields develop their vevs, the $5\times 5$ fermions mass matrices are  generated. Since their upper $3\times 3$ blocks contain only zeros (we recall that the SM Yukawa couplings are forbidden by the U$(1)_X$ symmetry), one has the freedom to rotate the first three families. It can easily be shown~\cite{King:2018fcg} that this allows one to choose a flavor basis in which the fermion mass matrices acquire the following form:
\begin{equation}\label{eq:massmatrix}
\mathcal{M}_\psi=\left(
\begin{array}{c|ccccc}
& \psi_{1R} & \psi_{2R} & \psi_{3R} & \psi_{4R} & \widetilde{\psi}_{4R}\\
\hline
\psi_{1L} & 0 & 0 & 0 & (y^\psi_{14}\langle H^0\rangle) & 0 \\
\psi_{2L} & 0 & 0 & 0 & y^\psi_{24}\langle H^0\rangle & 0\\
\psi_{3L} & 0 & 0 & 0 & y^\psi_{34}\langle H^0\rangle & x^\psi_{34}\langle\phi\rangle \\
\psi_{4L} & 0 & 0 & y^\psi_{43}\langle H^0\rangle & 0 & M_4^\psi\\
\widetilde{\psi}_{4L} & 0 & x^\psi_{42}\langle\phi\rangle & x^\psi_{43}\langle\phi\rangle & M_4^{\widetilde{\psi}} & 0\\
\end{array}\right)\,.
\end{equation}
In the above, the term in parentheses assumes a non-zero value in the mass matrix of the down-type quarks, while it is zero for the up-type quarks and charged leptons. The exact forms of the matrices $\mathcal{M}_u$, $\mathcal{M}_d$ and $\mathcal{M}_e$ are presented in Appendix~\ref{sec:App_diag}.

In order to calculate the masses of the physical quarks and leptons, the $5\times 5$ matrices of Eq.~(\ref{eq:massmatrix}) need to be diagonalized. Due to a large number of free parameters in the Yukawa sector one may expect that the resulting functional dependence of the eigenvalues of $\mathcal{M}_\psi$ on the couplings $y^\psi_{i4},\, y^\psi_{43},\,x^\psi_{4i},\,x^\psi_{34}$ and the masses $M_4^{\psi(\widetilde{\psi})}$ is highly nontrivial. It turns out, however, that it is not necessarily the case and that simplified  expressions for the fermion masses can be derived. Denoting the scalar vevs as
\begin{equation}\label{eq:ewsbmin}
\langle H_u^0\rangle=v_u/\sqrt{2},\qquad \langle H_d^0\rangle=v_d/\sqrt{2},\qquad \langle\phi\rangle=v_\phi/\sqrt{2}
\end{equation}
and defining $\tan\beta=v_u/v_d$,
the masses of the third and second family quarks and leptons are approximately given by (see also Ref.~\cite{King:2018fcg} for a related derivation) 
\begin{eqnarray}
m_t&\approx & \frac{1}{\sqrt{2}}\frac{y^u_{43}x^Q_{34}v_\phi v_u}{\sqrt{(x^Q_{34}v_\phi)^2+2(M_4^Q)^2}},\qquad m_c\approx \frac{y^u_{24}x^u_{42}v_\phi v_u}{2\,M_4^u}\label{eq:masstop}\\
m_b&\approx & \frac{1}{\sqrt{2}}\frac{y^d_{43}x^Q_{34}v_\phi v_d}{\sqrt{(x^Q_{34}v_\phi)^2+2(M_4^Q)^2}},\qquad m_s\approx \frac{y^d_{24}x^d_{42}v_\phi v_d}{2\,M_4^d}\label{eq:massbot}\\
m_\tau &\approx & \frac{1}{\sqrt{2}}\frac{y^e_{43}x^L_{34}v_\phi v_d}{\sqrt{(x^L_{34}v_\phi)^2+2(M_4^L)^2}},\qquad m_\mu\approx \frac{y^e_{24}x^e_{42}v_\phi v_d}{2\,M_4^e}\,.\label{eq:masstau}
\end{eqnarray}
While Eqs.~(\ref{eq:masstop})-(\ref{eq:masstau}) allow determination of the SM fermion masses with an accuracy within a factor of $2-3$ only, they can be used to gain intuition of which NP Yukawa couplings play a dominant role in establishing the correct masses of particular fermions. For example, large $x^Q_{34}$ and $y^u_{43}$ are expected to fit $m_t$, while $y^u_{24}\ll 1$ or $x^u_{42}\ll 1$ would be required to suppress the charm mass. Similarly, large $y^d_{43}$ is needed to generate $m_b=4.18$ GeV. Additionally, in order to obtain the correct value of the top quark mass, the singlet scalar vev $v_\phi$ should be of the same order as the VL mass parameter $M_4^Q$. Note, however, that in our phenomenological analysis we always perform the numerical diagonalization of the mass matrices (\ref{eqn:mm_lep_1}), (\ref{eqn:mm_up_1}) and (\ref{eqn:mm_down_1}).

One important observation which can be deduced from Eqs.~(\ref{eq:masstop}) and (\ref{eq:massbot}) is that the ratio of the top and bottom masses, $m_t/m_b\approx 34$, puts relevant constraints on the allowed parameter space of the model. In fact, we have
\begin{equation}\label{eq:mtmbrat}
\frac{m_t}{m_b}\approx \frac{y_{43}^u}{y_{43}^d}\tan\beta\,.    
\end{equation}
The relation (\ref{eq:mtmbrat}) leads to two distinct classes of solutions. In the first one, with both the Yukawa couplings of order one, $\tan\beta\sim\mathcal{O}(10)$ is required. In the other one, with $\tan\beta\sim\mathcal{O}(1)$,  a large hierarchy between the up and down sector couplings, $y^d_{43}\ll y^u_{43}$, must be imposed.

The masses of VL fermions are given, to a very good approximation, by the corresponding VL mass parameters with small contributions stemming from their mixing with the second and the third family,
\begin{eqnarray}M_{U_1}&\approx&\sqrt{ (M^Q_4)^2 + \frac{1}{2}(v_\phi x^Q_{34})^2-\frac{(M_4^Q y^u_{43} v_u)^2}{(x^Q_{34}v_\phi)^2+2(M_4^Q)^2}}\label{eq:massbsmu1}\\
M_{U_2}&\approx&\sqrt{(M^u_4)^2 + \frac{1}{2}(v_\phi x^u_{43})^2+\frac{1}{2}(v_\phi x^u_{42})^2+\frac{2(M^u_4 y^u_{43} v_u)^2}{2(M^u_4)^2 + (v_\phi x^u_{43})^2+(v_\phi x^u_{42})^2}}\label{eq:massbsmu2}\\
M_{D_1}&\approx&\sqrt{(M^Q_4)^2 + \frac{1}{2}(v_\phi x^Q_{34})^2},\qquad M_{D_2}=\sqrt{(M^d_4)^2 + \frac{1}{2}(v_\phi x^d_{43})^2+\frac{1}{2}(v_\phi x^d_{42})^2}\label{eq:massbsmd}\\
M_{E_1}&\approx&\sqrt{(M^L_4)^2 + \frac{1}{2}(v_\phi x^L_{34})^2},\qquad M_{E_2}=\sqrt{(M^e_4)^2 + \frac{1}{2}(v_\phi x^e_{43})^2+\frac{1}{2}(v_\phi x^e_{42})^2}.\label{eq:massbsme}
\end{eqnarray}
In the neutrino sector, the corresponding mass matrix is $7\times 7$ and its explicit form can be found in Eq.~(\ref{eqn:mm_neutrino_1}). The resulting  masses of the heavy neutrinos read 
\begin{equation}
M_{N_1}=M_{N_2}\approx M^\nu_4,\qquad 
M_{N_3}=M_{N_4} \approx \sqrt{(M^L_4)^2 + \frac{1}{2}(v_\phi x^L_{34})^2}.\label{eq:massbsmn}
\end{equation}
By comparing Eq.~(\ref{eq:massbsmn}) with Eq.~(\ref{eq:massbsme}), we can pinpoint two generic features of the model considered in this study: heavy neutrinos $N_{1,2}$ are the lightest VL leptons in the spectrum, while the pair $N_{3,4}$ is mass-degenerate (at the tree level) with the charged VL lepton $E_1$. We will later see that this mass pattern has important consequences for the resulting phenomenology. 

As a final remark, let us notice that one complete VL family allows us to give masses to the second and third family of the SM fermions only. To generate the masses for the first family as well, one extra VL family is required (for an example of such a construction, see Ref.~~\cite{Hernandez:2021tii}). Since such an extension would only increase the number of free parameters in the model without affecting any phenomenological findings, in this study we limit ourselves to its most economical version.

\subsection{CKM mixing matrix} \label{seq:ckm}

The full $5\times 5$ mixing matrix takes the following form~\cite{CarcamoHernandez:2021yev}:
\begingroup
\begin{equation}
    V_{\func{mixing}} = V_{L}^{u}. \func{diag}\left( 1, 1, 1, 1, 0 \right).V_{L}^{d\dagger}
    \label{eqn:CKM_Analytic_definition}
\end{equation}
\endgroup
where $V_{L}^{u}$ and $V_{L}^{d}$ are the left-handed mixing matrices of Eqs.~(\ref{eq:mudiag}) and~(\ref{eq:mddiag}) which diagonalize the up- and down-type quark mass matrices $\mathcal{M}_u$ and $\mathcal{M}_d$. The zero element of the matrix (\ref{eqn:CKM_Analytic_definition}) indicates the fact that the singlet VL quarks do not interact with the SM gauge bosons $W^{\pm}$. Following the strategy of Ref.~\cite{CarcamoHernandez:2021yev} and working under the assumption that $v_{u,d}/M_4^{Q,u,d}\ll 1$, we can approximate the $3\times3$ CKM matrix as 
\begingroup
\begin{equation} 
    V_{\func{CKM}}^{3\times3} \approx \left(
\begin{array}{ccc}
 1-x_{ud}^2/2 & x_{ud} & x_{ud} x_d \\
 -x_{ud} & 1- x_{ud}^2 /2 & x_d-x_u \\
 -x_u x_{ud} & x_u-x_d & 1 \\
\end{array}
\right)\,, 
    \label{eqn:CKM_Analytic}
\end{equation}
\endgroup
where 
\begingroup
\begin{equation} 
 x_d= \frac{y^d_{24} x^d_{43} M_4^Q}{y^d_{43} x^Q_{34} M_4^d}\,, \qquad \qquad \qquad x_u= \frac{y^u_{24} x^u_{43} M_4^Q}{y^u_{43} x^Q_{34} M_4^u }\,, \qquad \qquad \qquad x_{ud}= \frac{y^d_{14}}{y^d_{24}} \, .
\end{equation}
\endgroup
Based on the conclusions from Sec.~\ref{sec:hirer} one expects $x_u,x_d\ll 1$. Note also that:
\begin{itemize}
    \item The element $V_{us}$ of the CKM matrix is given by  $y_{14}^d/y^d_{24}$ in our model. The presence of a non-zero coupling $y_{14}^d$ is thus crucial to generate the Cabibbo angle of the right size. We also expect $y^d_{14}\approx 0.22\, y^d_{24}.$
    \item The correct value of the element $V_{ud}$ is generated automatically once the Cabibbo angle is set.  
    \item To reproduce the correct value of the element $V_{ub}$, one needs $x_d\approx 0.017$. It then follows that $x_u\approx -0.023$ is required in order to fit the element $V_{cb}$ (it also implies $x^u_{43}$ of order one).
    \item The only element of the CKM matrix that can not be accurately reproduced is $V_{td}$.
\end{itemize}

To analzye this issue more quantitatively, it is convenient to rewrite the CKM matrix~(\ref{eqn:CKM_Analytic}) in terms of the Wolfenstein parameters \cite{Wolfenstein:1983yz}. Defining, for example, 
\begingroup
\begin{equation}\label{eq:xuwolf}
 x_d = A \, \lambda^3 \, \sqrt{\eta^2 + \rho^2}\,, \qquad \qquad
 x_u = A \, \lambda^2 (\sqrt{\eta^2 + \rho^2} \, -1)\,, \qquad\qquad x_{ud} = \frac{x_u \, x_d}{\lambda} \, ,
\end{equation}
\endgroup
one obtains
\begin{equation}
    |V_{\func{CKM}}^{3\times3}| = \left(
\begin{array}{ccc}
 1 - \lambda^2/2 & \lambda & A \, \lambda^3 \, \sqrt{\eta^2 + \rho^2} \\
 \lambda & 1 - \lambda^2/2 & A \, \lambda^2 \\
 A \, \lambda^3 \left( 1- \sqrt{\eta^2 + \rho^2} \right) & A \, \lambda^2 & 1 \\
\end{array}
\right)
+ \mathcal{O}(\lambda^4)\,.
    \label{eqn:CKM_Analytic_Wolfenstein}
\end{equation}
Plugging the Wolfenstein parameters extracted from the global fit~\cite{ParticleDataGroup:2022pth} 
into Eq.~(\ref{eqn:CKM_Analytic_Wolfenstein}) and comparing it with the experimental determination of the CKM matrix elements reported in Ref.~\cite{ParticleDataGroup:2022pth}, one can estimate to what extent the measured structure of the CKM matrix can be reproduced in our model. One obtains
\begingroup
\begin{equation}    \frac{|V_{\func{CKM}}^{\func{exp}}|-|V_{\func{CKM}}^{3\times3}|}{\delta |V_{\func{CKM}}^{\func{exp}}|} = \left(
\begin{array}{ccc}
 0 & 0 & 0 \\
 0 & 0.04 & 0 \\
 8.88 & 0.23 & 0.01 \\
\end{array}
\right)\,.
    \label{eqn:chi_CKM_Analytic_Wolfenstein}
\end{equation}
\endgroup
It results from Eq.~(\ref{eqn:chi_CKM_Analytic_Wolfenstein}) that in the framework of our model we may not be able to correctly reproduce all the elements of the CKM matrix (this observation will be later confirmed by our numerical scan). 
\footnote{Note that modifying the definitions of the parameters $x_u$, $x_d$ and $x_{ud}$ in Eq.~(\ref{eq:xuwolf}), one could be able to fit better the element $V_{td}$, but at the price of losing the accuracy in reproducing $V_{cb}$.} Once more, this issue could be solved by introducing an extra VL family.

To conclude this section, we would like to stress again that the approximation adopted in the foregoing discussion hinges on the assumption of the specific mass hierarchy in the NP sector, which may not be entirely fulfilled. Therefore, in the phenomenological analysis we will be always calculating all the elements of the CKM matrix numerically.

\section{Scalar potential constraints}\label{sec:scalar}

In this section, we discuss the constraints stemming from the scalar potential of the model. In particular, we define the alignment limit of the SM-like Higgs boson, we derive the conditions for the scalar potential to be bounded from below in the presence of three independent scalar fields, and we verify whether the electroweak (EW) vacuum is stable.

\subsection{Scalar masses in the alignment limit}\label{sec:App_sp}

In the interaction basis, the most generic renormalizable scalar potential of the model defined in Table~\ref{tab:BSM_model} takes the form \cite{Hernandez:2021tii}:
\begin{equation}\label{eq:scalpot}
\begin{split}
V &= \mu_{u}^{2} (H_{u}^{\dagger} H_{u}) + \mu_{d}^{2} (H_{d}^{\dagger} H_{d}) + \mu_{\phi}^{2} (\phi^{*} \phi)-\frac{1}{2} \mu_{\func{sb}}^{2} \left( \phi^{2} + \phi^{*2} \right)
\\
&+\frac{1}{2} \lambda_{1} (H_{u}^{\dagger} H_{u})^{2} + \frac{1}{2} \lambda_{2} (H_{d}^{\dagger} H_{d})^{2} + \lambda_{3} (H_{u}^{\dagger} H_{u}) (H_{d}^{\dagger} H_{d}) + \lambda_{4} (H_{u}^{\dagger} H_{d}) (H_{d}^{\dagger} H_{u})
\\
&-\frac{1}{2} \lambda_{5} (\epsilon_{ij} H_{u}^{i} H_{d}^{j} \phi^{2} + \func{H.c.})+\frac{1}{2} \lambda_{6} (\phi^{*}\phi)^{2} + \lambda_{7} (\phi^{*}\phi)(H_{u}^{\dagger}H_{u}) + \lambda_{8} (\phi^{*}\phi)(H_{d}^{\dagger}H_{d}),
\end{split}
\end{equation}
where $\mu_{u,d,\phi}^{2}$ are dimensionful mass parameters, $\lambda_{1,2,\cdots,8}$ denote dimensionless quartic coupling constants, 
and $\mu^2_{\func{sb}}$ is an extra mass term which softly violates the global U(1)$_X$ symmetry. The main reason to introduce the latter is to prevent
a massless Goldstone boson of the spontaneously broken U(1)$_X$ to appear in the spectrum. As we will see below, the soft-breaking term does not affect the CP-even and the charged scalar masses since it only enters the mass matrix of the pseudoscalars. 

Expanding the fields $H_u$, $H_d$ and $\phi$ around their vacuum states, 
\begin{eqnarray}\label{eq:higgses}
H_{u} = 
\begin{pmatrix}
H_{u}^{+} \\
\frac{1}{\sqrt{2}} \left( v_{u} + \func{Re} H_{u}^{0} + i\func{Im} H_{u}^{0} \right)
\end{pmatrix},\qquad\nonumber
H_{d} = 
\begin{pmatrix}
\frac{1}{\sqrt{2}} \left( v_{d} + \func{Re} H_{d}^{0} + i\func{Im} H_{d}^{0} \right) \\
H_{d}^{-}
\end{pmatrix},\\
\phi = \frac{1}{\sqrt{2}} \left( v_{\phi} + \func{Re} \phi + i\func{Im} \phi \right),\qquad\qquad\qquad\qquad\qquad\qquad
\end{eqnarray}
where the vevs are defined in Eq.~(\ref{eq:ewsbmin}), one can use the minimization conditions for the scalar potential (\ref{eq:scalpot}) to express the dimensionful mass parameters in terms of the quartic couplings and the vevs,
\begin{equation}
\begin{split}
\mu_{u}^{2} &= -\frac{1}{2} \left(\lambda_{1} v_{u}^{2} +  \lambda_{3} v_{d}^{2} + \lambda_{7} v_{\phi}^{2} \right) -\frac{1}{4} \lambda_{5} \left( \frac{v_{d}}{v_{u}} \right) v_{\phi}^{2},\\
\mu_{d}^{2} &= -\frac{1}{2} \left(\lambda_{2} v_{d}^{2} + \lambda_{3} v_{u}^{2} + \lambda_{8} v_{\phi}^{2} \right) -\frac{1}{4} \lambda_{5} \left( \frac{v_{u}}{v_{d}} \right) v_{\phi}^{2},
\\
\mu_{\phi}^{2} &= -\frac{1}{2} \left( \lambda_{6} v_{\phi}^{2}+  \lambda_{5} v_d v_u + \lambda_{7}v^2_{u} + \lambda_{8} v_{d}^{2} \right) + \mu_{\func{sb}}^{2}\,.
\label{eqn:dimensionful_parameters_global}
\end{split}
\end{equation}
One must have 
\begin{equation}\label{eq:mucond}
\mu^2_u<0\,,\qquad \mu_d^2<0\,,\qquad \mu_\phi^2<0\,.
\end{equation}
in order to generate the non-zero vevs for all the scalar fields.

The explicit forms of the scalar mass matrices derived from the potential (\ref{eq:scalpot}) are collected in Appendix~\ref{sec:App_scal}.
The real parts of the scalar fields, $\func{Re} H_{u}^{0}$, $\func{Re} H_{d}^{0}$ and $\func{Re} \phi$, account for three CP-even Higgs bosons.  The corresponding mass matrix $\mathbf{M}_{\func{CP-even}}^{2}$ (see Eq.~(\ref{eq:m2even})) can be diagonalized with a mixing matrix $R_h$ defined in Eq.~(\ref{eq:rh}). 
The masses of three physical neutral scalars, $h_1$, $h_2$ and $h_3$, correspond to the eigenvalues of $\mathbf{M}_{\func{CP-even}}^{2}$,
\begin{equation}\label{eq:rh1}
{\rm diag}\{M_{h_1}^2,M_{h_2}^2,M_{h_3}^2\}=R_h (\mathbf{M}_{\func{CP-even}}^{2}) R_h^T\,.
\end{equation}

In the following, we will want to identify the SM Higgs boson with the lightest neutral scalar  $h_1$. To this end, we choose to work in the so-called alignment limit, defined as a set of constraints on the quartic couplings $\lambda_i$ under which $h_1$ features the same tree-level couplings with the SM particles as the SM Higgs. We show in Appendix~\ref{sec:App_scal} that this assumption requires
\begin{eqnarray}
\lambda_8\,\cos^2\beta+\lambda_7\,\sin^2\beta+\lambda_5\,\sin\beta \cos\beta&=&0\label{eq:al1}\\
\lambda_2\,\cos^2\beta-\lambda_1\,\sin^2\beta-\lambda_3(\cos^2\beta-\sin^2\beta)&=&0\,,\label{eq:al2}
\end{eqnarray}
where the equality imposes a perfect alignment condition.
The masses of the CP-even scalars in the alignment limit read
\begin{eqnarray}
M^2_{h_1}&=&v^2\left(\lambda_1\, \sin^2\beta+\lambda_3\,\cos^2\beta \right)\label{eq:mh1}\\
M^2_{h_2}&=&\lambda_6\,v^2_\phi-\frac{1}{8\sin\beta \cos\beta}\left(B_{23}+\sqrt{4A_{23}^2+B_{23}^2}\right)  \\
M^2_{h_3}&=&\lambda_6\,v^2_\phi-\frac{1}{8\sin\beta \cos\beta}\left(B_{23}-\sqrt{4A_{23}^2+B_{23}^2}\right)\,,
\end{eqnarray}
with $A_{23}$ and $B_{23}$ defined in Eq.~(\ref{eq:a23}) and Eq.~(\ref{eq:b23}), respectively, and $v=\sqrt{v_u^2+v_d^2}=246$ GeV.

The CP-odd scalar mass matrix in the basis $\left( \func{Im} H_{u}^{0},\, \func{Im} H_{d}^{0},\, \func{Im} \phi \right)$, $\mathbf{M}_{\func{CP-odd}}^{2}$, 
is defined in Eq.~(\ref{eq:m2odd}). 
After the diagonalization, the physical CP-odd spectrum consists of one massless Goldstone boson and two massive pseudoscalars, $a_1$ and $a_2$,
\begin{equation}\label{eq:Ra}
{\rm diag}\{0,M_{a_1}^2,M_{a_2}^2\}=R_a (\mathbf{M}_{\func{CP-odd}}^{2}) R_a^T\,,
\end{equation}
with the masses given by
\begin{eqnarray}\label{eq:ra1}
M^2_{a_1}&=&-\frac{\lambda_5}{2\sin 2\beta}\left(v^2\sin^2 2\beta+v^2_\phi\right)\label{eq:massa1}\\
M^2_{a_2}&=&2\,\mu_{\func{sb}}^{2}\,.
\label{eq:massa2}
\end{eqnarray}
Note that $\lambda_5<0$ and $\mu_{\func{sb}}^{2}>0$ are required to guarantee the positivity of $M^2_{a_1}$ and $M^2_{a_2}$. 

Finally, the charged scalar mass matrix in the basis $\left( H_{u}^{\pm}, H_{d}^{\pm} \right)$, $\mathbf{M}_{\func{Charged}}^{2}$, is defined in Eq.~(\ref{eq:m2char}).
After the diagonalization with a mixing matrix $R_\beta$, one is left with a massless charged Goldstone boson and a charged Higgs boson, 
\begin{equation}
{\rm diag}\{0,M_{h^\pm}^2\}=R_\beta (\mathbf{M}_{\func{Charged}}^{2}) R_\beta^T\,.
\end{equation}
The corresponding mass reads in this case
\begin{equation}
M^2_{h^\pm}=\frac{\lambda_4 v^2}{2}-\frac{\lambda_5 v_\phi^2}{2\sin 2\beta}\,.
\end{equation}

As a closing remark, let us notice that the alignment condition (\ref{eq:al2}) indicates 
\begin{equation}\label{alignment_lambda2}
\lambda_2=\lambda_3+\tan^2\beta(\lambda_1-\lambda_3)\,.
\end{equation}
In order to preserve the perturbativity of $\lambda_2$ (more on this in Sec.~\ref{sec:RG}), the term in parentheses needs to be fine-tuned with a precision $\mathcal{O}(1/\tan^2\beta)$ or better, effectively fixing $\lambda_3\approx \lambda_1$ with the same accuracy. On the other hand, Eq.~(\ref{eq:mh1}) implies that we can identify $\lambda_1$ with the quartic coupling of the SM, $\lambda_1=0.258$, as long as $\tan\beta\gtrsim 3$. 
Similarly, the alignment condition (\ref{eq:al1}) gives
\begin{equation}\label{alignment_lambda8}
\lambda_8=-\tan \beta\left( \lambda_{7}\tan\beta  + \lambda_{5}\right)\,.
\end{equation}
Perturbativity of $\lambda_8$ then requires $\lambda_7\sim\mathcal{O}(1/\tan^2\beta)$ and $\lambda_5\sim\mathcal{O}(1/\tan\beta)$. 

\subsection{Bounded-from-below limits} \label{subsec:bound}

To guarantee that the minimum around which we expand the scalar potential (\ref{eq:scalpot}) is physically meaningful, we must ensure that the  potential is bounded from below, which means that it cannot tend to negative infinity along any direction in the field space. This requirement puts additional restrictions on the allowed values of the couplings $\lambda_i$. To derive  the `bounded-from-below' constraints, one should analyze all possible directions along which the scalar fields $H_u$, $H_d$ and $\phi$ can flow towards arbitrarily large values. The details of our derivation are presented in Appendix~\ref{sec:App_vs}. Here we summarize our findings in the form of inequality conditions which need to be satisfied by the quartic couplings of the potential (\ref{eq:scalpot}): 
\begin{equation}
    \begin{split}
        \lambda_{8} + \sqrt{\lambda_{2} \lambda_{6}} > 0 \\
        \lambda_{7} + \sqrt{\lambda_{1} \lambda_{6}} > 0 \\
        \lambda_{3} + \sqrt{\lambda_{2} \lambda_{1}} > 0 \\
        \lambda_{3} + \lambda_{4} + \sqrt{\lambda_{2} \lambda_{1}} > 0 \\
        -\frac{1}{4} \frac{(\func{Re} \lambda_{5})^{2} +(\func{Im} \lambda_{5})^{2}}{\lambda_{a}} + \lambda_{4} > 0 \\
        4 \lambda_{b}^{2} - (\func{Re} \lambda_{5})^{2} + \func{Re} \lambda_{5} \func{Im} \lambda_{5} > 0 \\
        4 \lambda_{b}^{2} - (\func{Im} \lambda_{5})^{2} + \func{Re} \lambda_{5} \func{Im} \lambda_{5} > 0
        \label{eqn:VS_conditions}
    \end{split}
\end{equation}
where $\lambda_{a} = \frac{3}{2} \lambda_{6} + \lambda_{3} \frac{\lambda_{6}}{\sqrt{\lambda_{1} \lambda_{2}}} + \lambda_{7} \sqrt{\frac{\lambda_{6}}{\lambda_{1}}} + \lambda_{8}\sqrt{\frac{\lambda_{6}}{\lambda_{2}}}$ and $\lambda_{b} = \sqrt{\lambda_{a} \lambda_{4}}$. Since in this study we do not investigate the CP violation, we assume that all the parameters of the lagrangian are real, indicating $ \func{Im} \lambda_{5} =0$. Note also that several novel conditions w.r.t.~the findings of Refs.~\cite{Hernandez:2021tii,Lee:2022sic} are identified in Eq.~(\ref{eqn:VS_conditions}).

\subsection{Vacuum stability} \label{subsec:stab}

In theories which feature an extended scalar sector, the scalar potential can easily develop more than one local minimum. As a result, the theory may tunnel from one minimum to another. In principle, color and charge breaking minima deeper than the EWSB minimum of Eq.~(\ref{eq:ewsbmin}) can arise in our model (see, e.g.~\cite{Muhlleitner:2016mzt}). Moreover, several charge and color conserving minima can coexist, in which case we do not know {\it a priori} which of them  corresponds to the desired EWSB minimum. 

The strong vacuum stability condition for the scalar potential requires that the EWSB vacuum corresponds to a global minimum. In such a case the potential is said to be stable. If, on the other hand, the EWSB minimum is a local minimum but the tunneling time to a true global minimum exceeds the age of the Universe, the potential is said to be metastable. In this study we employ the publicly available numerical package \texttt{Vevacious++}~\cite{Camargo-Molina-OLeary:2014} (the C++ version of~\cite{Camargo-Molina:2013qva}) to find all tree- and one-loop level minima of the scalar potential defined in Eq.~(\ref{eq:scalpot}) and to calculate the tunneling time from the EWSB minimum to the deepest minimum found. 

\section{Flavor physics constraints} \label{sec:lepton}


In this section, we review additional constraints which may affect the allowed parameter space of the analyzed model. These extra restrictions come from the experimental measurements of several flavor observables, including the anomalous magnetic moment of the muon, the lepton flavor violating decays of the tau lepton, and the elements of the CKM matrix. We discuss them in the following one by one.

\subsection{Muon anomalous magnetic moment}\label{sec:gm2}

The discrepancy between the SM prediction~\cite{Czarnecki:2002nt,Melnikov:2003xd,Aoyama:2012wk,Kurz:2014wya,Davier:2010nc,Gnendiger:2013pva,Colangelo:2014qya,Davier:2017zfy,Masjuan:2017tvw,Colangelo:2017fiz,Hoferichter:2018kwz,Keshavarzi:2018mgv,Colangelo:2018mtw,Hoferichter:2019mqg,Davier:2019can,Keshavarzi:2019abf,Gerardin:2019vio,Bijnens:2019ghy,Colangelo:2019uex,Blum:2019ugy,Aoyama:2019ryr,Aoyama:2020ynm} and the experimental measurement of the anomalous magnetic moment of the muon has been confirmed separately by the Brookhaven National Laboratory~\cite{Muong-2:2006rrc} and the Fermilab experimental groups~\cite{Muong-2:2021ojo,Muong-2:2023cdq}, giving rise to the combined $5.1\,\sigma$ anomaly: 
\begin{equation}\label{eq:gmexp}
    \Delta a_{\mu} = a_{\mu}^{\func{exp}} - a_{\mu}^{\func{SM}} = \left( 2.49 \pm 0.48 \right) \times 10^{-9}.
\end{equation}

In a generic NP model which features heavy scalars $\phi_i$ and fermions $\psi_j$ coupled to the SM muons
via the Yukawa-type interactions $y_{L}^{ij}\phi_i\, \bar{\psi}_j P_L\, \mu$ and  
$y_{R}^{ij}\phi_i\, \bar{\psi}_j P_R\, \mu$ (where $P_{L,R}=(1\mp \gamma^5)/2$ are the usual projection operators), a well-known one-loop contribution to the muon anomalous magnetic moment reads
\begin{multline}\label{eq:gm2}
\Delta a_{\mu} = \sum_{i,j}\left\{
-\frac{m_\mu^2}{16\pi^2 M_{\phi_i}^2}\left(|y^{ij}_L|^2+|y^{ij}_R|^2\right)\left[Q_j\mathcal{F}_1\left(x_{ij}\right)-Q_i\mathcal{G}_1\left(x_{ij}\right)\right]\right.\\
\left.-\,\frac{m_\mu\, M_{\psi_j}}{16 \pi^2 M_{\phi_i}^2}\textrm{Re}\left(y^{ij}_L y_R^{ij\ast}\right)\left[Q_j\mathcal{F}_2\left(x_{ij}\right)-Q_i\mathcal{G}_2\left(x_{ij}\right)\right]\right\},
\end{multline}
where $M_{\phi_i}$ is the physical mass of a heavy scalar,  $M_{\psi_j}$ is the physical mass of a heavy fermion, $x_{ij}=M_{\psi_j}^2/M_{\phi_i}^2$, and the electric charges of $\phi_i$ and $\psi_j$ are related as
$Q_i+Q_j=-1$. The loop functions are defined in the following way:
\begin{equation}
    \begin{split}
\mathcal{F}_1(x)&=\frac{1}{6\left(1-x\right)^4}\left(2+3x-6x^2+x^3+6x\ln x\right)   \\  
\mathcal{F}_2(x)&=\frac{1}{\left(1-x\right)^3}\left(-3+4x-x^2-2\ln x\right)\\
\mathcal{G}_1(x)&=\frac{1}{6\left(1-x\right)^4}\left(1-6x+3 x^2+2 x^3-6 x^2 \ln x\right)\\
\mathcal{G}_2(x)&=\frac{1}{\left(1-x\right)^3}\left(1-x^2+2 x \ln x\right)\,.
\end{split}
\end{equation}
The first addend in Eq.~(\ref{eq:gm2}) captures the loop chirality-conserving contributions to $\Delta a_\mu$. These are known to be generically too small to account for the anomaly (\ref{eq:gmexp}) when the most recent LHC bounds on the NP masses are taken into account \cite{Kowalska:2017iqv,Athron:2021iuf}. We will thus focus on the second addend in Eq.~(\ref{eq:gm2}), which corresponds to the loop chirality-flipping contributions to $\Delta a_\mu$. 

In the framework of the model defined in Table~\ref{tab:BSM_model}, two classes of contributions to the anomalous magnetic moment of the muon can arise, induced by one-loop diagrams with an exchange of neutral (pseudo)scalars and charged VL leptons, as shown in Fig.~\ref{fig:muon_mass_muong2}(a), or charged scalars and neutral VL leptons, as shown in Fig.~\ref{fig:muon_mass_muong2}(b). In the first case, the chirality-flipping contributions to $\Delta a_\mu$ read
\begin{equation}
\Delta a_\mu^{Eh^0}\approx\frac{1}{16\pi^2}\sum_{j=1}^2\sum_{i=1}^3\left[
\frac{m_\mu\, M_{E_j}}{ M_{h^0_i}^2}\textrm{Re}\left(c_L c_R^{\ast}\right)^{E_j,h_i^0}\mathcal{F}_2\left(M_{E_j}^2/M^2_{h^0_i}\right)\right]
\end{equation}
for the CP-even scalars and
\begin{equation}
\Delta a_\mu^{Ea}\approx\frac{1}{16\pi^2}\sum_{j=1}^2\sum_{i=1}^2\left[
\frac{m_\mu\, M_{E_j}}{M_{a_i}^2}\textrm{Re}\left(c_L c_R^{\ast}\right)^{E_j,a_i}\mathcal{F}_2\left(M_{E_j}^2/M^2_{a_i}\right)\right]
\end{equation}
for the CP-odd scalars. 
The one-loop contributions to  $\Delta a_\mu$ from the neutral leptons and charged scalars are given by
\begin{equation}
\Delta a_\mu^{Nh^\pm}\approx-\frac{1}{16\pi^2}\sum_{j=1}^4\left[
\frac{m_\mu\, M_{N_j}}{M_{h^\pm}^2}\textrm{Re}\left(c_L c_R^{\ast}\right)^{N_j,h^\pm}\mathcal{G}_2\left(M_{N_j}^2/M^2_{h^\pm}\right)\right].
\end{equation}

The parameters $c_{L/R}$ denote the effective couplings arising from the muon-(pseudo)scalar-VL fermion vertices in the mass basis. They depend on the lepton Yukawa couplings of Eqs.~(\ref{eqn:Lag_ren}) and (\ref{eqn:Lag_neutrino}), as well as on the elements of the mixing matrices $R_h$ (Eq.~(\ref{eq:rh1})), $R_a$ (Eq.~(\ref{eq:Ra})), and $V^e_{L/R}$ (Eq.~(\ref{eq:vle})). The explicit forms of $c_{L/R}$ are rather complex and we refrain from showing them here. Note, however, that in our numerical analysis we are going to compute all the contributions to $\Delta a_\mu$ with the numerical package \texttt{SPheno}~\cite{Porod:2003um,Porod:2011nf}. 

\begingroup
\begin{figure}[t]
    \centering
    \subfloat[]{%
    \includegraphics[width=0.40\textwidth]{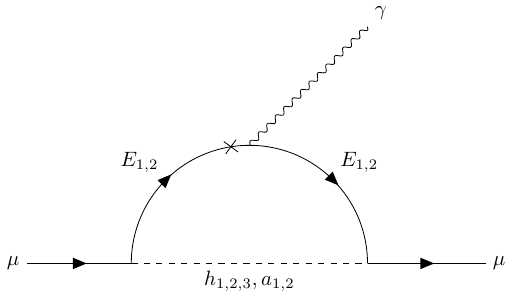}}
    \hspace{1cm}
    \subfloat[]{%
    \includegraphics[width=0.40\textwidth]{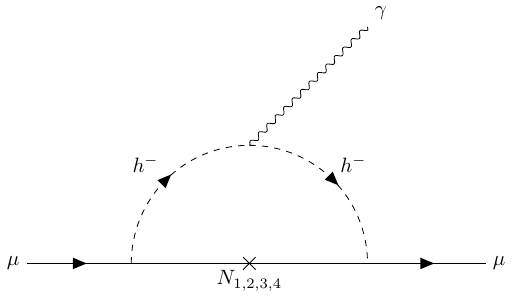}}
    \caption{The one-loop chirality-flipping contributions to $\Delta a_\mu$ mediated by (a) a neutral (pseudo)scalar/charged lepton exchange, and (b) a charged scalar/neutral lepton exchange.}\label{fig:muon_mass_muong2}
\end{figure}
\endgroup

\subsection{Lepton flavor violating decays}

Due to the non-zero mixing between the second and the third generation of fermions, charged lepton flavour violating processes may occur. The $\tau \rightarrow \mu \gamma$ decay receives contributions from the one-loop diagrams analogous to those of $\Delta a_\mu$. The corresponding branching ratio (BR) is given by~\cite{Mandal:2019gff}
\begin{equation}\label{eq:taumug}
{\rm BR}(\tau\to\mu\gamma) =\frac{\alpha_{\rm{em}}m_\tau^3}{4\,\Gamma_\tau} \sum_{i,j}\left(|A_L^{ij}|^2+|A_R^{ij}|^2\right)\,,
\end{equation}
where $\Gamma_\tau=2.3\times 10^{-12}$~\cite{ParticleDataGroup:2022pth} indicates the total decay width of the tau, $\alpha_{\rm{em}}$ is the fine structure constant, and the decay amplitude $A_L^{ij}$ reads
\begin{multline}\label{eq:ampA}
A_L^{ij} = \frac{1}{32\pi^2 M_{\phi_i}^2}\left\{m_\tau\left(y^{ij}_{\tau, L} y^{ij\ast}_{\mu,L}\right)\left[Q_j\mathcal{F}_1\left(x_{ij}\right)-Q_i\mathcal{G}_1\left(x_{ij}\right)\right]\right.\\
\left.+M_{\psi_j}\left(y^{ij}_{\tau,L} y^{ij\ast}_{\mu,R}\right)\left[Q_j\mathcal{F}_2\left(x_{ij}\right)-Q_i\mathcal{G}_2\left(x_{ij}\right)\right]\right\}\,.
\end{multline}
The corresponding amplitude $A_R^{ij}$ is obtained from Eq.~(\ref{eq:ampA}) by replacing $L\leftrightarrow R$. Just like it was in the $\Delta a_\mu$ case, the main contribution to ${\rm BR}(\tau\to\mu\gamma)$ originates from the second addend in Eq.~(\ref{eq:ampA}).
The current experimental 90\% confidence level (C.L.) upper bound on $ \func{BR}\left( \tau \rightarrow \mu \gamma \right)$ from the Belle collaboration reads~\cite{Belle:2021ysv}:
\begingroup
\begin{equation}
    \func{BR}\left( \tau \rightarrow \mu \gamma \right)_{\rm{exp}} <4.2 \times 10^{-8}\,.
\end{equation}
\endgroup

The $\tau\to 3\,\mu$ decay can proceed through the one-loop penguin and box diagrams. The latter are subdominant in our model as they do not receive the chiral enhancement. The corresponding formulae for the penguin-diagram BRs are lengthy and not particularly enlightening. They can be found, for example, in Eq.~(37) of Ref.~\cite{Mandal:2019gff}. The 90\%~C.L. upper bound on $\func{BR}\left( \tau \rightarrow 3\,\mu\right)$ by the Belle collaboration reads~\cite{Hayasaka:2010np}:
\begingroup
\begin{equation}
    \func{BR}\left( \tau \rightarrow 3\,\mu\right)_{\rm{exp}} <2.1 \times 10^{-8}\,.
\end{equation}
\endgroup

\subsection{CKM anomaly} \label{sec:ckmanomal}

Among the experimental puzzles which are not explained by the SM we should also mention various tensions between three different determinations of the Cabibbo angle. This observable can be extracted from the short distance radiative corrections to the $\beta$ decay, from the experimental data on kaon decays, and from the lattice calculations \cite{Grossman:2019bzp,Belfatto:2023tbv,Crivellin:2020lzu,Czarnecki:2019iwz}. All these measurements are in tension with each other, giving rise to two interesting anomalies. 

The first anomaly is related to the violation of the CKM matrix unitarity when one compares the values of $\bigl|V_{ud}\bigr|$ and $\bigl| V_{us}\bigr|$ resulting from the $\beta$ decay and from the kaon decays. The second anomaly originates from two different measurements of $\bigl|V_{us}\bigr|$: from the semileptonic $K\rightarrow \pi l \nu$ and the leptonic $K\rightarrow \mu\nu$ decay, respectively.  

The experimental upper bound on the CKM deviation from the unitarity reads~\cite{ParticleDataGroup:2022pth}
\begingroup
\begin{equation}
    \Delta_{\func{CKM}} = \sqrt{1-V_{ud}^{2}-V_{us}^{2}-V_{ub}^{2}} < 0.05.
\label{eqn:CKM_non_unitarity}
\end{equation}
\endgroup
To explain the anomaly of Eq.~(\ref{eqn:CKM_non_unitarity}), one can consider extensions of the SM in which the fermion sector is enlarged by VL quarks mixing at the tree level with the SM quarks~\cite{Belfatto:2023tbv,Cheung:2020vqm,Branco:2021vhs,Albergaria:2023vls}. In such a setting 
deviations from the unitarity of the three-dimensional CKM matrix can  arise quite naturally. Since the model defined in Table~\ref{tab:BSM_model} contains all the necessary ingredients to account for the CKM anomaly, we include it in our list of constraints. 

\section{Perturbativity constraints} \label{sec:RG}

The model defined in Table~\ref{tab:BSM_model} is intended as a phenomenological scenario which correctly describes the physics around the energy scale determined by the typical masses in the NP sector. Nevertheless, it is important to understand what is the range of validity of such a model or, in other words, what is the energy scale at which the model can not be trusted anymore and should be embedded in some more fundamental UV completion. While such a  ``cut-off'' scale lacks a truly rigorous definition, one can try to estimate it by simply requiring that whatever extra degrees of freedom emerge in the theory above this scale to make the model UV complete, they do not affect  its phenomenological predictions. 

As an example, let us consider the muon anomalous magnetic moment operator, which in the low-energy effective filed theory (EFT) reads
\begin{equation}\label{eq:eftg2}
\frac{e}{2\,m_\mu}\Delta a_\mu\left(\bar{\mu}\,\sigma_{\mu\nu}\,F^{\mu\nu}\mu\right)\equiv\frac{C}{\Lambda}\left(\bar{\mu}\,\sigma_{\mu\nu}\,F^{\mu\nu}\mu\right)\,.
\end{equation}
Here $\Lambda$ is a cut-off scale of the examined EFT while $C$ denotes a generic Wilson coefficient. Note that since the operator in Eq.~(\ref{eq:eftg2}) is chirality flipping, it is more convenient to define $C=\tilde{C}\,m_\mu/\Lambda$.
One can now derive from Eq.~(\ref{eq:eftg2}) rough estimates of the energy scale associated with a hypothetical NP contributing to $\Delta a_\mu$ at different loop orders,
\begin{eqnarray}
\textrm{tree level}: &\tilde{C}\approx 1,\qquad\quad\,\,\,\,&\Lambda\approx 3000\,\,\textrm{GeV}\\
\textrm{1 loop}: &\tilde{C}\approx 1/16\pi^2,\quad\,&\Lambda\approx 230\,\,\textrm{GeV}\\
\textrm{2 loop}:& \tilde{C}\approx (1/16\pi^2)^2,&\Lambda\approx 20\,\,\textrm{GeV}
\end{eqnarray}
and so on. 

Going beyond the EFT approximation, let us investigate a one-loop chirality flipping contribution to $\Delta a_\mu$ like the one in Eq.~(\ref{eq:gm2}). Assuming that it arises from an unspecified UV completion of our model above the scale $\Lambda$, it can be estimated by the corresponding UV mass $m_{\Lambda}$ and the UV Yukawa couplings $y_{L/R}(\Lambda)$ as
\begin{equation}\label{eq:g2uv}
\Delta a_\mu^\Lambda\sim\frac{1}{16\pi^2}\frac{m_\mu\,v}{m_{\Lambda}^2}y_L(\Lambda)\,y_R(\Lambda).
\end{equation}
By demanding that the new contribution (\ref{eq:g2uv}) does not shift our phenomenological predictions for $\Delta a_\mu$ by more than $3\,\sigma$, we can derive a lower bound on the UV mass,
\begin{equation}\label{eq:massrange}
m_{\Lambda}\gtrsim\sqrt{y_L(\Lambda)y_R(\Lambda)}\, 15\, \textrm{TeV}\,.
\end{equation}
For the Yukawa couplings at the upper edge of perturbativity, $y_L(\Lambda)=y_R(\Lambda)=\sqrt{4\pi}$, Eq.~(\ref{eq:massrange}) translates into a conservative estimation of the scale of validity of our phenomenological model, 
\begin{equation}\label{eq:valid}
m_{\Lambda}\gtrsim 50\,\,\textrm{TeV}. 
\end{equation}
In other words, the model can not be UV completed below $m_{\Lambda}$.

An immediate consequence of Eq.~(\ref{eq:valid}) is that the commonly employed perturbativity bounds, which read $\lesssim\sqrt{4\pi}$ for the gauge/Yukawa and $\lesssim4\pi$ for the quartic couplings, need to be imposed on the running parameters of the model evaluated at the scale $\Lambda$ rather than on the bare couplings of the lagrangian (as it was done, for example, in Ref.~\cite{Hernandez:2021tii}).

To implement the RG-based perturbativity constraints, we follow the RG flow of all the coupling constants from the scale $\mu_0=1.5$ TeV, which is a proxy for the NP scale in our model, to $\Lambda=50$ TeV.
The one-loop RG equations (RGEs) were computed using the publicly available numerical code \texttt{SARAH}~\cite{Staub:2013tta,Staub:2015kfa} and are summarized in Appendix~\ref{sec:App_RG}. Due to a large number of Yukawa and quartic interactions in our model, it is not possible to perform the perturbativity analysis in a generic way as the RGEs are non-linear differential equations that can not be solved analytically. On the other hand, the perturbativity bounds are expected to be relevant only for those couplings whose values must be of order $1$ (or larger) for phenomenological reasons. This observation allows us to reduce the RGE system and to simplify the analysis.

In the Yukawa sector, the couplings of interest are $x^Q_{34}$, $y^u_{43}$, $y^u_{34}$ and $x^u_{43}$ (see Sec.~\ref{sec:setup} for the discussion). We find that the modulus of their value cannot exceed $1.4$ at $\mu_0$ if they are to remain perturbative up to 50 TeV. This conclusion is derived under the assumption that all the other couplings (but two) are set to $1$ at the initial scale $\mu_0$. The two exceptions are $y^d_{14}$ and $y^e_{24}$ (expected to be much smaller than 1 as the Yukawas of the second generation), whose values at $\mu_0$ are set to $0.7$.

In the scalar sector, the perturbativity bounds are presumably most relevant for the couplings $\lambda_1$, $\lambda_6$ and $\lambda_7$, whose RGEs feature a power-four dependence on the large Yukawa couplings $y^u_{43}$, $x^u_{43}$ and $x^Q_{34}$ (cf. Eq.~(\ref{eq:betal1}), Eq.~(\ref{eq:betal6}) and Eq.~(\ref{eq:betal7}), respectively). 
In Fig.~\ref{fig:RG_1st} we illustrate the RG running of $\lambda_1$, $\lambda_6$ and $\lambda_7$ for a randomly chosen benchmark point which satisfies all the constraints discussed in Secs.~\ref{sec:setup} and~\ref{sec:scalar}. The running of all the remaining quartic couplings is very slow in the considered energy range and does not pose any danger from the point of view of their perturbativity. Once the whole system is analyzed with the alignment conditions (\ref{eq:al1}) and~(\ref{eq:al2}) in place, it turns out that the perturbativity requires the modulus of the quartic couplings to be smaller than 2. 
A straightforward consequence of this result is that all the benchmark points found previously in Ref.~\cite{Hernandez:2021tii} are disfavored. 

\begingroup
\begin{figure}[t]
\centering
\includegraphics[width=0.6\textwidth]{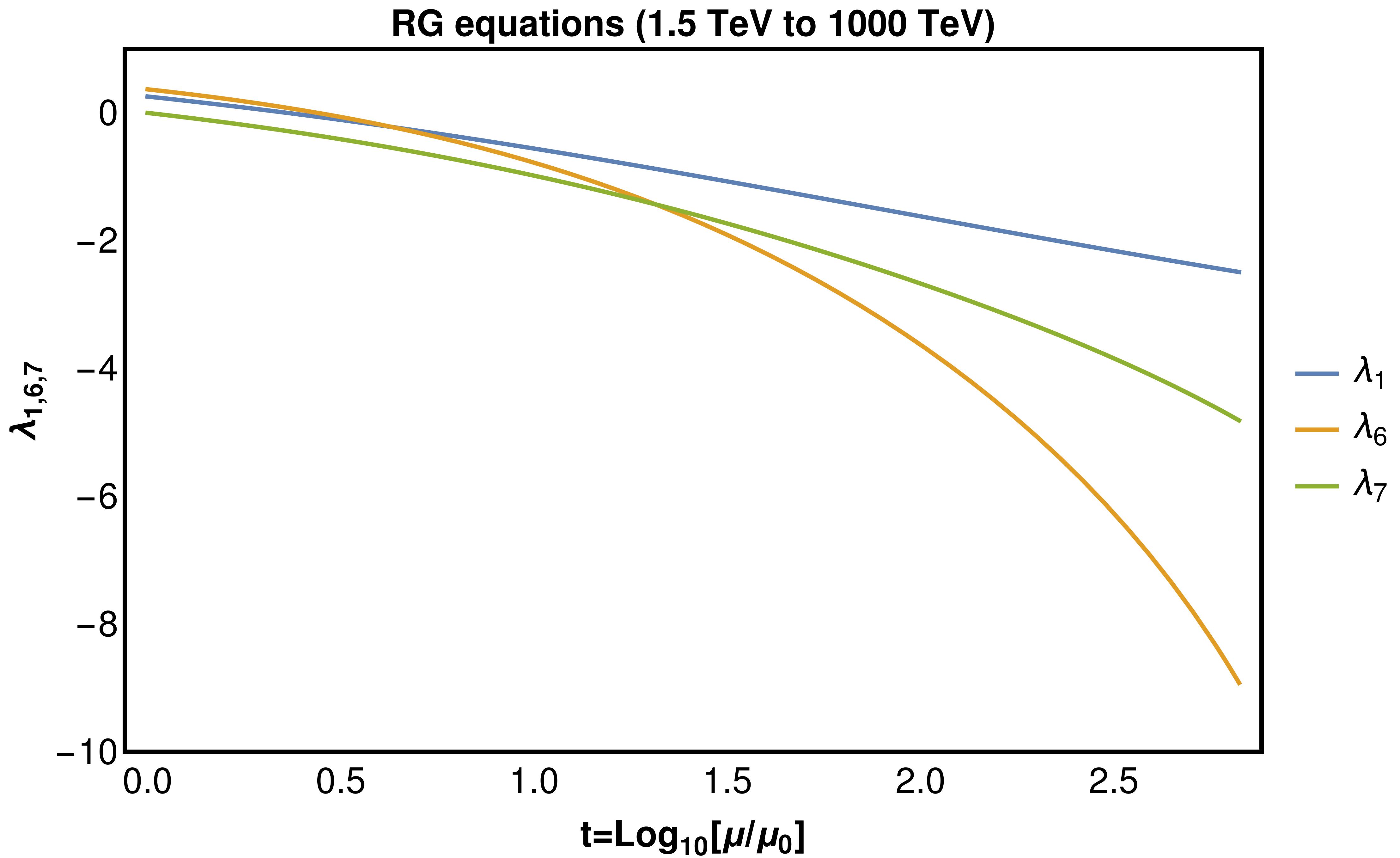}
\caption{The RG running of the quartic couplings $\lambda_1$, $\lambda_6$ and $\lambda_7$ for a randomly chosen benchmark point which satisfies all the constraints discussed in Secs.~\ref{sec:setup} and~\ref{sec:scalar}.  The renormalization scale $\mu$ ranges from $1.5$ TeV to $1000\func{TeV}$. $\mu_0=1.5$ TeV is a reference scale. We do not show the RG evolution of other quartic and Yukawa couplings as it is very slow in the considered energy range.}
\label{fig:RG_1st}
\end{figure}
\endgroup

Finally, let us comment on another constraint which may arise in our model, the so called perturbative unitarity. Although the $S$-matrix for a scattering process must be unitary in the full theory, it may happen that at some order in the perturbative expansion the unitarity is violated, signaling the breakdown of the expansion. This is usually related to some of the couplings becoming too large. The perturbative unitarity translates into conditions for the partial wave amplitudes, which have to be smaller than 1/2. To examine such constraints in our model we use \texttt{SPheno}, which computes the maximal eigenvalue of a $2\rightarrow 2$ scattering matrix at the tree-level. 
On the other hand, since we already require all the quartic couplings to remain perturbative up to the energy scale of 50 TeV, we may suspect that the perturbative unitarity bounds are automatically satisfied. As we will see in the next section, this is indeed the case.

\section{Numerical analysis and benchmark scenarios} \label{sec:num}

In this section we perform a global numerical analysis of the model. We begin by discussing the employed scanning methodology, the definition of the  chi-square ($\chi ^2$) statistics and the initial ranges for all the model's parameters. Next, we present three best-fit benchmark scenarios which arise from the minimization of the $\chi ^2$ function. Finally, we provide a discussion of some experimental signatures that these benchmark scenarios  could produce.

\subsection{Scanning methodology} \label{subsec:num_1}

\begin{table}[t]
\centering
\renewcommand{\arraystretch}{1.3}
\resizebox{\textwidth}{!}{$
\begin{tabular}{|p{0.8cm}p{2.3cm}|p{0.8cm}p{2.3cm}|p{0.8cm}p{3.2cm}|p{0.8cm}p{2.3cm}|p{0.8cm}p{2.3cm}|}
\toprule
\toprule
\multicolumn{10}{c}{Scalar sector} \\
\midrule
$\tan\beta$ & $\left[ 2, 50 \right]$ & $v_{\phi}$ & $\left[ 1000, 1500 \right]$ & $\mu_{sb}^{2}$ & $\left[ 4, 64 \right] \times 10^{4}$ & $\lambda_{2}$ & $\left[ -2.0, +2.0 \right]$ & $\lambda_{3}$ & $\left[ 0.24, 0.28 \right]$ \\ 
$\lambda_{4}$ & $\left[ -2.0, +2.0 \right]$ & $\lambda_{5}$ & $\left[ -0.2, 0.0 \right]$ & $\lambda_{6}$ & $\left[ -2.0, +2.0 \right]$ & $\lambda_{7}$ & $\left[ -0.01, +0.01 \right]$ & $\lambda_{8}$ & $\left[ -1.0, +1.0 \right]$ \\
\midrule
\midrule
\multicolumn{10}{c}{Lepton sector} \\
\midrule
\midrule
$y_{24}^{e}$ & $\left[ -0.7, +0.7 \right]$ & $y_{43}^{e}$ & $\left[ -1.0, +1.0 \right]$ & $y_{14}^{\nu}$& $\left[ -1.0, +1.0 \right]\times 10^{-10}$ & $y_{14}^{\prime\nu}$ & $\left[ -1.0, +1.0 \right]$ & $M_{4}^{e}$ & $\pm \left[ 200, 1000 \right]$ \\
$y_{34}^{e}$ & $\left[ -1.0, +1.0 \right]$ & $x_{42}^{e}$ & $\left[ -1.0, +1.0 \right]$ & $y_{24}^{\nu}$ & $\left[ -1.0, +1.0 \right] \times 10^{-10}$ & $y_{24}^{\prime\nu}$ & $\left[ -1.0, +1.0 \right]$ & $M_{4}^{\nu}$ & $\pm \left[ 200, 1000 \right]$ \\
$x_{34}^{L}$ & $\left[ -1.0, +1.0 \right]$ & $x_{43}^{e}$ & $\left[ -1.0, +1.0 \right]$ & $y_{34}^{\nu}$ & $\left[ -1.0, +1.0 \right] \times 10^{-10}$ & $y_{34}^{\prime\nu}$ & $\left[ -1.0, +1.0 \right]$ & $M_{4}^{L}$ & $\pm \left[ 200, 1000 \right]$ \\
\midrule
\midrule
\multicolumn{10}{c}{Quark sector} \\
\midrule
\midrule
$y_{24}^{u}$ & $\left[ -1.0, +1.0 \right]$ & $y_{43}^{u}$ & $\left[ -1.4, +1.4 \right]$ & $y_{14}^{d}$ & $\left[ -0.7, +0.7 \right]$ & $y_{43}^{d}$ & $\left[ -1.0, +1.0 \right]$ & $M_{4}^{d}$ & $\pm \left[ 1200, 4000 \right]$ \\
$y_{34}^{u}$ & $\left[ -1.4, +1.4 \right]$ & $x_{42}^{u}$ & $\left[ -1.0, +1.0 \right]$ & $y_{24}^{d}$ & $\left[ -1.0, +1.0 \right]$ & $x_{42}^{d}$ & $\left[ -1.0, +1.0 \right]$ & $M_{4}^{u}$ & $\pm \left[ 1200, 4000 \right]$ \\
$x_{34}^{Q}$ & $\left[ -1.0, +1.0 \right]$ & $x_{43}^{u}$ & $\left[ -1.4, +1.4 \right]$ & $y_{34}^{d}$ & $\left[ -1.0, +1.0 \right]$ & $x_{43}^{d}$ & $\left[ -1.0, +1.0 \right]$ & $M_{4}^{Q}$ & $\pm \left[ 1200, 4000 \right]$ \\
\bottomrule
\bottomrule
\end{tabular}
$}
\caption{Scanning ranges for the input parameters of the model defined in Table~\ref{tab:BSM_model}. The alignment limit (cf. Sec.~\ref{sec:scalar}), the RGE  perturbativity constraints (cf. Sec.~\ref{sec:RG}) and a tentative lower bound on the VL mass parameters (see the text) are imposed. In the Yukawa sector only the non-zero couplings are shown. Dimensionful quantities are given in GeV and GeV$^2$.}
\label{tab:parameter_spaces}
\end{table}

In Table~\ref{tab:parameter_spaces} we summarize the scanning ranges for all the parameters of the model. These include the quartic couplings and the soft-breaking term of the scalar potential (\ref{eq:scalpot}), the non-zero Yukawa couplings and the mass parameters of the lagrangian (\ref{eqn:Lag_ren}), the vev of the singlet scalar, and $\tan\beta$. 

In the scalar sector the alignment conditions (\ref{eq:al1}) and~(\ref{eq:al2}) are imposed, leading to the limited scanning ranges for  $\lambda_3$, $\lambda_5$ and $\lambda_7$ (cf.~Eqs.~(\ref{alignment_lambda2}) and~(\ref{alignment_lambda8})). For all the other quartic couplings the perturbativity bounds discussed in Sec.~\ref{sec:RG} are enforced.
 Similarly, the Yukawa couplings are scanned in the ranges consistent with their RGE perturbativity constraints. Finally, small values of some of the neutrino coupling constants are  necessary to generate tiny masses for the SM neutrinos. 

A tentative lower bound of $1200 \func{GeV}$ is imposed on the VL quark mass parameters. This is a rough (and conservative) approximation of the constraints from the direct NP searches at the LHC, which will be discussed in more details in Sec.~\ref{subsec:num_3}. The scanning range for $v_\phi$ then follows from the requirement of reproducing the correct mass of the top quark, as discussed in Sec.~\ref{sec:hirer}.
Similarly, we adopt $200$ GeV lower bounds on the VL lepton mass parameters in order to be roughly consistent with the corresponding LHC constraints, which we examine in Sec.~\ref{subsec:num_4}.
Finally, the range for $\mu^2_{sb}$ was chosen to make sure that the mass of the associated CP-odd state (cf. Eq.~(\ref{eq:massa2})) is not excluded by the current experimental searches~\cite{ParticleDataGroup:2022pth}.

The experimental constraints employed in our numerical scan are listed in Table~\ref{tab:numerical_values}. 
The central values and the experimental errors for the quark and lepton masses and for the CKM matrix elements are quoted after the PDG report~\cite{ParticleDataGroup:2022pth}. 
Since the uncertainties for $m_\mu$ and  $m_\tau$ are very small, rendering the fitting procedure numerically challenging, we adopt an error of 10\% for these two observables. The experimental constraints from the flavor physics were discussed in Sec.~\ref{sec:lepton}.

We construct the $\chi^2$-statistic function as
\begin{equation}\label{eq:chi2}
\chi^{2} = \sum_i \frac{\left(\mathcal{O}_i^{\func{model}} - \mathcal{O}_i^{\func{cen}} \right)^2}{(\mathcal{O}_i^{\func{err}})^{2}}\,,
\end{equation}
where $\mathcal{O}_i^{\func{model}}$ indicates the value of an observable calculated in our model, $\mathcal{O}_i^{\func{cen}}$ is the central value of its experimental measurement, $\mathcal{O}_i^{\func{err}}$ is the corresponding experimental error, and the sum runs over all the measured observables listed in Table~\ref{tab:numerical_values}. The upper bounds, corresponding to the last three rows of Table~\ref{tab:numerical_values}, are not included in the $\chi^2$ function, but applied as hard-cuts instead (a point in the parameter space is rejected if such a condition is not satisfied). 

\begingroup
\begin{table}[t]
\centering
\renewcommand{\arraystretch}{1.3}
\begin{tabular}{P{0.13\textwidth}|P{0.13\textwidth}|P{0.13\textwidth}||P{0.13\textwidth}|P{0.13\textwidth}|P{0.13\textwidth}}
\toprule
\toprule
Measurement & Central Value & Exp. Error & Measurement & Central Value & Exp. Error \\
\toprule
\toprule
$m_\mu$ & $0.10566$ & $10\%$ & $|V_{ud}|$ & $0.97370$ & $0.00014$ \\
$m_\tau$ & $1.77686$ & $10\%$ & $|V_{us}|$ & $0.22450$ & $0.00080$ \\
$m_c$ & $1.270$ & $0.020$ & $|V_{ub}|$ & $0.00382$ & $0.00024$ \\
$m_s$ & $0.0934$ & $0.0034$ & $|V_{cd}|$ & $0.22100$ & $0.00400$ \\
$m_b$ & $4.18$ & $0.02$ & $|V_{cs}|$ & $0.98700$ & $0.01100$ \\
$m_t$ & $172.76$ & $0.30$ & $|V_{cb}|$ & $0.04100$ & $0.00140$ \\
$\Delta a_\mu$ & $2.49 \times 10^{-9}$ & $0.48 \times 10^{-9}$ & $|V_{td}|$ & $0.00800$ & $0.00030$ \\
 &  & & $|V_{ts}|$ & $0.03880$ & $0.00110$ \\
 &  & & $|V_{tb}|$ & $1.01300$ & $0.03000$ \\ \midrule
Measurement & Upper bound &  &  &  &  \\ \midrule
$\func{BR}\left( \tau \rightarrow \mu \gamma \right)$ & $<4.2 \times 10^{-8}$ & &  &  \\ 
$\func{BR}\left( \tau \rightarrow 3\mu \right)$ & $<2.1 \times 10^{-8}$ & &  &  &  \\ 
$\Delta_{\func{CKM}}$ & $<$0.05 & &  &  &  \\
\bottomrule
\bottomrule
\end{tabular}
\caption{The experimental measurements which we employ in our numerical scan. \newline Masses are in GeV.}
\label{tab:numerical_values}
\end{table}
\endgroup

To minimize the $\chi^{2}$ function, we adopt the following strategy. First, we perform an initial scan of the parameter space consistent with Table~\ref{tab:parameter_spaces}. As a result, we obtain a seed which is then used to minimize the $\chi^{2}$ function by iterating a random walk algorithm with an adaptive step function. The step function is chosen such that at each iteration all input parameters are updated by less than $\kappa \%$, and $\kappa$ reduces with an exponential decay law throughout the minimization procedure. During each iteration, we discard all the points that do not satisfy the upper bounds on $\Delta_{\text{CKM}}$,  $\textrm{BR}(\tau \rightarrow \mu \gamma)$ and $\textrm{BR}(\tau \rightarrow 3\mu)$, as well as the boundedness constraints on the scalar potential given in Eq.~(\ref{eqn:VS_conditions}) and the perturbative unitarity. Moreover, we investigate the vacuum stability with \texttt{Vevacious++}~\cite{Camargo-Molina-OLeary:2014} and we keep only those points whose vacuum is identified as ``stable".

\subsection{Benchmark scenarios} \label{subsec:num_2}

\begingroup
\begin{table}[t]
\centering
\resizebox{0.95\textwidth}{!}{$
\renewcommand{\arraystretch}{1.3}
\begin{tabular}{P{0.07\textwidth}|P{0.12\textwidth}|P{0.12\textwidth}|P{0.12\textwidth}||P{0.07\textwidth}|P{0.12\textwidth}|P{0.12\textwidth}|P{0.12\textwidth}}
\toprule
\toprule
\multicolumn{8}{c}{Scalar sector} \\
\toprule
   & BP1 & BP2 & BP3 &  & BP1 & BP2 & BP3 \\
\midrule
$\tan\beta$ & $13$ & $8$ & $12$ & $\lambda_{1}$ & $0.258$ & $0.258$ & $0.258$ \\
$v_{u}$ & $245.3$ & $244.3$ & $245.2$ & $\lambda_{2}$ & $0.514$ & $0.153$ & $0.623$ \\
$v_{d}$ & $18.9$ & $30.5$ & $20.4$ & $\lambda_{3}$ & $0.257$ & $0.260$ & $0.256$ \\
$v_{\phi}$ & $1015$ & $1077$ & $1012$ & $\lambda_{4}$ & $0.552$ & $0.304$ & $0.167$ \\
$\mu_{u}^{2}$ & $-7.8 \times 10^3$ & $-6.6 \times 10^3$ & $-7.6 \times 10^3$ & $\lambda_{5}$  & $-0.039$ & $-0.072$ & $-0.061$\\
$\mu_{d}^{2}$ & $-8.2 \times 10^3$ & $-8.6 \times 10^4$ & $-3.4 \times 10^4$ & $\lambda_{6}$ & $0.370$ & $0.487$ & $0.663$ \\
$\mu_{\phi}^{2}$ & $-4.9 \times 10^4$ & $-9.4\times 10^4$ & $-2.3 \times 10^5$ & $\lambda_{7}$ & $0.001$ & $0.002$ & $0.002$\\
$\mu_{\func{sb}}^{2}$ & $1.4 \times 10^{5}$ & $1.9 \times 10^{5}$ & $1.1 \times 10^{5}$ & $\lambda_{8}$ & $0.254$ & $0.423$ & $0.417$\\
\toprule
\toprule
\multicolumn{4}{c||}{Quark sector} & \multicolumn{4}{c}{Lepton sector} \\
\toprule
  & BP1 & BP2 & BP3 &  & BP1 & BP2 & BP3 \\
\toprule
$y_{24}^{u}$ & $-0.051$ & $-0.049$ & $0.050$ & $y_{24}^{e}$ & $0.028$ & $-0.015$ & $0.022$ \\
$y_{34}^{u}$ & $-0.980$ & $1.185$ & $-1.024$ & $y_{34}^{e}$ & $-0.895$ & $0.612$ & $0.790$ \\
$x_{34}^{Q}$ & $0.924$ & $-0.842$ & $-0.877$ & $x_{34}^{L}$ & $0.616$ & $-0.729$ & $0.724$ \\
$y_{43}^{u}$ & $1.382$ & $1.093$ & $-1.337$ & $y_{43}^{e}$ & $-0.223$ & $0.144$ & $-0.191$ \\
$x_{42}^{u}$ & $0.550$ & $0.821$ & $-0.595$ & $x_{42}^{e}$ & $0.156$ & $0.165$ & $0.188$ \\
$x_{43}^{u}$ & $1.286$ & $1.261$ & $1.263$ & $x_{43}^{e}$ & $-0.168$ & $0.228$ & $-0.205$ \\
\midrule
$y_{14}^{d}$ & $-0.022$ & $0.035$ & $0.026$ & $y_{14}^{\nu}$ & $-2\times 10^{-11} $ &  $5\times10^{-11} $ & $3\times10^{-11}$ \\
$y_{24}^{d}$ & $0.096$ & $0.151$ & $-0.113$ & $y_{24}^{\nu}$ & $3\times10^{-11}$ & $8\times10^{-12}$ & $6\times10^{-11}$ \\
$y_{34}^{d}$ & $-0.684$ & $0.274$ & $0.267$ & $y_{34}^{\nu}$ & $-5\times10^{-11}$ & $9\times10^{-11}$ & $9\times10^{-11}$ \\
$y_{43}^{d}$ & $-0.672$ & $-0.489$ & $0.656$ & $y_{14}^{\prime\nu}$ & $-0.824$ & $-0.674$ & $-0.674$ \\
$x_{42}^{d}$ & $-0.371$ & $-0.110$ & $0.225$ & $y_{24}^{\prime\nu}$ & $-0.895$ & $-0.874$ & $-0.896$ \\
$x_{43}^{d}$ & $-0.160$ & $0.072$ & $-0.127$ & $y_{34}^{\prime\nu}$ & $0.701$ & $0.744$ & $-0.812$ \\
\midrule
\multicolumn{8}{c}{Mass parameters} \\
\midrule
  & BP1 & BP2 & BP3 &  & BP1 & BP2 & BP3 \\
\midrule
$M_{4}^{u}$ & $-1317$ & $1405$ & $1334$ & $M_{4}^{e}$ & $-517$ & $-575$ & $533$ \\
$M_{4}^{d}$ & $-3644$ & $3068$ & $-2882$ & $M_{4}^{\nu}$ & $204$ & $-212$ & $217$ \\
$M_{4}^{Q}$ & $-1384$ & $1443$ & $1322$ & $M_{4}^{L}$ & $-206$ & $-222$ & $-202$ \\
\bottomrule
\bottomrule
\end{tabular}
$}
\caption{Input parameters for three best-fit benchmark scenarios. Dimensionful quantities are given in GeV and GeV$^{2}$.}
\label{tab:good_benchmark}
\end{table}
\endgroup
 
In Table~\ref{tab:good_benchmark} we present input parameters for three best-fit benchmark scenarios identified by performing the numerical scan discussed in Sec.~\ref{subsec:num_1}. The corresponding mass spectra are summarized in Table~\ref{tab:mass_spectrum} while the breakdown of individual contributions to the $\chi^{2}$ function is shown in Table~\ref{tab:contributions_chi}. 

\begingroup
\begin{table}[t]
\centering
\resizebox{0.99\textwidth}{!}{$
\renewcommand{\arraystretch}{1.3}
\begin{tabular}{P{0.12\textwidth}|P{0.12\textwidth}|P{0.12\textwidth}|P{0.12\textwidth}||P{0.12\textwidth}|P{0.12\textwidth}|P{0.12\textwidth}|P{0.12\textwidth}}
\toprule
\toprule
\multicolumn{8}{c}{SM fermions} \\
\toprule
   & BP1 & BP2 & BP3 &  & BP1 & BP2 & BP3 \\
\midrule
$m_{c}$ & $1.262$ & $1.282$ & $1.259$ & $m_{\mu}$ & $0.110$ & $0.110$ & $0.110$ \\
$m_{t}$ & $172.7$ & $172.8$ & $172.6$ & $m_{\tau}$ & $1.864$ & $1.756$ & $1.765$ \\
$m_{s}$ & $0.089$ & $0.093$ & $0.091$ & $m_{\nu_{2}} \left[ 10^{-10} \right]$ & $4.659$ & $6.587$ & $0.252$ \\
$m_{b}$ & $4.169$ & $4.196$ & $4.175$ & $m_{\nu_{3}} \left[ 10^{-10} \right]$ & $8.253$ & $18.38$ & $20.95$ \\
\midrule
\multicolumn{8}{c}{NP fermions} \\
\toprule
\multicolumn{4}{c}{Quark sector} & \multicolumn{4}{c}{Lepton sector} \\
\toprule
   & BP1 & BP2 & BP3 &  & BP1 & BP2 & BP3 \\
\midrule
$M_{U_{1}}$ & $1495$ & $1561$ & $1440$ & $M_{E_{1}}$ & $487$ & $596$ & $554$ \\
$M_{U_{2}}$ & $1708$ & $1842$ & $1704$ & $M_{E_{2}}$ & $543$ & $615$ & $570$ \\
$M_{D_{1}}$ & $1534$ & $1579$ & $1464$ & $M_{N_{1,2}}$ & $205$ & $214$ & $218$ \\
$M_{D_{2}}$ & $3655$ & $3070$ & $2888$ & $M_{N_{3,4}}$ & $488$ & $598$ & $556$ \\
\midrule
\multicolumn{8}{c}{Scalars} \\
\midrule
   & BP1 & BP2 & BP3 &  & BP1 & BP2 & BP3 \\
\midrule
$M_{h_{1}}$ & $125$ & $125$ & $125$ & $M_{a_{1}}$ & $362$ & $411$ & $433$ \\
$M_{h_{2}}$ & $362$ & $412$ & $435$ & $M_{a_{2}}$ & $532$ & $614$ & $469$ \\
$M_{h_{3}}$ & $617$ & $752$ & $824$ & $M_{h^{\pm}}$ & $384$ & $423$ & $440$ \\ 
\bottomrule
\bottomrule
\end{tabular}
$}
\caption{Mass spectra for three best-fit benchmark scenarios. All masses are in GeV.}
\label{tab:mass_spectrum}
\end{table}
\endgroup

In general, the three benchmark scenarios demonstrate quite similar features, both in terms of the input parameters and of the resulting NP spectra. This is largely due to the fact that we aim at reproducing masses and mixings of the SM fermions and this, as we discussed in Sec.~\ref{sec:setup}, puts strong constraints on (some of) the model's parameters. 

Let us first notice that the masses of all the SM fermions of the third and second generation can be fitted very precisely. Each individual contribution to the $\chi^2$ function is smaller than $0.7$, with an exception of $m_s$ in BP1, in which case we have $\chi^2_b=1.6$. We can also observe that, as we anticipated in Sec.~\ref{sec:setup}, the Yukawa couplings which link the VL sector with the SM fermions of the third generation are in general larger than those associated with the second generation. 

On the other hand, fitting the CKM matrix is a little bit more tricky and the bulk of the total $\chi^2$ stems from this very  sector. As anticipated in Sec.~\ref{seq:ckm}, the main contribution to the $\chi^2$ function is given by the element $|V_{td}|$, with the corresponding  $\chi_{V_{td}}^{2}$ ranging from $16$ for BP1 to $25$ for BP3. Smaller yet still relevant contributions come from the entries $|V_{ub}|$ and $|V_{ts}|$. Finally, an order 10 contribution to the $\chi^2$ function from the element $|V_{us}|$ is mainly due to a very small experimental error associated with this particular observable. All other elements of the CKM matrix are fitted within their $1\sigma$ experimental ranges. As an illustration, we present below the full $5\times 5$ CKM matrix for the benchmark scenario BP1, 
\begingroup
\begin{equation}
|V_{\func{CKM}}| ^{(\textrm{BP1})} = 
    \left(
\begin{array}{ccc|cc}
 0.97394\,\,\,\, & 0.22679\,\,\,\, & 0.00298\,\, & 1.4\times 10^{-7} & 0.00008 \\
 0.22671 & 0.97301 & 0.04296 & 0.00003 & 0.00042 \\
 0.00681 & 0.04236 & 0.99821 & 0.00968 & 0.00221\\
 \hline
 0.00054 & 0.00270 & 0.02853 & 0.86532 & 0.00114 \\
 0.00082 & 0.00390 & 0.02996 & 0.50113 & 0.00076\\
\end{array}
\right)\,.
\end{equation}
\endgroup
The two remaining best-fit points follow the same pattern. Incidentally, note that the CKM anomaly is $\mathcal{O}(10^{-4})$ in our setup, well  below the experimental upper bound of Eq.~(\ref{eqn:CKM_non_unitarity}).

Interestingly, in all three cases each quartic (Yukawa) coupling remains smaller than $4\pi$ ($\sqrt{4\pi}$) up to $1000\func{TeV}$. We can therefore conclude that the validity range of our model extends well beyond the putative scale of $50\func{TeV}$. 
We also checked that the maximal eigenvalue of the scattering matrix computed by \texttt{SPheno} is $\mathcal{O}(10^{-2})$ for all the benchmark scenarios, indicating that the perturbative unitarity bound is satisfied as well.

\begingroup
\begin{table}[t]
\centering
\resizebox{1.0\textwidth}{!}{$
\renewcommand{\arraystretch}{1.3}
\begin{tabular}{P{0.12\textwidth}|P{0.12\textwidth}|P{0.12\textwidth}|P{0.12\textwidth}||P{0.12\textwidth}|P{0.12\textwidth}|P{0.12\textwidth}|P{0.12\textwidth}}
\toprule
\toprule
\multicolumn{4}{c||}{Quarks masses} & \multicolumn{4}{c}{CKM elements} \\
\toprule
  & BP1 & BP2 & BP3 &  & BP1 & BP2 & BP3 \\
\midrule
$\chi_{c}^{2}$ & $0.154$ & $0.360$ & $0.280$ & $\chi_{V_{us}}^{2}$ & $8.225$ & $8.290$ & $5.986$ \\
$\chi_{t}^{2}$ & $0.022$ & $0.018$ & $0.119$ & $\chi_{V_{ub}}^{2}$ & $12.33$ & $10.36$ & $9.327$ \\
$\chi_{s}^{2}$ & $1.569$ & $0.014$ & $0.450$ & $\chi_{V_{td}}^{2}$ & $15.69$ & $18.30$ & $24.58$ \\
$\chi_{b}^{2}$ & $0.330$ & $0.640$ & $0.052$ & $\chi_{V_{ts}}^{2}$ & $10.45$ & $9.796$ & $6.559$ \\
\midrule
$\chi_{Q}^{2}$ & $2.075$ & $1.031$ & $0.901$ & $\chi_{V}^{2}$ & $55.45$ & $55.26$ & $55.92$ \\
\toprule
\multicolumn{4}{c||}{Charged leptons masses} & \multicolumn{4}{c}{$\Delta a_\mu$} \\
\toprule
$\chi_{\mu}^{2}$ & $0.210$ & $0.170$ & $0.207$ & $\chi_{\Delta a_{\mu}}^{2}$ & $0.328$ & $0.657$ & $0.375$ \\
\cmidrule{5-8}
$\chi_{\tau}^{2}$ & $0.241$ & $0.014$ & $0.004$ & \multicolumn{4}{c}{Total} \\
\midrule
$\chi_{L}^{2}$ & $0.451$ & $0.183$ & $0.211$ & $\chi^{2}_{\textrm{TOT}}$ & $58.30$ & $57.31$ & $57.41$ \\
\bottomrule
\bottomrule
\end{tabular}
$}
\caption{Breakdown of the $\chi^2$ contributions from various observables implemented in the $\chi^2$ function of Eq.~(\ref{eq:chi2}). The CKM  contributions which are smaller than 3 are not shown. $\chi^2_Q$, $\chi^2_L$ and $\chi^2_V$ indicate total $\chi^2$ contributions from the quark masses, lepton masses, and the CKM matrix elements, respectively. $\chi^{2}_{\textrm{TOT}}$ stands for the total $\chi^{2}$ function of each best-fit scenario.}
\label{tab:contributions_chi}
\end{table}
\endgroup
\begingroup
\begin{table}[b]
\centering
\resizebox{0.95\textwidth}{!}{$
\renewcommand{\arraystretch}{1.3}
\begin{tabular}{P{0.12\textwidth}|P{0.12\textwidth}|P{0.12\textwidth}|P{0.12\textwidth}||P{0.12\textwidth}|P{0.12\textwidth}|P{0.12\textwidth}|P{0.12\textwidth}}
\toprule
\toprule
\multicolumn{8}{c}{ Contributions to $\Delta a_\mu\times 10^{9}$} \\
\toprule
\multicolumn{4}{c}{Charged scalars} & \multicolumn{4}{c}{CP-even scalars} \\
\toprule
Loop  & BP1 & BP2 & BP3 &  Loop & BP1 & BP2 & BP3 \\
\midrule
$h^{\pm}, N_{1,2}$ & $-1.076\,\,\,\,$ & $-0.792\,\,\,\,$ & $-0.942\,\,\,\,$ & $h_{1}, E_{1}$ & $-0.003\,\,\,\,$ & $-0.001\,\,\,\,$ & $-0.009\,\,\,\,$ \\
$h^{\pm}, N_{3,4}$ & $3.300 $ & $2.898$ & $3.153 $ & $ h_{1}, E_{2}$ & $0.003$ & $0.001$ & $0.009$ \\
\cmidrule{1-4}
$h^{\pm}, N_{\text{tot}}$ & $2.225 $ & $2.106 $ & $2.211 $ & $h_{2}, E_{1}$ & $-0.409\,\,\,\,$ & $-0.520\,\,\,\,$ & $-0.969\,\,\,\,$ \\
\cmidrule{1-4}
\multicolumn{4}{c||}{CP-odd scalars} & $ h_{2}, E_{2}$ & $0.437$ & $0.548$ & $0.994$ \\
\cmidrule{1-4}
$a_1, E_{1}$ & $0.425$ & $0.528$ & $0.938$ & $h_{3}, E_{1}$ & $0.018$ & $0.115$ & $0.076$ \\
$a_1, E_{2}$ & $-0.544\,\,\,\,$ & $-0.611\,\,\,\,$ & $-1.529\,\,\,\,$ & $h_{3}, E_{2} $ & $-0.017\,\,\,\,$ & $-0.127\,\,\,\,$ & $-0.076\,\,\,\,$ \\
\cmidrule{5-8}
$a_2, E_{1}$ & $-0.033\,\,\,\,$ & $-0.135\,\,\,\,$ & $-0.071\,\,\,\,$ & $h, E_{\rm{tot}} $ & $0.032$ & $0.027$ & $0.025$ \\
\cmidrule{5-8}
$a_2, E_{2}$ & $0.110$ & $0.196$ & $0.621$ & \multicolumn{4}{c}{Total} \\
\midrule
$a, E_{\rm{tot}}$ & $-0.015\,\,\,\,$ & $-0.023\,\,\,\,$ & $-0.041\,\,\,\,$ & $\Delta a_{\mu}$ & $2.215 $ & $2.101 $ & $2.196 $ \\
\toprule
\toprule
\end{tabular}
$}
\caption{Contributions to $\Delta a_\mu$ from the individual one-loop diagrams shown in Fig.~(\ref{fig:muon_mass_muong2}). The subscript "tot" indicates the sum of all the contributions of a given type.}\label{tab:contributions_g2}
\end{table}
\endgroup

Masses of the NP leptons are determined, to a large extent, by correctly fitting the experimental value of $\Delta a_\mu$ (an overall contribution  from this observable to the total $\chi^2$ function does not exceed 0.7 in all the benchmark scenarios). Contributions to $\Delta a_\mu$ from the individual one-loop diagrams of Fig.~(\ref{fig:muon_mass_muong2}) are summarized in Table~\ref{tab:contributions_g2}. We present separately fractions of $\Delta a_\mu$ generated by the charged scalars $h^\pm$ and the neutral leptons $N_{1,2,3,4}$, by the CP-odd scalars $a_{1,2}$ and the charged leptons $E_{1,2}$, and by the CP-even scalars $h_{1,2,3}$ and the charged leptons $E_{1,2}$.  We also show the sum of all the contributions of a given type, indicated by a subscript "tot".  

We observe that the largest contributions to $\Delta a_\mu$ arise from the charged scalar/heavy neutrino loops. We thus disprove the conclusions of Refs.~\cite{Hernandez:2021tii,Lee:2022sic} where  
it was assumed that the charged lepton loops were the only NP contributions to muon $(g-2)$ present in the model. As we show in our analysis, all possible one-loop diagrams contributing to $\Delta a_\mu$ should be treated at equal footing and none of them should be discarded {\it a priori}. 

Even more interestingly, the observed dominance of the heavy neutrino contributions to the anomalous magnetic moment of the muon seems to be a generic feature of the model which does not pertain exclusively to the identified benchmark scenarios. The charged scalar/heavy neutrino loops are determined, among other parameters, by combinations of the $y^{\prime\nu}$ couplings which are not constrained by any SM masses and mixing and thus can become relatively large. Contrarily, the same Yukawa coupling is responsible for the generation of the neutral (pseudo)scalar/charged lepton loops and for the correct tree-level mass of the muon. It is thus required to be small and the corresponding contributions to $\Delta a_\mu$ are suppressed. 
Additionally, one also observes cancellations between the individual contributions to $\Delta a_\mu$ stemming from the (pseudo)scalar diagrams with different VL leptons, which is a known and common feature of many NP models with the VL fermions (see, e.g., \cite{Darme:2018hqg,Kowalska:2020zve} for a discussion). 

Finally, we should mention the size of the BRs for the lepton flavor violating decays  $\tau \rightarrow \mu \gamma$ and $\tau\to 3\mu$ in our model, which are of the order $(3-4) \times 10^{-8}$ for the former and $(6-9)\times 10^{-10}$ for the latter. Given that in the future the Belle-II collaboration is expected to improve their current exclusion bounds by an order of magnitude or more~\cite{10.1093/ptep/ptz106,Belle-II:2022cgf}, it may turn out that the tau leptonic decays offer the best experimental way of verifying the predictions of the NP model analyzed in this study. 

\section{LHC study of the benchmark scenarios} \label{sec:checkGBP}

In this section we confront the benchmark scenarios identified in Sec.~\ref{sec:num} with the null results of the direct NP searches at the LHC. We analyze, one by one, the constraints originating from considering the production of VL quarks, VL leptons, and exotic scalars.

\subsection{Vector-like quarks} \label{subsec:num_3}

The VL quarks (VLQs) can be copiously produced at the LHC, either in pairs through the strong interactions or singly through an
exchange of the EW gauge bosons. In the former case, the dominant production channels at the leading order are 
gluon fusion and quark-antiquark annihilation, whose production cross sections depend on the VLQ mass and its SU(3)$_C$ quantum numbers only (see Refs.~\cite{Kowalska:2019qxm,Olivas:2021nft} for analytical formulae). Therefore, the experimental lower bounds on VLQ masses are expected to be, to a large extent, model independent, baring only a slight dependence on the relative strength of the individual VLQ decay channels.

\begingroup
\begin{figure}[t]
    \begin{subfigure}{0.4\textwidth}    \includegraphics[keepaspectratio,width=\textwidth]{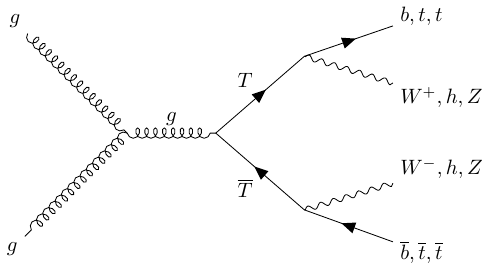}
    \end{subfigure}
    \hspace{1cm}
    \begin{subfigure}{0.4\textwidth}    \includegraphics[keepaspectratio,width=\textwidth]{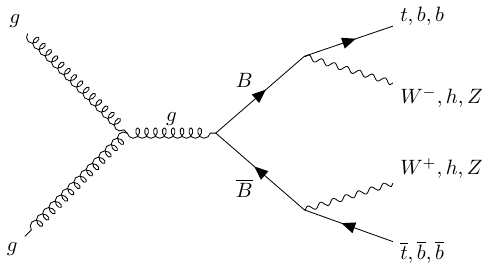}
    \end{subfigure}
    \caption{Pair production of the VLQs $T$ and $B$ via the gluon fusion at the LHC considered by the ATLAS collaboration in Ref.~\cite{ATLAS:2022tla}. }
    \label{fig:VLQ_doublet}
\end{figure}
\endgroup

The most recent analysis from ATLAS, based on the data from proton–proton collisions at a centre-of-mass energy of $\sqrt{s}=13$ TeV, corresponding to an integrated luminosity of $139$ fb$^{-1}$~\cite{ATLAS:2022tla}, considered the pair production of VL top partners $T$ and VL bottom partners $B$ with the decay channels $T\to Zt,\,ht,\,Wb$ and $B\to Zb,\, hb,\,Wt$ and with large missing
transverse momentum. The corresponding Feyman diagrams are depicted  in Fig.~\ref{fig:VLQ_doublet}. The strongest  95\%~C.L. lower bounds on the VLQ mass derived in Ref.~\cite{ATLAS:2022tla} read
\begin{equation}
    M_{T/B}> 1.41\func{TeV}
\end{equation}
for the EW doublets\footnote{For the VLQ mass larger than 800 GeV this indicates BR$(T\to Zt)$=BR$(T\to ht)=50\%$ and BR$(B\to Wt)=100\%$~\cite{ATLAS:2022tla}.} and 
\begin{equation}
    M_{T}> 1.26\func{TeV},\qquad M_{B}> 1.33\func{TeV} 
\end{equation}
for the EW singlets.\footnote{For the VLQ mass larger than 800 GeV this indicates BR$(T\to Zt)=25\%$, BR$(T\to ht)=25\%$, BR$(T\to Wb)=50\%$, BR$(B\to Zb)=25\%$, BR$(B\to hb)=25\%$ and BR$(B\to Wt)=50\%$~\cite{ATLAS:2022tla}.} 
The analogous results from CMS can be found in Ref.~\cite{CMS:2022fck}.
Similarly, by assuming that at least one of the VLQs decays into a $Z$ boson with the BR=100\%, the $13$ TeV ATLAS search~\cite{ATLAS:2022hnn} obtained even stronger bounds,
\begin{equation}
    M_{T}> 1.60\func{TeV},\qquad M_{B}> 1.42\func{TeV}\,. 
\end{equation}

At first glance it may seem that all our benchmark scenarios are consistent with the VLQ exclusion bounds. On the other hand, in the model considered in this study the couplings of the physical heavy quarks $U_{1,2}$ and $D_{1,2}$ with the third-generation SM quarks and the EW gauge (Higgs) bosons are generated via tree-level mixing after the EW and U(1)$_X$ symmetries are spontaneously broken (cf. Sec.~\ref{sec:setup} and Appendix~\ref{sec:App_diag}). Since there is {\it a priori} no reason for the resulting BRs to correspond to any of the benchmark cases considered by ATLAS and CMS in their analyses (exotic decays to the charged Higgs are possible, for example), we need to reexamine the experimental results in the framework of our model. 

To this end, we calculate with \texttt{MadGraph5 MC@NLO}~\cite{Alwall:2014hca} the cross sections for the pair production of $U_1$, $U_2$, $D_1$ and $D_2$. The results are presented in Table~\ref{tab:VLQ_csBPs}. By comparing these numbers with the observed experimental 95\%~C.L. upper bounds on the signal cross section from Ref.~\cite{ATLAS:2022tla} (to give an example, $\sigma^{\textrm{exp}}_{95\%}(p\,p\to \bar{T}T)=4\times 10^{-3}$ pb for $M_{\textrm{VLQ}}=1.5$ TeV) we conclude that our benchmark scenarios are indeed not excluded by the current LHC searches for the VLQs, irrespectively of the actual sizes of their 
BRs.

\begin{table}[t]
\centering
\resizebox{0.95\textwidth}{!}{$
\renewcommand{\arraystretch}{1.3}
\begin{tabular}{P{0.06\textwidth}|P{0.07\textwidth}|P{0.14\textwidth}|P{0.07\textwidth}|P{0.14\textwidth}|P{0.07\textwidth}|P{0.14\textwidth}|P{0.07\textwidth}|P{0.14\textwidth}} 
\toprule
\toprule
& $M_{U_{1}}$ & $\sigma (p p \rightarrow U_{1} U_{1})$ & $M_{U_{2}}$ & $\sigma (p p \rightarrow U_{2} U_{2})$  & $M_{D_{1}}$ & $\sigma (p p \rightarrow D_{1} D_{1})$  & $M_{D_{2}}$ & $\sigma (p p \rightarrow D_{2} D_{2})$  \\
\midrule
BP1 & $1495$ & $1.3 \times 10^{-3}$ & $1708$ & $3.9 \times 10^{-4}$ & $1534$ & $1.0 \times 10^{-3}$ & $3655$ & $4.5 \times 10^{-9}$  \\
BP2 & $1561$ & $8.9 \times 10^{-4}$ & $1842$ & $1.9 \times 10^{-4}$ & $1579$ & $8.1 \times 10^{-4}$ & $3070$ & $1.6 \times 10^{-7}$ \\
BP3 & $1440$ & $1.8 \times 10^{-3}$ & $1704$ & $4.0 \times 10^{-4}$ & $1464$ & $1.6 \times 10^{-3}$ &  $2888$ & $5.0 \times 10^{-7}$\\
\bottomrule
\bottomrule
\end{tabular}
$}
\caption{Cross sections (in pb) for the pair production of the VLQs for our three benchmark scenarios. Masses are in GeV.}
\label{tab:VLQ_csBPs}
\end{table}

It is instructive to investigate the prospects of testing our model in future runs at the LHC. The total cross section for the VLQ pair production, followed by a decay into the third generation quarks and the EW gauge/Higgs bosons, can be expressed using the narrow width approximation (NWA) as
\begingroup
\begin{equation}
    \widetilde{\sigma}\left( p p \rightarrow \overline{Q} Q \rightarrow f \overline{f}\, V V \right) \approx \sigma\left( p p \rightarrow Q \overline{Q} \right) \func{BR}\left( Q \rightarrow f V \right) \func{BR}\left( \overline{Q} \rightarrow \overline{f} V \right),
    \label{eqn:VLQ_cs}
\end{equation}
\endgroup
where $Q = U_{1,2}, D_{1,2}$, $f = t, b$ and $V = W, Z, h_1$. Under the assumption that $\func{BR}\left( Q\rightarrow fV\right)=\func{BR}\left( Q \rightarrow h_1 t/b \right) + \func{BR}\left( Q \rightarrow Zt/b \right) + \func{BR}\left(Q \rightarrow Wb/t \right) = 1$, the cross section~(\ref{eqn:VLQ_cs}) reduces to the signal cross section constrained by the experimental collaborations, 
$\sigma^{\textrm{exp}}_{95\%}\approx \sigma\left( p p \rightarrow Q \overline{Q}\right)$. If, on the other hand, the three BRs do not sum to one, we expect the resulting exclusion bounds to be weaker than the bounds reported by ATLAS and CMS.

The lightest VLQ in our model, $U_1$, is characterized by the following BRs:
\begin{eqnarray}
\func{BP1}:\quad&    \textrm{BR}\left( U_{1} \rightarrow h_1t \right)=0.188,\quad  \textrm{BR}\left( U_{1} \rightarrow Zt \right) =0.146, \quad \textrm{BR}\left( U_{1} \rightarrow Wb \right)=0.040\nonumber\\
\func{BP2}:\quad&    \textrm{BR}\left( U_{1} \rightarrow h_1t \right)=0.135,\quad  \textrm{BR}\left( U_{1} \rightarrow Zt \right) =0.101, \quad \textrm{BR}\left( U_{1} \rightarrow Wb \right)=0.044\\
\func{BP3}:\quad&    \textrm{BR}\left( U_{1} \rightarrow h_1t \right)=0.209,\quad  \textrm{BR}\left( U_{1} \rightarrow Zt \right) =0.165, \quad \textrm{BR}\left( U_{1} \rightarrow Wb \right)=0.037\nonumber
\end{eqnarray}
with the resulting cross sections $\widetilde{\sigma}^{\textrm{BP1}}=1.8 \times 10^{-4}$ pb, $\widetilde{\sigma}^{\textrm{BP2}}=7.0 \times 10^{-5}$ pb and $\widetilde{\sigma}^{\textrm{BP3}}=3.0 \times 10^{-4}$ pb.  The second-to-the-lightest VLQ, $D_1$, has
\begin{eqnarray}
\func{BP1}:\quad&    \textrm{BR}\left( D_{1} \rightarrow h_1b \right)=0.001,\quad  \textrm{BR}\left( D_{1} \rightarrow Zb \right) =0.001, \quad \textrm{BR}\left( D_{1} \rightarrow Wt \right)=0.375\nonumber\\
\func{BP2}:\quad&    \textrm{BR}\left( D_{1} \rightarrow h_1b \right)=0.001,\quad  \textrm{BR}\left( D_{1} \rightarrow Zb \right) =0.001, \quad \textrm{BR}\left( D_{1} \rightarrow Wt \right)=0.221\\
\func{BP3}:\quad&    \textrm{BR}\left( D_{1} \rightarrow h_1b \right)=0.001,\quad  \textrm{BR}\left( D_{1} \rightarrow Zb \right) =0.001, \quad \textrm{BR}\left( D_{1} \rightarrow Wt \right)=0.387\nonumber
\end{eqnarray}
with  $\widetilde{\sigma}^{\textrm{BP1}}=1.4 \times 10^{-4}$ pb, $\widetilde{\sigma}^{\textrm{BP2}}=4.1 \times 10^{-5}$ pb and $\widetilde{\sigma}^{\textrm{BP3}}=2.4 \times 10^{-4}$ pb.
We can thus conclude that in order to probe the VL masses featured by our benchmark scenarios, at least one order of magnitude enhancement of the experimental sensitivity in the VLQ searches is required. 

Finally, we analyze the possibility of testing the model via processes in which the VLQs are produced one at the time. The single VL $T$ quark production was analyzed by ATLAS in 
Refs.~\cite{ATLAS:2023bfh,ATLAS:2023pja,ATLAS:2022ozf}, while the single VL  $B$ quark production in Ref.~\cite{ATLAS:2023qqf}. The corresponding Feynman diagrams are shown in Fig.~\ref{fig:VLQ_SP}. The 95\%~C.L. experimental upper bounds on the relevant signal cross sections are of the order $(10^{-2}-10^{-1})$ pb. 

We calculated the cross sections for the VLQ single productions of our three benchmark points using the NWA. The hadronic cross sections were obtained with \texttt{MadGraph5 MC@NLO} and the BRs with \texttt{SPheno}. We found that the cross section for a single production of the VLQs $U_1$ is $\mathcal{O}(10^{-5})$ pb, while for $D_1$ it amounts to $\mathcal{O}(10^{-7})$ pb. We can thus conclude that in our model the single production is a less promising search strategy than the pair production. This was to be expected as the single production is in general less competitive than the pair production for the Yukawa couplings smaller than 1, see e.g.~\cite{Faroughy:2016osc,Kowalska:2018ulj}.

\begingroup
\begin{figure}[t]
    \begin{subfigure}{0.4\textwidth}    \includegraphics[keepaspectratio,width=\textwidth]{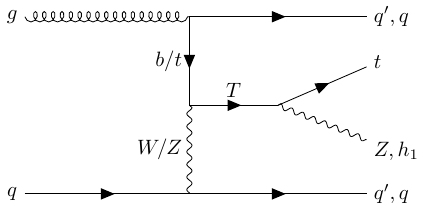}
    \end{subfigure}
    \hspace{1cm}
    \begin{subfigure}{0.4\textwidth}    \includegraphics[keepaspectratio,width=\textwidth]{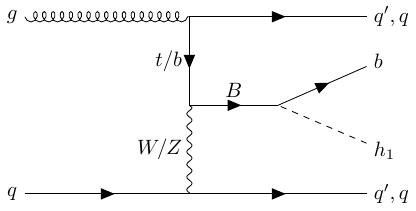}
    \end{subfigure}
    \caption{Single production of the VLQs $T$ and $B$ via the EW gauge boson exchange at the LHC, as considered by the ATLAS collaboration in Refs.~\cite{ATLAS:2023bfh,ATLAS:2023pja,ATLAS:2022ozf} for $T$ and in Ref.~\cite{ATLAS:2023qqf} for $B$.}
    \label{fig:VLQ_SP}
\end{figure}
\endgroup

\subsection{Vector-like leptons} \label{subsec:num_4}

At the tree level, the VL leptons (VLLs) are pair produced at the LHC via the Drell-Yan processes. The corresponding cross sections for our three benchmark scenarios are collected in Table~\ref{tab:VLL_csBPs}. The analysis of all the  possible experimental signatures is in this case much more involved than for the VLQs as the lepton decay BRs strongly depend on the presence in the spectrum of the exotic scalars lighter than the VLLs.  The following mass hierarchies are observed:
\begin{eqnarray}
    \func{BP1}:& M_{N_{1,2}}<M_{h_2},M_{a_1},M_{h^\pm} < M_{E_1},M_{N_{3,4}}<M_{a_2}<M_{E_2}<M_{h_3}\nonumber\\
    \func{BP2}:& M_{N_{1,2}}<M_{h_2},M_{a_1},M_{h^\pm} < M_{E_1},M_{N_{3,4}}<M_{a_2}<M_{E_2}<M_{h_3}\\   
    \func{BP3}:& \,\,M_{N_{1,2}}<M_{h_2},M_{a_1},M_{h^\pm} < M_{a_2}<M_{E_1},M_{N_{3,4}}<M_{E_2}<M_{h_3}\,.\nonumber
\end{eqnarray}
\begin{table}[t]
\centering
\resizebox{0.92\textwidth}{!}{$
\renewcommand{\arraystretch}{1.3}
\begin{tabular}{P{0.06\textwidth}|P{0.07\textwidth}|P{0.14\textwidth}|P{0.07\textwidth}|P{0.14\textwidth}|P{0.07\textwidth}|P{0.17\textwidth}|P{0.17\textwidth}}
\toprule
\toprule
& $M_{E_{2}}$ & $\sigma (p p \rightarrow \bar{E}_{2} E_{2})$ & $M_{E_{1}}$ & $\sigma (p p \rightarrow \bar{E}_{1} E_{1})$  & $M_{N_{3,4}}$ & $\sigma (p p \rightarrow N_{3,4} N_{3,4})$ & $\sigma (p p \rightarrow E_{1} N_{3,4})$ \\
\midrule
BP1 & $543$ & $1.5 \times 10^{-3}$ & $487$ & $5.7 \times 10^{-3}$ & $488$ & $2.7 \times 10^{-5}$ & $3.6 \times 10^{-3}$ \\
BP2 & $615$ & $8.0 \times 10^{-4}$ & $596$ & $1.8 \times 10^{-3}$ & $598$ & $8.5 \times 10^{-6}$ & $1.2 \times 10^{-3}$ \\
BP3 & $570$ & $1.2 \times 10^{-3}$ & $554$ & $2.8 \times 10^{-3}$ & $556$ & $2.8 \times 10^{-5}$ & $1.8 \times 10^{-3}$ \\
\bottomrule
\bottomrule
\end{tabular}
$}
\caption{Cross sections (in pb) for the pair production of the VLLs for our three benchmark scenarios. Masses are in GeV.}
\label{tab:VLL_csBPs}
\end{table}
In all the cases the lightest VLLs, neutrinos $N_{1,2}$, originate predominantly from the SU(2)$_L$ singlets and their production cross section at the LHC is suppressed, $\mathcal{O}(10^{-5})$ pb.

The second-to-the-lightest VLLs, $E_1$ and heavy neutrinos $N_{3,4}$, come from the same SU(2)$_L$ doublets and are almost degenerate in mass. Therefore, three production channels should be considered simultaneously: $p\,p\rightarrow Z/\gamma\rightarrow E_1\bar{E}_1$, $p\,p\rightarrow Z/\gamma\rightarrow N_{3,4}N_{3,4}$ and
$p\,p\rightarrow W^\pm \rightarrow E_1 N_{3,4}$. The dominant branching ratios for the subsequent decays of $E_1$ and $N_{3,4}$, evaluated with \texttt{SPheno}, are collected in Table~\ref{tab:VLL_e1NBPs}. In all three cases the VLLs decay predominantly to the SM muons, which is a direct consequence of the fact that we impose $\Delta a_\mu$ as a constraint in our likelihood function and the largish muon-lepton-scalar Yukawa couplings are preferred. 

\begin{table}[b]
\centering
\resizebox{0.75\textwidth}{!}{$
\renewcommand{\arraystretch}{1.2}
\begin{tabular}{P{0.16\textwidth}|P{0.08\textwidth}|P{0.16\textwidth}|P{0.08\textwidth}|P{0.16\textwidth}|P{0.08\textwidth}}
\toprule
\toprule
BR & BP1 & BR & BP2 & BR & BP3 \\
\midrule
$E_1\to\mu\,a_1$ & 37\% & $E_1\to\mu\,a_1$ & 23\% & $E_1\to\mu\,a_1$ & 21\%\\
$E_1\to\mu\,h_2$ & 37\% & $E_1\to\mu\,h_2$ & 24\% & $E_1\to\mu\,h_2$ & 25\%\\
 & & $E_1\to N_{1,2}\,W^\pm$ & 26\% & $E_1\to\tau\,a_1$ & 11\%\\
\midrule
BR & BP1 & BR & BP2 & BR & BP3 \\
\midrule
$N_{3,4}\to \mu\, h^\pm$ & 70\% & $N_{3,4}\to \mu\, h^\pm$ & 51\% & $N_{3,4}\to \mu\, h^\pm$ & 56\% \\
 & & $N_{3,4}\to N_{1,2}\, Z $ & 12\% & & \\
 & & $N_{3,4}\to N_{1,2}\, h_1$ & 11\% & & \\ 
\bottomrule
\bottomrule
\end{tabular}
$}
\caption{Dominant BRs for the decays of $E_1$ and $N_{3,4}$. }
\label{tab:VLL_e1NBPs}
\end{table}

A closer look at Table~\ref{tab:VLL_e1NBPs} reveals that the relative strengths of various VLL decay channels are, to some extent, scenario dependent. Moreover, the final experimental signatures hinge on the subsequent decay channels of the scalar particles, which are also pretty complex (we discuss it in more details in Sec.~\ref{subsec:num_5}). 
As an example, let us consider a process $p\,p\rightarrow E_1\,\bar{E}_1\to \mu\,\bar{\mu}\, a_1\, a_1$ for the benchmark scenario BP1. The lightest pseudoscalar can decay in this case either to a $b\,\bar{b}$ pair (with the BR of 28\%) or to  $\nu\,N_{1,2}$ (with the BR of 69\%). The decay of the heavy neutrinos then proceeds as $N_{1,2}\to e^\pm/\mu^\pm\,W^\pm$ with the BR of 56\%, or $N_{1,2}\to \nu\,Z$ with the BR of 28\%.
We can thus expect the following distinctive experimental signatures emerging from the $p\,p\rightarrow E_1\,\bar{E}_1\to \mu\,\bar{\mu}\, a_1\, a_1$ process: a) 2 muons + (b)-jets, b) multileptons + missing energy, c) multileptons + jets + missing energy, with the total signal cross section reduced w.r.t.~the production cross section reported in Table~\ref{tab:VLL_csBPs} by the product of the subsequent BRs.

To make the things even worse, there are not many LHC analysis that would explicitly look for the VLLs. The only dedicated ATLAS search, based on the 139 $\func{fb}^{-1}$ of  data from the 13 TeV run~\cite{ATLAS:2023sbu}, looks for the VLLs coupled predominantly to taus. The analogous CMS analysis based on the 77.4 $\func{fb}^{-1}$ of data can be found in Ref.~\cite{PhysRevD.100.052003}. The decay chains considered by the two collaborations are shown in Fig.~\ref{fig:VLL_doublet}. In both cases, the total cross sections for the VLL production can be probed down to $10^{-3}$ pb (of course, the actual value is mass dependent). 

\begingroup
\begin{figure}[t]
    \centering
    \begin{subfigure}{0.4\textwidth}    \includegraphics[keepaspectratio,width=\textwidth]{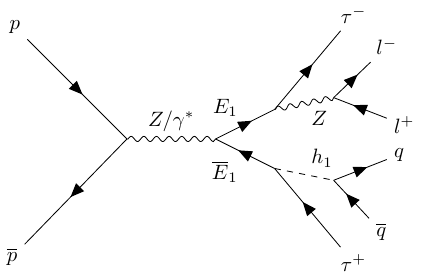}
    \end{subfigure}
        \hspace{1cm}
    \begin{subfigure}{0.4\textwidth}    \includegraphics[keepaspectratio,width=\textwidth]{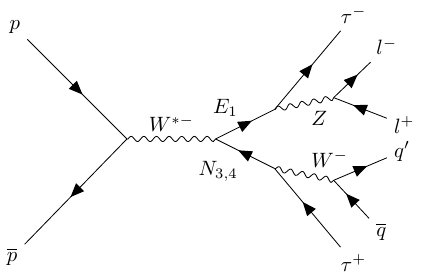}
    \end{subfigure}
    \caption{Pair production of the VLLs $E_1$ and $N_{3,4}$ at the LHC, as considered by the ATLAS~\cite{ATLAS:2023sbu} and CMS~\cite{PhysRevD.100.052003} collaborations.}
    \label{fig:VLL_doublet}
\end{figure}
\endgroup

In our model all three benchmark scenarios feature very low BRs for $E_1$ and $N_{3,4}$ decaying to taus, which do not exceed 10\%. We can thus expect a strong suppression of the resulting signal w.r.t.~the experimental analysis. Indeed, using the NWA the total cross section for the process considered in Refs.~\cite{ATLAS:2023sbu} and~\cite{PhysRevD.100.052003} can be written as follows:
\begingroup
\begin{eqnarray}
    \sigma\left( p\,p \rightarrow \tau^{-}\tau^{+} l^{-}l^{+} q\,\overline{q} \right)& \approx & \sigma\left( p\,p \rightarrow E_{1} \overline{E}_{1} \right) \func{BR}\left( E_{1} \rightarrow \tau^{-} l^{-} l^{+} \right) \func{BR}\left( \overline{E}_{1} \rightarrow \tau^{+} q\,\overline{q} \right)\nonumber\\
    & + &\sigma\left( p\,p \rightarrow E_{1} N_{3,4} \right) \func{BR}\left( E_{1} \rightarrow \tau^{-} l^{-} l^{+} \right) \func{BR}\left(N_{3,4} \rightarrow \tau^{+} q\,\overline{q} \right)\,.
\end{eqnarray}
\endgroup
Combining the cross sections from Table~\ref{tab:VLL_csBPs} with the relevant BRs calculated with  \texttt{SPheno}, we obtain
\begin{eqnarray}\label{eq:boundsVL}
    \func{BP1}:& \sigma\left( p\,p \rightarrow \tau^{-}\tau^{+} l^{-}l^{+} q\,\overline{q} \right) &= 8.3\times 10^{-7}\,\func{pb} \nonumber   \\
    \func{BP2}:&  \sigma\left( p\,p \rightarrow \tau^{-}\tau^{+} l^{-}l^{+} q\,\overline{q} \right) &= 8.1\times 10^{-6}\,\func{pb}   \\
    \func{BP3}:&  \sigma\left( p\,p \rightarrow \tau^{-}\tau^{+} l^{-}l^{+} q\,\overline{q} \right) &= 5.3\times 10^{-7}\,\func{pb}.    \nonumber        
\end{eqnarray}
If we now compare the  predictions of Eq.~(\ref{eq:boundsVL}) with the corresponding experimental 95\% C.L. exclusion cross sections from Ref.~\cite{ATLAS:2023sbu}, we can conclude that the benchmark scenarios identified in Sec.~\ref{sec:num} are not excluded by the dedicated LHC searches for the VLLs.\footnote{In principle, some of the SUSY searches looking for the chargino/neutralino production, e.g. Ref.~\cite{ATLAS:2023lfr}, analyze signatures that could be generated in our model. However, the resulting cross section for such a process is way too low to allow the derivation of any constraints.} Moreover, it may also be challenging to test the VLL sector of our model in future runs at the LHC, if no dedicated experimental strategies for the muon final state signatures are proposed.

\subsection{Exotic scalars} \label{subsec:num_5}

\begingroup
\begin{table}[t]
\centering
\renewcommand{\arraystretch}{1.3}
\begin{tabular}{P{0.16\textwidth}|P{0.08\textwidth}|P{0.08\textwidth}|P{0.08\textwidth}}
\toprule
\toprule
Process & BP1 & BP2 & BP3 \\
\midrule
$a_1\to \nu\,N_{1,2}$  & 69\% & 82\% & 72\%\\
$a_1\to \bar{b}\,b$  & 28\%  & 14\% & 25\%\\
$a_2\to \bar{t}\,t$ & 73\% & 73\%  & 36\%\\
$a_2\to \bar{b}\,b$  & & & 15\%\\
$a_2\to \nu\,N_{1,2}$  & & & 46\%\\
\midrule
$h_2\to \nu\,N_{1,2}$  & 69\% & 84\% & 72\%\\
$h_2\to \bar{b}\,b$  & 28\% & 14\% & 25\%\\
$h_3\to \bar{t}\,t$  & 50\% & 43\% & 34\% \\
$h_3\to \bar{\tau}\,E_1$  & 10\% & 9\% & 20\%\\
\midrule
$h^\pm\to \bar{\mu}\,N_{1,2}$ & 41\% & 51\% & 48\%\\
$h^\pm\to \bar{e}\,N_{1,2}$  & 34\% & 30\% & 26\% \\
$h^\pm\to \bar{b}\,t$  & 19\% & 13\% & 19\% \\
\bottomrule
\bottomrule
\end{tabular}
\caption{Dominant BRs ($>5\%$) for the decays of the exotic scalars.}
\label{tab:cross_sections_BRs}
\end{table}
\endgroup

Finally, we investigate the possibility of testing the predictions of our model via the LHC searches in the scalar sector. There is a plethora of experimental analyses, both by ATLAS and CMS, that look for the non-SM Higgs bosons (see Ref.~\cite{Kling:2020hmi} for a recent review in the framework of the 2HDM). At the same time, in our  benchmark scenarios the exotic scalars can decay through a variety of channels (the dominant BRs, obtained with \texttt{SPheno}, are collected in Table~\ref{tab:cross_sections_BRs}). 

To facilitate the analysis, we use the publicly available code 
\texttt{HiggsTools}~\cite{Bahl:2022igd}, a toolbox 
 for evaluating bounds from the direct searches for the exotic scalar particles at LEP and the LHC, whose database contains 258 different limits. We find that all our benchmark scenarios are tagged as ``allowed'' by \texttt{HiggsTools}.

It is instructive to take a closer look at the output of \texttt{HiggsTools}, as it indicates which searches are most sensitive to the spectra featured by our best-fit scenarios. This is quantified by a parameter called ``observed ratio'', $R_{\textrm{obs}}$, which is the ratio of the predicted cross section and the experimental
limit at the 95\%~C.L. The point in the parameter space is excluded if $R_{\textrm{obs}}>1$. We observe that the highest values of $R_{\textrm{obs}}$ (0.6 for BP1 and BP3, 0.14 for BP2) are reached for the $h_2\to\tau^+\tau^-$ and $a_1\to\tau^+\tau^-$ decays constrained by the ATLAS $139\func{fb}^{-1}$ analysis~\cite{ATLAS:2020zms}, despite very low decay BRs in this channel.

To investigate it in more details, we calculated the $a_1/h_2\to\tau^+\tau^-$ cross sections with  \texttt{MadGraph5 MC@NLO}. The results are reported in the last three columns of Table~\ref{tab:conventional_scalar_1}. These are to be compared with the 95\%~C.L. experimental lower bounds on the cross section reported in the third column of Table~\ref{tab:conventional_scalar_1}. We find a very good agreement with the output of \texttt{HiggsTools} in terms of the parameter $R_{\textrm{obs}}$, thus confirming that 
the decays of the exotic scalars into taus are going to be the most promising way of testing the predictions of the model at the LHC. 

In Table~\ref{tab:conventional_scalar_1} we also present other decay channels of $a_1$ and $h_2$ that feature the high sensitivity. While the current experimental bounds on those searches are weaker that those relative to the $\tau^-\tau^+$ final state, they may offer complementary signatures of the model in future LHC runs.

Incidentally, note that the BRs for the decays of $h_2$ and $a_1$ to the EW gauge bosons, $\gamma\gamma$, $ZZ$ and $WW$, are $\mathcal{O}(10^{-8})$ and $\mathcal{O}(10^{-9})$, respectively, which is orders of magnitude below the current experimental bounds. 

Finally, let us comment on the possibility of testing the 
$a_1/h_2 \rightarrow t^+t^-$ decay through the measurement of an effective coupling $g_{a_1/h_2tt}$. This scenario was investigated by CMS in Ref.~\cite{CMS:2019pzc}. Comparing the values of the $g_{a_1/h_2tt}$ coupling evaluated with \texttt{SPheno} (0.084 for BP2, 0.096 for BP3) with the experimental 95\%~C.L. upper bounds (0.80 for BP2 and 0.70 for BP3)\footnote{The CMS analysis~\cite{CMS:2019pzc} does not cover the scalar masses below 400 GeV.} we conclude that no additional constraint on our model arises from this particular search.

\begin{table}[t]
\centering
\renewcommand{\arraystretch}{1.3}
\begin{tabular}{P{0.15\textwidth}|P{0.12\textwidth}|P{0.24\textwidth}|P{0.11\textwidth}|P{0.11\textwidth}|P{0.11\textwidth}}
\toprule
\toprule
Channel & Experiment 
& $\sigma^{\func{exp}}_{95\%}$ (BP1, BP2, BP3) & $\sigma^{\textrm{BP1}}$ & $\sigma^{\textrm{BP2}}$  & $\sigma^{\textrm{BP3}}$  \\
\midrule
\multirow{2}{*}{$a_1/h_2 \rightarrow \tau^+\tau^-$} & CMS~\cite{CMS:2018rmh} 
& $0.060 \quad  0.030 \quad 0.020$ & \multirow{2}{*}{$0.037$} & \multirow{2}{*}{$0.004$} & \multirow{2}{*}{$0.011$} \\ \cmidrule{2-3}
\multirow{2}{*}{} & ATLAS~\cite{ATLAS:2020zms}
& $0.050 \quad 0.020 \quad 0.016$ &  &  &  \\ 
\midrule
\multirow{2}{*}{$a_1/h_2 \rightarrow \mu^+\mu^-$} & CMS~\cite{CMS:2019mij}
& $0.007\quad 0.006 \quad 0.005 $ & \multirow{2}{*}{$1.3 \times 10^{-4}$} & \multirow{2}{*}{$1.4 \times 10^{-5}$} & \multirow{2}{*}{$4.4 \times 10^{-5}$} \\ \cmidrule{2-3}
\multirow{2}{*}{} & ATLAS~\cite{ATLAS:2019odt} 
& $0.009 \quad 0.004 \quad 0.003$ &  &  \\ \midrule
\multirow{2}{*}{$a_1/h_2 \rightarrow b^+b^-$} & CMS~\cite{CMS:2018hir} 
& $6.0 \qquad 3.5 \qquad 3.0$ & \multirow{2}{*}{$0.554$} & \multirow{2}{*}{$0.061$} & \multirow{2}{*}{$0.061$} \\ \cmidrule{2-3}
\multirow{2}{*}{} & ATLAS~\cite{ATLAS:2019tpq} 
& $- \qquad - \qquad -$ &  &  &  \\ \midrule
\midrule
\multirow{2}{*}{$h^{\pm} \rightarrow \bar{t}b$} & CMS~\cite{CMS:2019rlz} 
& $0.40 \qquad 0.30 \qquad 0.25$ & \multirow{2}{*}{$7.6 \times 10^{-3}$} & \multirow{2}{*}{$2.1 \times 10^{-3}$} & \multirow{2}{*}{$4.5 \times 10^{-3}$} \\ \cmidrule{2-3}
\multirow{2}{*}{} & ATLAS~\cite{ATLAS:2021upq} 
& $0.45 \qquad 0.30 \qquad 0.25$ & \multirow{2}{*}{} & \multirow{2}{*}{} & \multirow{2}{*}{} \\ 
\bottomrule
\bottomrule
\end{tabular}
\caption{An overview of the LHC scalar searches which present the highest sensitivity to the benchmark scenarios identified in Sec.~\ref{sec:num}. The columns show, respectively, the decay channel, the experimental analyses investigating this channel, the experimental 95\%~C.L. upper bound on the cross section for the mass corresponding to the mass of the scalar in a benchmark scenario, the actual cross section calculated in the benchmark scenario.}
\label{tab:conventional_scalar_1}
\end{table}

\section{Conclusions} \label{sec:conclusion}

In this study, we performed a global analysis of an extension of the SM which contains one full family of VL fermions, an extra SU(2)$_L$ scalar doublet and an SU(2)$_L$ scalar singlet.  It also features a  U(1)$_X$ global symmetry spontaneously broken by the singlet scalar vev. This scenario was originally proposed in Ref.~\cite{King:2018fcg} to generate the masses of the third and the second family of the SM fermions, as well as to account correctly for their mixing patterns. 

In our analysis we confronted the model with  the experimental bounds from the flavor physics observables, which include the anomalous magnetic moment of the muon and the rare decays of the tau lepton. Additionally, the model was subjected to the theoretical constraints stemming from the stability of the scalar potential and from the perturbativity of the renormalized couplings. Importantly, we revisited and corrected the bounded-from-below and the alignment limits, which in the context of the same model were previously discussed in Refs.~\cite{Hernandez:2021tii,Lee:2022sic}. In particular, we showed that additional constraints on the quartic couplings arise if the  
full three-scalar potential is considered. We also argued that the perturbativity bounds should not be imposed on the low-scale parameters of the lagrangian but on the running couplings evaluated at the renormalization scale which sets an upper limit of the model's validity. These RG-based perturbativity conditions require the low-scale scalar couplings to be smaller than 2 and the Yukawa couplings smaller than 1.5. 

With all the constraints in place, we performed a numerical scan of the model's parameter space and we identified three benchmark scenarios that satisfied all the theoretical and experimental requirements. 
One distinctive feature of these solutions is that the charged scalar/heavy neutrino loops provide  dominant contributions to the observable $\Delta a_\mu$. This finding is qualitatively different from the conclusions obtained in Refs.~\cite{Hernandez:2021tii,Lee:2022sic} where only the charged lepton loops were considered. 
We would like to emphasize that the dominance of the heavy neutrino contribution to $\Delta a_\mu$ is a generic characteristic of the model and not a mere artifact of the specific benchmark scenarios.  The main reason behind this feature is that the same coupling which generates the neutral scalar/charged lepton loops is also responsible for the correct tree-level mass of the muon and thus it is required to be small. 

We also performed a detailed LHC analysis of our three best-fit scenarios. We investigated the experimental constraints stemming from the direct searches for VLQs, VLLs and exotic scalars. We found that none of the currently available exclusion bounds can test the spectra featured by the benchmark scenarios. This provides a proof of concept that the model in study is feasible as an explanation of both the SM masses and mixings and of the relevant experimental phenomena.  

Regarding future prospects for experimental verification of the model, several observations can be made. Firstly, both charged and neutral VLLs decay predominantly to muons in our framework. On the other hand, all currently available LHC analyses focus on taus in the final state, for which the cross sections obtained in our model are several orders of magnitude below the experimental sensitivity.  
Therefore, we would like to encourage the experimental collaborations to provide dedicated analyses of the VLLs coupled to the second family of the SM fermions.
Such a study would not only allow to test the predictions of our model,  but it would prove very useful in any phenomenological research that aims at explaining the muon $(g-2)$ anomaly in a NP framework with the VLLs. 

Secondly, the experimental searches for the VLQs can become a fruitful testing ground for our model already in the current run of the LHC. The cross sections for the pair production of VLQs featured by the benchmark scenarios are one order of magnitude smaller than the current experimental upper bounds and should be in reach of the dedicated VLQs searches based on the larger data samples.
 
Finally, we observed that the most constraining decay channel for the exotic scalars is $a_1/h_2 \rightarrow \tau^+\tau^-$, for which the ratio of the predicted to the experimental cross sections is close to 1. It may thus provide complementary signatures of our model in future runs at the LHC.

However, the ultimate verification of the NP scenario considered in this study may come from the flavor physics. The Belle-II collaboration plans on  improving, by at least one order of magnitude, their experimental bounds on the rare leptonic decays of the tau lepton. As the corresponding branching ratios featured by our three benchmark scenarios are very close to the current 90\%~C.L. exclusion limits, the rare decays could be the first experimental signatures to be tested. 

\section*{Acknowledgements}

K.K. would like to thank Enrico Sessolo for discussions and comments on the manuscript. A.E.C.H is supported by ANID-Chile FONDECYT 1210378, ANID PIA/APOYO AFB180002 and Milenio-ANID-ICN2019\_044. K.K., H.L. and D.R. are supported by the National Science Centre (Poland) under the research Grant No.~2017/26/E/ST2/00470. The use of the CIS computer cluster at the National Centre for Nuclear Research in Warsaw is gratefully
acknowledged.

\appendix
\section{Fermion mass matrices} \label{sec:App_diag}

\subsection{Charged leptons}\label{sec:charmass}

The mass matrix for the charged leptons, $\mathcal{M}_e$, can be derived from Eq.~(\ref{eqn:Lag_ren}) after identifying the generic fermions $\psi$ with the corresponding lepton fields from Table~\ref{tab:BSM_model} and the generic scalar $H$ with $H_d$. One thus has
\begin{equation}
\psi_{iR}=e_{iR},\qquad \psi_{iL}=L_{iL}, \qquad \psi_{4R}=e_{4R},\qquad \psi_{4L}=L_{4L}, \qquad
\widetilde{\psi}_{4R}=\widetilde{L}_{4R},\qquad \widetilde{\psi}_{4L}=\widetilde{e}_{4L}\,
\end{equation}
with the following components of the SU(2)$_L$ doublets: $L_{iL}=(\nu_{iL}, e_{iL})^T$, $L_{4L}=(\nu_{4L}, e_{4L})^T$ and $\widetilde{L}_{4R}=(\widetilde{\nu}_{4R}, \widetilde{e}_{4R})^T$. As a result, the mass matrix reads
\begingroup
\begin{equation}\label{eqn:mm_lep_1}
\mathcal{M}_{e }=
\left( 
\begin{array}{c|ccccc}
& e _{1R} & e _{2R} & e _{3R} & e _{4R} & \widetilde{e }_{4R} \\[0.5ex] \hline
e_{1L} & 0 & 0 & 0 & 0 & 0 \\[1ex]
e_{2L} & 0 & 0 & 0 & y_{24}^{e } \frac{v_{d}}{\sqrt{2}} & 0 \\[1ex] 
e_{3L} & 0 & 0 & 0 & y_{34}^{e } \frac{v_{d}}{\sqrt{2}} & -x_{34}^{L } \frac{v_{\phi}}{\sqrt{2}} \\[1ex]
e_{4L} & 0 & 0
& y_{43}^{e } \frac{v_{d}}{\sqrt{2}} & 0 & 
-M_{4}^{L } \\[1ex]
\widetilde{e }_{4L} & 0 & x_{42}^{e} \frac{v_{\phi}}{\sqrt{2}}
& x_{43}^{e} \frac{v_{\phi}}{\sqrt{2}} & M_{4}^{e} & 0 \\ 
\end{array}%
\right)\,,
\end{equation}
\endgroup
where $M_{4}^{L}$ ($M_{4}^{e}$) denotes the mass of the VL lepton doublet (singlet) and $x^L_{34}\equiv x^e_{34}$. To facilitate the comparison with the corresponding mass matrix defined in \texttt{SARAH}, we adopt the sign convention used in the code. Note, however, that such a choice does not affect the conclusions drawn in our study as we allow all the Yukawa couplings and all the VL mass parameters to assume both positive and negative values in our numerical scan.

The $5\times 5$ charged lepton mass matrix $\mathcal{M}_e$ can be diagonalized by means of two unitary transformations $V_{L}^{e}$ and $V_{R}^{e}$,
\begingroup
\begin{equation}\label{eq:vle}
    V_{L}^{e} \mathcal{M}_{e} V_{R}^{e\dagger} = \func{diag}\left( 0, m_{\mu}, m_{\tau}, M_{E_1}, M_{E_{2}} \right).
\end{equation}
\endgroup
In Sec.~\ref{sec:setup} the approximate expressions for the eigenvalues $m_{\mu}$ and $m_{\tau}$ were provided in Eq.~(\ref{eq:masstau}), whereas the analogous formulae for the eigenvalues $M_{E_1}$ and $M_{E_2}$ were given in Eq.~(\ref{eq:massbsme}). While those equations are very useful to get a general idea on which lagrangian parameters are relevant for generating the physical charged lepton masses, in our numerical analysis we diagonalize all the fermion mass matrices numerically, employing  the \texttt{SPheno} code generated by \texttt{SARAH}.

\subsection{Up-type quarks}

In analogy to the charged lepton sector, the mass matrix for the up-type quarks, $\mathcal{M}_u$, can be derived from Eq.~(\ref{eqn:Lag_ren}) after taking $H=H_u$ and making the following identification:
\begin{equation}\label{eq:mq55}
\psi_{iR}=u_{iR},\qquad \psi_{iL}=Q_{iL}, \qquad \psi_{4R}=u_{4R},\qquad \psi_{4L}=Q_{4L}, \qquad
\widetilde{\psi}_{4R}=\widetilde{Q}_{4R},\quad\,\, \widetilde{\psi}_{4L}=\widetilde{u}_{4L}\,.
\end{equation}
In Eq.~(\ref{eq:mq55}) the SU(2)$_L$ doublets have the following components: $Q_{iL}=(u_{iL}, d_{iL})^T$, $Q_{4L}=(u_{4L}, d_{4L})^T$ and $\widetilde{Q}_{4R}=(\widetilde{u}_{4R}, \widetilde{d}_{4R})^T$. The corresponding mass matrix with the \texttt{SARAH} sign convention reads\begingroup
\begin{equation}
\mathcal{M}_{u }=\left( 
\begin{array}{c|ccccc}
& u _{1R} & u _{2R} & u _{3R} & u _{4R} & \widetilde{u }_{4R} \\[0.5ex] \hline
u_{1L} & 0 & 0 & 0 & 0 & 0 \\[1ex]
u_{2L} & 0 & 0 & 0 & -y_{24}^{u } \frac{v_{u}}{\sqrt{2}} & 0 \\[1ex] 
u_{3L} & 0 & 0 & 0 & -y_{34}^{u } \frac{v_{u}}{\sqrt{2}} & x_{34}^{Q } \frac{v_{\phi}}{\sqrt{2}} \\[1ex]
u_{4L} & 0 & 0
& -y_{43}^{u } \frac{v_{u}}{\sqrt{2}} & 0 & 
-M_{4}^{Q } \\[1ex]
\widetilde{u }_{4L} & 0 & x_{42}^{u} \frac{v_{\phi}}{\sqrt{2}}
& x_{43}^{u} \frac{v_{\phi}}{\sqrt{2}} & M_{4}^{u} & 0 \\ 
\end{array}%
\right)\,,
\label{eqn:mm_up_1}
\end{equation}
\endgroup
with $x^Q_{34}\equiv x^u_{34}$.
The up-type quark mass matrix $\mathcal{M}_u$ can be diagonalized via the mixing matrices $V_{L}^{u}$ and $V_{R}^{u}$ as
\begingroup
\begin{equation}\label{eq:mudiag}
    V_{L}^{u} \mathcal{M}_{u} V_{R}^{u\dagger} = \func{diag}\left( 0, m_{c}, m_{t}, M_{U_1}, M_{U_2} \right).
\end{equation}
\endgroup
The approximate expressions for the eigenvalues $m_{c}$ and $m_{t}$ can be found in Eq.~(\ref{eq:masstop}), and for the eigenvalues $M_{U_1}$ and $M_{U_2}$ in Eqs.~(\ref{eq:massbsmu1}) and (\ref{eq:massbsmu2}), respectively.

\subsection{Down-type quarks}

The down-type quark lagrangian can be obtained from the generic lagrangian~(\ref{eqn:Lag_ren}) by taking $H=H_d$ and making the following replacements of the fermion fields,
\begin{equation}
\psi_{iR}=d_{iR},\qquad \psi_{iL}=Q_{iL}, \qquad \psi_{4R}=d_{4R},\qquad \psi_{4L}=Q_{4L}, \qquad
\widetilde{\psi}_{4R}=\widetilde{Q}_{4R},\quad\,\, \widetilde{\psi}_{4L}=\widetilde{d}_{4L}\,.
\end{equation}
The corresponding down-type quark mass matrix with the \texttt{SARAH} sign convention reads 
\begingroup
\begin{equation}
\mathcal{M}_{d }=\left( 
\begin{array}{c|ccccc}
& d _{1R} & d _{2R} & d _{3R} & d _{4R} & \widetilde{d }_{4R} \\[0.5ex] \hline
d_{1L} & 0 & 0 & 0 & y_{14}^{d} \frac{v_{d}}{\sqrt{2}} & 0 \\[1ex]
d_{2L} & 0 & 0 & 0 & y_{24}^{d } \frac{v_{d}}{\sqrt{2}} & 0 \\[1ex] 
d_{3L} & 0 & 0 & 0 & y_{34}^{d } \frac{v_{d}}{\sqrt{2}} & -x_{34}^{Q } \frac{v_{\phi}}{\sqrt{2}} \\[1ex]
d_{4L} & 0 & 0
& y_{43}^{d } \frac{v_{d}}{\sqrt{2}} & 0 & 
M_{4}^{Q } \\[1ex]
\widetilde{d }_{4L} & 0 & x_{42}^{d} \frac{v_{\phi}}{\sqrt{2}}
& x_{43}^{d} \frac{v_{\phi}}{\sqrt{2}} & M_{4}^{d} & 0 \\ 
\end{array}%
\right)\,.
\label{eqn:mm_down_1}
\end{equation}
\endgroup
Note that, unlike in the case of the up-type quarks and charged leptons, it is impossible to rotate away the $(1,4)$ element of the matrix $\mathcal{M}_{d }$. The reason is that the mixing between the SM doublets $Q_{1L}$ and $Q_{2L}$ has already been used in the up-quark sector to rotate away the corresponding entry of $\mathcal{M}_u$~\cite{King:2018fcg}. As a result, the Yukawa coupling $y_{14}^{d}$ is present in $\mathcal{M}_d$. The down-type quark mass matrix can be diagonalized by the unitary matrices $V_{L}^{d}$ and $V_{R}^{d}$,
\begingroup
\begin{equation}\label{eq:mddiag}
    V_{L}^{d} \mathcal{M}_{d} V_{R}^{d\dagger} = \func{diag}\left( 0, m_{s}, m_{b}, M_{D_1}, M_{D_2} \right).
\end{equation}
\endgroup
The approximate formulae for the eigenvalues $m_{s}$ and $m_{b}$ can be found in Eq.~(\ref{eq:massbot}), and for the eigenvalues $M_{D_1}$ and $M_{D_2}$ in Eq.~(\ref{eq:massbsmd}).

Incidentally, the presence of the matrix element $y_{14}^{d} v_{d}$ has important consequences for the phenomenology of the model defined in Table~\ref{tab:BSM_model}. As it was discussed in Sec.~\ref{sec:setup}, the first generation of the SM fermions remains massless if only one complete VL family is added to the spectrum. 
On the other hand, the mixing of the $d$ quark with the strange and bottom quarks is mediated by $y_{14}^{d} v_{d}$. As a result, the full CKM matrix can be generated in this setup and one needs to include its elements in the global fit. 

\subsection{Neutrino sector}

Finally, we discuss the neutrino mass matrix which emerges  from the particle content given in Table~\ref{tab:BSM_model}. The corresponding lagrangian can be deduced from Eq.~(\ref{eqn:Lag_ren}) after the following identification:
\begin{equation}
\psi_{iL}=L_{iL}, \qquad \psi_{4R}=\nu_{4R},\qquad \psi_{4L}=L_{4L}, \qquad
\widetilde{\psi}_{4R}=\widetilde{L}_{4R},\qquad \widetilde{\psi}_{4L}=\widetilde{\nu}_{4L},\qquad
H=H_u\,,
\end{equation}
where the SU(2)$_L$ doublets $L_{iL}$, $L_{4L}$ and $\widetilde{L}_{4R}$ are defined in Sec.~\ref{sec:charmass}. Note that since there is no $\nu_{iR}$ field in our model, the couplings $y^\nu_{4j}$ and $x^\nu_{4j}$ vanish. On the other hand, the VL neutrino $\nu_{4R}$ is a singlet under the SM gauge symmetry, so an extra term with $H_d^\ast$ replacing $H_u$ arises. In the end, the neutrino lagrangian reads:
\begingroup
\begin{equation}\label{eqn:Lag_neutrino}
\mathcal{L}_{\func{ren},\nu}^{\func{Yukawa}} = 
y_{i4}^{\nu} L_{iL} H_{u} \nu_{4R} + 
x_{i4}^L L_{iL} \phi \widetilde{L}_{4R} 
+ y_{i4}^{\prime\nu} L_{iL} H_{d}^\ast \widetilde{\nu}_{4L} +
M_{4}^{L} L_{4L} \widetilde{L}_{4R} +
M_{4}^{\nu}\widetilde{\nu}_{4L} \nu_{4R} 
+ \func{h.c.}\,.
\end{equation}
\endgroup
Eq.~(\ref{eqn:Lag_neutrino}) defines a mixed Majorana-Dirac neutrino sector, which after the EWSB gives rise to a $7 \times 7$ Majorana neutrino mass matrix
\begingroup
\begin{equation}
\mathcal{M}_{\nu }=\left( 
\begin{array}{c|ccccccc}
& \nu_{1L} & \nu_{2L} & \nu_{3L} & \nu_{4L} & \nu_{4R} & \widetilde{\nu}_{4L} & \widetilde{\nu}_{4R} \\[0.5ex] \hline
\nu_{1L} & 0 & 0 & 0 & 0 & -y_{14}^{\nu} \frac{v_{u}}{\sqrt{2}} & y_{14}^{\prime\nu} \frac{v_{d}}{\sqrt{2}} & 0 \\[1ex]
\nu_{2L} & 0 & 0 & 0 & 0 & -y_{24}^{\nu} \frac{v_{u}}{\sqrt{2}} & y_{24}^{\prime\nu} \frac{v_{d}}{\sqrt{2}} & 0 \\[1ex] 
\nu_{3L} & 0 & 0 & 0 & 0 & -y_{34}^{\nu} \frac{v_{u}}{\sqrt{2}} & y_{34}^{\prime\nu} \frac{v_{d}}{\sqrt{2}} & x_{34}^{L} \frac{v_{\phi}}{\sqrt{2}} \\[1ex]
\nu_{4L} & 0 & 0 & 0 & 0 & 0 & 0 & M_{4}^{L} \\[1ex]
\nu_{4R} & -y_{14}^{\nu} \frac{v_{u}}{\sqrt{2}} & -y_{24}^{\nu} \frac{v_{u}}{\sqrt{2}}
& -y_{34}^{\nu} \frac{v_{u}}{\sqrt{2}} & 0 & 0 & 
M_{4}^{\nu} & 0 \\[1ex]
\widetilde{\nu}_{4L} & y_{14}^{\prime\nu} \frac{v_{d}}{\sqrt{2}} & y_{24}^{\prime\nu} \frac{v_{d}}{\sqrt{2}}
& y_{34}^{\prime\nu} \frac{v_{d}}{\sqrt{2}} & 0 & M_{4}^{\nu} & 0 & 0 \\[1ex]
\widetilde{\nu}_{4R} & 0 & 0 & x_{34}^{L} \frac{v_{\phi}}{\sqrt{2}} & M_{4}^{L} & 0 & 0 & 0 \\
\end{array}%
\right),
\label{eqn:mm_neutrino_1}
\end{equation}
\endgroup
where once again we chose to work with the \texttt{SARAH} sign convention. 

The neutrino mass matrix is symmetric, it can thus be diagonalized via an orthogonal mixing matrix $V^{\nu}$,
\begingroup
\begin{equation}
    V^{\nu} \mathcal{M}_{\nu } V^{\nu\dagger} = \func{diag}\left( 0, m_{\nu_{2}}, m_{\nu_{3}}, M_{N_1},  M_{N_2}, M_{N_3}, M_{N_4} \right).
\end{equation}
\endgroup

\section{Scalar mass matrices} \label{sec:App_scal}

In this Appendix we collect the explicit formulae for the scalar mass matrices derived from the scalar potential~(\ref{eq:scalpot}) under the spontaneous symmetry breaking conditions~(\ref{eq:ewsbmin}).

The CP-even scalar mass matrix in the basis $(\func{Re} H_{u}^{0},\, \func{Re} H_{d}^{0},\,\func{Re} \phi)$ evaluated at the vacuum reads
\begin{equation}\label{eq:m2even}
\mathbf{M}_{\func{CP-even}}^{2} = 
\left(
\begin{array}{ccc}
\lambda_1 v_u^2-\lambda_5\frac{v_d v_\phi^2 }{4 v_u} & \lambda_3 v_u v_d +\lambda_5 \frac{v^2_\phi}{4} & \lambda_7 v_u v_\phi+\lambda_5\frac{v_d v_\phi}{2} \\
\lambda_3 v_u v_d+\lambda_5 \frac{v^2_\phi}{4}  & \lambda_2 v_d^2-\lambda _5\frac{v_u v_\phi^2}{4 v_d} & \lambda_8v_d v_\phi + \lambda_5\frac{v_u v_\phi}{2}\\
 \lambda_7 v_u v_\phi+ \lambda_5\frac{v_d v_\phi}{2} & \lambda_8 v_d v_\phi+\lambda_5  \frac{v_u v_\phi}{2} & \lambda _6 v_\phi^2\\
\end{array}
\right).
\end{equation}
The matrix~(\refeq{eq:m2even}) can be diagonalized by an orthogonal matrix $R_h$ parameterized with three mixing angles. We will denote them as $\alpha_{12}$ for the $(H_u,H_d)$ mixing, $\alpha_{13}$ for the $(H_u,\phi)$ mixing, and $\alpha_{23}$ for the $(H_d,\phi)$ mixing. In this parametrization, the mixing matrix $R_h$ is given by
\begin{equation}\label{eq:rh}
R_h= 
\left(
\begin{array}{ccc}
c_{12} c_{13} & s_{12} c_{13} & s_{13} \\
-s_{12} c_{23} - c_{12} s_{13} s_{23} & c_{12} c_{23} - s_{12} s_{13} s_{23} & c_{13} s_{23} \\  
s_{12} s_{23} - c_{12} s_{13} c_{23} & -c_{12} s_{23} - s_{12} s_{13} c_{23} & c_{13} c_{23}
\end{array}
\right),
\end{equation}
with the standard notation $s_{ij}=\sin\alpha_{ij}$ and $c_{ij}=\cos\alpha_{ij}$.

The elements of the matrix $R_h$ determine the couplings of the physical Higgs bosons with the SM particles. It is convenient to define a reduced coupling as the ratio between the coupling of the physical Higgs scalar $h_i$ and the corresponding coupling of the SM Higgs,
\begin{equation}
c_{h_iXX}=\frac{g_{h_iXX}}{g_{h_{\rm{SM}}XX}}\,,
\end{equation}
where $X$ stands for the SM fermions and gauge bosons. For the model defined in Table~\ref{tab:BSM_model}, the reduced couplings to quarks and charged leptons are given by
\begin{equation}
c_{h_itt}=\frac{(R_h)_{i1}}{\sin\beta},\qquad c_{h_ibb}=\frac{(R_h)_{i2}}{\cos\beta} \qquad c_{h_i\tau\tau}=\frac{(R_h)_{i2}}{\cos\beta}\,,
\end{equation}
while the reduced couplings to the EW gauge bosons read
\begin{equation}
c_{h_iZZ}=c_{h_iWW}=(R_h)_{i1}\sin\beta+(R_h)_{i2}\cos\beta\,.
\end{equation}

In this study we choose to work in the alignment limit, which is defined as a set of constraints on the quartic couplings $\lambda_i$ under which the lightest CP-even scalar $h_1$ has the same tree-level couplings with the SM particles as the SM Higgs. This means that the reduced couplings to fermions should be very close to 1,
\begin{equation}\label{eq:alignment}
\frac{\cos\alpha_{12} \cos\alpha_{13}}{\sin\beta}\approx 1,\qquad\quad \frac{\sin\alpha_{12} \cos\alpha_{13}}{\cos\beta}\approx 1\,.
\end{equation}  
It can be easily verify that Eq.~(\ref{eq:alignment}) leads to the following conditions on the CP-even scalars mixing angles,
\begin{equation}
\alpha_{12}+\beta=\frac{\pi}{2}+n\,\pi,\qquad \alpha_{13}=2\,n\pi\,,\quad \textrm{with}\,\,n=0,1,2\dots
\end{equation}
indicating no mixing between the doublet $H_u$ and the singlet $\phi$. In this setting, the two SU(2)$_L$ scalar doublets mix with the mixing angle $\frac{\pi}{2}-\beta$, while the doublet $H_d$ mixes with the singlet $\phi$ with the mixing angle $\alpha_{23}$. The CP-even scalars mixing matrix thus reduces to
\begin{equation}
R_h^{\text{alignment}}=
\left(
\begin{array}{ccc}
s_\beta & c_\beta & 0 \\
-c_\beta c_{23} & s_\beta c_{23} & s_{23} \\  
c_\beta s_{23}  & -s_\beta s_{23} & c_{23}
\end{array}
\right)=
\left(
\begin{array}{ccc}
1 & 0 & 0 \\
0 & c_{23} & s_{23} \\
0 & -s_{23} & c_{23}
\end{array}
\right)\times
\left(
\begin{array}{ccc}
s_\beta & c_\beta & 0 \\
-c_\beta & s_\beta & 0 \\
0 & 0 & 1
\end{array}
\right)\,.
\end{equation}
The alignment conditions~(\ref{eq:alignment}) translates into the non-trivial relations between the scalar potential couplings,
\begin{eqnarray}
\lambda_8\,c^2_\beta+\lambda_7\,s^2_\beta+\lambda_5\,s_\beta c_\beta&=&0\\
\lambda_2\,c^2_\beta-\lambda_1\,s^2_\beta-\lambda_3(c^2_\beta-s^2_\beta)&=&0\,.
\end{eqnarray}
 One can also express the mixing angle $\alpha_{23}$ in terms of the parameters of the scalar potential,
\begin{equation}
\cos(2\alpha_{23})=-\frac{B_{23}}{\sqrt{4A_{23}^2+B_{23}^2}}\,,\qquad \sin(2\alpha_{23})=-\frac{A_{23}}{\sqrt{4A_{23}^2+B_{23}^2}}\,,
\end{equation}
where we define
\begin{eqnarray}
A_{23}&=&2\, v\,v_\phi s_\beta (c_\beta \lambda_5 + 2 \lambda_7 s_\beta)\label{eq:a23}\\
B_{23}&=&\lambda_5 v_\phi^2 + 4 s_\beta c_\beta\left(\lambda_6 v_\phi^2 - (\lambda_1 - \lambda_3) v^2 s^2_\beta \right)\,.\label{eq:b23}
\end{eqnarray}

The CP-odd mass matrix in the basis $\left( \func{Im} H_{u}^{0}, \func{Im} H_{d}^{0}, \func{Im} \phi \right)$ reads
\begin{equation}\label{eq:m2odd}
\mathbf{M}_{\func{CP-odd}}^{2}
=
-\lambda_5 \left(
\begin{array}{ccc}
\frac{v_d v_\phi^2}{4 v_u} & \frac{v^2_\phi}{4} & \frac{v_d v_\phi}{2}\\
\frac{v^2_\phi}{4} & \frac{v_u v_\phi^2}{4 v_d} & \frac{v_u v_\phi}{2}\\
\frac{v_d v_\phi}{2}& \frac{v_u v_\phi}{2} & v_u v_d-2\frac{\mu_{\func{sb}}^{2}}{\lambda_5} \\
\end{array}
\right)\,.
\end{equation}

Finally, the charged scalar mass matrix is given by
\begin{equation}\label{eq:m2char}
\mathbf{M}_{\func{Charged}}^{2} 
=
\left(
\begin{array}{cc}
\lambda_4\frac{v_d^2}{2}-\lambda_5\frac{v_d v^2_\phi}{4v_u} & \lambda_4 \frac{v_u v_d}{2}-\lambda_5\frac{v_\phi^2}{4}\\
\lambda_4 \frac{v_u v_d}{2}-\lambda_5\frac{v_\phi^2}{4} & \lambda_4\frac{v_u^2}{2}-\lambda_5\frac{v_uv^2_\phi}{4v_d}
\end{array}
\right).
\end{equation}

\section{Derivation of the bounded-from-below conditions} \label{sec:App_vs}

In this Appendix, we derive the scalar potential bounded-from-below conditions shown in Eq.~(\ref{eqn:VS_conditions}). In doing so, we follow the approach of Ref.~\cite{Bhattacharyya:2015nca} and we extend it to the three-field case.

In order to determine the shape of the scalar potential  (\ref{eq:scalpot}) in the limit of the large fields, it is enough to investigate the behavior of the quartic terms,
\begin{equation}\label{eq:scal4}
\begin{split}
V_{4}&= \frac{1}{2} \lambda_{1}(H_{u}^{\dagger} H_{u})^{2} + \frac{1}{2} \lambda_{2}(H_{d}^{\dagger} H_{d})^{2} + \lambda_{3} (H_{u}^{\dagger} H_{u}) (H_{d}^{\dagger} H_{d}) + \lambda_{4} (H_{u}^{\dagger} H_{d}) (H_{d}^{\dagger} H_{u})
\\
&- \frac{1}{2}\lambda_{5} (\epsilon_{ij} H_{u}^{i} H_{d}^{j} \phi^{2} + \func{H.c.}) + \frac{1}{2}\lambda_{6} (\phi^{*}\phi)^{2} + \lambda_{7} (\phi^{*}\phi)(H_{u}^{\dagger}H_{u}) + \lambda_{8} (\phi^{*}\phi)(H_{d}^{\dagger}H_{d}).
\end{split}
\end{equation}
It is convenient to parameterize each quartic term in the following way,
\begin{equation}
    \begin{split}
        a &= H_{u}^{\dagger} H_{u} \\
        b &= H_{d}^{\dagger} H_{d} \\
        c &= \phi^{*} \phi \\
        d &= \func{Re} H_{u}^{\dagger} H_{d} \\
        e &= \func{Im} H_{u}^{\dagger} H_{d} \\
        f &= \func{Re} \epsilon_{ij} H_{u}^{i} H_{d}^{j} \phi^{2} \\
        g &= \func{Im} \epsilon_{ij} H_{u}^{i} H_{d}^{j} \phi^{2}\,.
    \end{split}
\end{equation}
To make our results more general, we allow $\lambda_{5}$ to be complex. Note that $a,b,c \geq 0$ by definition, and 
\begin{equation}
    \begin{split}
        a\, b &\geq d^{2} + e^{2} \\
        a\, b\, c^{2} &\geq f^{2} + g^{2} \geq 2 f g.
        \label{eqn:VS_relation}
    \end{split}
\end{equation}
In terms of the new parameters, the scalar potential~(\ref{eq:scal4}) can be rewritten as
\begin{equation}\label{eq:v4par}
\begin{split}
V_{4} &= \frac{1}{4} \left( \sqrt{\lambda_{1}} a - \sqrt{\lambda_{2}} b \right)^{2} + \left( \frac{1}{2} \sqrt{\lambda_{1} \lambda_{2}} + \lambda_{3} \right) a\, b \\
&+ \frac{1}{4} \left( \sqrt{\lambda_{1}} a - \sqrt{\lambda_{6}} c \right)^{2} + \left( \frac{1}{2} \sqrt{\lambda_{1} \lambda_{6}} + \lambda_{7} \right) a\, c \\
&+ \frac{1}{4} \left( \sqrt{\lambda_{2}} b - \sqrt{\lambda_{6}} c \right)^{2} + \left( \frac{1}{2} \sqrt{\lambda_{2} \lambda_{6}} + \lambda_{8} \right) b\, c \\
&+ \lambda_{4} \left( d^{2} + e^{2} \right) - \left( \func{Re} \lambda_{5} f - \func{Im} \lambda_{5} g \right)\,.
\end{split}
\end{equation}

We are now ready to analyze the asymptotic behaviour of the potential~(\ref{eq:v4par}) in different field directions.
\smallskip

\textbf{$\mathbf{a=0}$.} 

The parameters $d,e,f,g$ automatically vanish, see Eq.~(\ref{eqn:VS_relation}), and the global potential reduces to
\begingroup
\begin{equation}
    V_{4} \left( a = d = e = f = g = 0 \right) = \frac{1}{2} \left( \sqrt{\lambda_{2}} b - \sqrt{\lambda_{6}} c \right)^{2} + \left( \lambda_{8} + \sqrt{\lambda_{2} \lambda_{6}} \right) b c \,,
\end{equation}
\endgroup
giving rise to the condition
\begingroup
\begin{equation}
    \lambda_{8} + \sqrt{\lambda_{2} \lambda_{6}} > 0.
\end{equation}
\endgroup

\smallskip
\textbf{$\mathbf{b=0}$.} 

In analogy to the previous case, one obtains
\begingroup
\begin{equation}
    V_{4} \left( b = d = e = f = g = 0 \right) = \frac{1}{2} \left( \sqrt{\lambda_{1}} a - \sqrt{\lambda_{6}} c \right)^{2} + \left( \lambda_{7} + \sqrt{\lambda_{1} \lambda_{6}} \right) a c ,
\end{equation}
\endgroup
which gives
\begingroup
\begin{equation}
    \lambda_{7} + \sqrt{\lambda_{1} \lambda_{6}} > 0.
\end{equation}
\endgroup

\smallskip
\textbf{$\mathbf{c=0}$.} 

This time, only the parameters $f$ and $g$ vanish and the reduced scalar potential reads
\begingroup
\begin{equation}\label{eq:v4czero}
    V_{4} \left( c = f = g = 0 \right) = \frac{1}{2} \left( \sqrt{\lambda_{1}} a - \sqrt{\lambda_{2}} b \right)^{2} + \left( \lambda_{3} + \sqrt{\lambda_{1} \lambda_{2}} \right) a b + \lambda_{4} \left( d^{2} + e^{2} \right)\,.
\end{equation}
\endgroup
In order to determine the fate of the scalar potential $V_4$ at the large field values, we need to analyze additional directions in the field space. We first choose a direction along which $a = \sqrt{\frac{\lambda_{2}}{\lambda_{1}}} b$ and $d = e = 0$. Inserting these expressions into Eq.~(\ref{eq:v4czero}), we arrive to the following condition
\begingroup
\begin{equation}
    \lambda_{3} + \sqrt{\lambda_{1} \lambda_{2}} > 0.
\end{equation}
\endgroup
Choosing another direction, $a = \sqrt{\frac{\lambda_{2}}{\lambda_{1}}} b$ and $a b = d^{2} + e^{2}$, we obtain
\begingroup
\begin{equation}
    \lambda_{3} + \lambda_{4} + \sqrt{\lambda_{1} \lambda_{2}} > 0.
\end{equation}
\endgroup

\smallskip
\textbf{$\mathbf{a = \sqrt{\frac{\lambda_{6}}{\lambda_{1}}} c}$, $\mathbf{b = \sqrt{\frac{\lambda_{6}}{\lambda_{2}}} c}$.} 

Under this assumption the scalar potential~(\ref{eq:v4par}) reduces to 
\begingroup
\begin{equation}
    \begin{split}
        V_{4} 
        &= \lambda_{a} c^{2} + \lambda_{4} \left( d^{2} + e^{2} \right) - \left( \func{Re} \lambda_{5} f - \func{Im} \lambda_{5} g \right),
    \end{split}
\end{equation}
\endgroup
where one defines
\begingroup
\begin{equation}
    \lambda_{a} = \frac{3}{2} \lambda_{6} + \lambda_{3} \frac{\lambda_{6}}{\sqrt{\lambda_{1} \lambda_{2}}} + \lambda_{7} \frac{\lambda_{6}}{\lambda_{1}} + \lambda_{8} \frac{\lambda_{6}}{\lambda_{2}}.
\end{equation}
\endgroup
From Eq.~(\ref{eqn:VS_relation}) one has
\begingroup
\begin{equation}
    c^{2} \geq \frac{f^{2} + g^{2}}{d^{2} + e^{2}}\,,
\end{equation}
\endgroup
leading to
\begingroup
\begin{equation}\label{eq:v4ineq}
    \begin{split}
        V_{4}\geq \lambda_{a}\frac{ f^{2} + g^{2} }{d^{2} + e^{2}}  + \lambda_{4} \left( d^{2} + e^{2} \right) - \left( \func{Re} \lambda_{5}\, f - \func{Im} \lambda_{5}', g \right).
    \end{split}
\end{equation}
\endgroup
The r.h.s. of Eq.~(\ref{eq:v4ineq}) can now be rewritten as
\begin{equation}
\textrm{R.H.S}=       \left( f - \frac{\func{Re} \lambda_{5}}{c_{1}} \right)^{2}+ \left( g + \frac{\func{Im} \lambda_{5}}{c_{1}} \right)^{2}- \frac{1}{4c_{1}} \left( (\func{Re} \lambda_{5})^{2}+  (\func{Im} \lambda_{5})^{2} \right) + \lambda_{4} \left( d^{2} + e^{2} \right)\,,
\end{equation}
where
\begin{equation}
    c_{1} = \frac{\lambda_{a}}{d^{2} + e^{2}}.
\end{equation}
Choosing an additional direction in the field space, $f = \frac{\func{Re} \lambda_{5}}{c_{1}}$ and $ g = -\frac{\func{Im} \lambda_{5}}{c_{1}} $, we can derive the following condition,
\begin{equation}
 -\frac{1}{4} \frac{(\func{Re} \lambda_{5})^{2} + (\func{Im} \lambda_{5})^{2}}{\lambda_{a}} + \lambda_{4}  > 0\,.
\end{equation}

Finally, let us rewrite the r.h.s. of Eq.~(\ref{eq:v4ineq}) in yet another way,
\begin{equation}
\textrm{R.H.S}=\left( \frac{\sqrt{c_{2}}}{\sqrt{d^{2} + e^{2}}} - \sqrt{\lambda_{4} \left( d^{2} + e^{2} \right)} \right)^{2} + 2\sqrt{c_{2} \lambda_{4}} - \left( \func{Re} \lambda_{5} f - \func{Im} \lambda_{5} g \right) \,,
\end{equation}
where 
\begin{equation}
    c_{2} = \lambda_{a} \left( f^{2} + g^{2} \right), \quad \lambda_{b} = \sqrt{\lambda_{a} \lambda_{4}}\,.
\end{equation}
Analyzing the quartic potential along the direction $\sqrt{c_2}=\sqrt{\lambda_4}(d^2+f^2)$, we obtain 
\begingroup
\begin{equation}
    V_{4}\geq 
        \left( 4\lambda_{b}^{2} -(\func{Re} \lambda_{5})^{2} + \func{Re} \lambda_{5} \func{Im} \lambda_{5} \right) f^{2} + \left( 4\lambda_{b}^{2} - (\func{Im} \lambda_{5})^{2} + \func{Re} \lambda_{5} \func{Im} \lambda_{5} \right) g^{2}\,, 
\end{equation}
\endgroup
leading straightforwardly to the last two conditions,
\begingroup
\begin{equation}
    \begin{split}
    4\lambda_{b}^{2} - (\func{Re} \lambda_{5})^{2} + \func{Re} \lambda_{5} \func{Im} \lambda_{5} &> 0 \\    
    4\lambda_{b}^{2} - (\func{Im} \lambda_{5})^{2} + \func{Re} \lambda_{5} \func{Im} \lambda_{5} &> 0.
    \end{split}
\end{equation}
\endgroup


\section{Renormalization group equations} \label{sec:App_RG}

In this Appendix, we collect the one-loop RGEs of our model computed with \texttt{SARAH}~\cite{Staub:2013tta,Staub:2015kfa}. We denote
\begin{equation}
\beta(X)\equiv \mu\frac{dX}{d\mu}\equiv \frac{1}{16\pi^2}\beta^{(1)}(X)\,.
\end{equation}

\begingroup
{\allowdisplaybreaks
\begin{align}
    \beta^{(1)}(g_{1}) &= \frac{103 g_1^3}{15} \\
    \beta^{(1)}(g_{2}) &= -\frac{g_2^3}{3} \\
    \beta^{(1)}(g_{3}) &= -\frac{13 g_3^3}{3} \\
    \beta^{(1)}(\lambda_{1}) &= -\frac{9}{5} g_1^2 \lambda _1-9 g_2^2 \lambda _1+\frac{27
    g_1^4}{100}+\frac{9 g_2^4}{4}+\frac{9}{10} g_1^2 g_2^2+12
    \lambda _1^2+4 \lambda _3^2+2 \lambda _4^2+2 \lambda _7^2+4
    \lambda _3 \lambda _4 \nonumber\\
    &+12 \lambda _1 (y_{43}^{u})^2+12 \lambda
    _1 \left[(y_{24}^ {u})^2+(y_{34}^{u})^2\right]-12 (y_{43}^{
    u})^4-12 \left[(y_{24}^{u})^2+(y_{34}^{ u})^2\right]{}^2\nonumber\\
    &+4 \lambda
    _1 \left[(y_{14}^{\nu })^2+(y_{24}^{\nu })^2+(y_{34}^{\nu
    })^2\right]-4 \left[(y_{14}^{\nu })^2+(y_{24}^{\nu})^2+(y_{34}^{\nu })^2\right]^2 \label{eq:betal1}\\
    \beta^{(1)}(\lambda_{2}) &=-\frac{9}{5} g_1^2 \lambda _2-9
    g_2^2 \lambda _2+\frac{27 g_1^4}{100}+\frac{9
    g_2^4}{4}+\frac{9}{10} g_1^2 g_2^2+12 \lambda _2^2+4    \lambda _3^2+2 \lambda _4^2+2 \lambda _8^2+4 \lambda _3    \lambda _4 \nonumber\\
    &+ 12 \lambda _2 (y_{43}^{d})^2+12 \lambda _2 \left[(y_{14}^{d})^2+(y_{24}^{d})^2+(y_{34}^{d})^2\right]-12( y_{43}^{d})^4-12\left[(y_{14}^{d})^2+(y_{24}^{d})^2+(y_{34}^{d})^2\right]{}^2\nonumber\\
    &+4\lambda_2 (y_{43}^{e})^2+4 \lambda _2 \left[(y_{24}^{e})^2+(y_{34}^{e})^2\right]-4 (y_{43}^{e})^4 -4 \left[(y_{24}^{e})^2+(y_{34}^{e})^2\right]{}^2\nonumber\\
    &+4 \lambda _2 \left[(x_{14}^{\nu })^2+(x_{24}^{\nu})^2+(x_{34}^{\nu })^2\right]-4 \left[(x_{14}^{\nu })^2+(x_{24}^{\nu})^2+(x_{34}^{\nu })^2\right]{}^2 \\
    \beta^{(1)}(\lambda_{3}) &=
    -\frac{9}{5} g_1^2 \lambda _3-9 g_2^2 \lambda
    _3+\frac{27 g_1^4}{100}+\frac{9 g_2^4}{4}+\frac{9}{10}
    g_1^2 g_2^2 +4 \lambda _3^2+2 \lambda _4^2+\lambda _5^2+6 \lambda _1 \lambda _3+6 \lambda _2 \lambda _3+2 \lambda _1\lambda _4\nonumber\\
    &+2 \lambda _2 \lambda _4+2 \lambda _7 \lambda
    _8+6 \lambda _3 (y_{43}^{d})^2+6 \lambda _3 \left[(y_{14}^{d})^2+(y_{24}^{d})^2+(y_{34}^{ d})^2\right]+2 \lambda _3 \left[(y_{24}^{e})^2+(y_{34}^{   e})^2\right]\nonumber\\
    &+2 \lambda _3 (y_{43}^{e})^2+6 \lambda _3 (y_{43}^{u})^2+6 \lambda _3 \left[(y_{24}^{u})^2+(y_{34}^{u})^2\right] +2 \lambda _3 \left[(x_{14}^{\nu})^2+(x_{24}^{\nu })^2+(x_{34}^{\nu })^2\right]\nonumber\\
    &-4 \left[x_{14}^{\nu } y_{14}^{\nu }+x_{24}^{\nu } y_{24}^{\nu}+x_{34}^{\nu } y_{34}^{\nu }\right]{}^2+2 \lambda _3
    \left[(y_{14}^{\nu })^2+(y_{24}^{\nu })^2+(y_{34}^{\nu})^2\right]\\ 
    \beta^{(1)}(\lambda_{4}) &= 
    -\frac{9}{5} g_1^2 \lambda _4-9 g_2^2 \lambda
    _4-\frac{9}{5} g_1^2 g_2^2+4 \lambda _4^2-\lambda _5^2+2 \lambda _1 \lambda _4+2 \lambda _2 \lambda _4+8 \lambda _3 \lambda _4+6 \lambda _4 (y_{43}^{u})^2+6 \lambda _4
    (y_{43}^{d})^2\nonumber\\
   & +6 \lambda _4
    \left[(y_{24}^{u})^2+(y_{34}^{u})^2\right] -12 (y_{43}^{d})^2 (y_{43}^{u})^2-12
    \left[y_{24}^{d} y_{24}^{u}+y_{34}^{d} y_{34}^{u}\right]{}^2+6 \lambda _4 \left[(y_{14}^{ d})^2+(y_{24}^{d})^2+(y_{34}^{d})^2\right]\nonumber\\
    &+2 \lambda _4 (y_{43}^{e})^2+2 \lambda
    _4 \left[(y_{24}^{e})^2+(y_{34}^{e})^2\right]-4\left[y_{24}^{e} y_{24}^{\nu }+y_{34}^{e} y_{34}^{\nu}\right]{}^2+2 \lambda _4
    \left[(x_{14}^{ \nu })^2+(x_{24}^{\nu })^2+(x_{34}^{\nu})^2\right]\nonumber\\
    &+4 \left[x_{14}^{\nu } y_{14}^{\nu }+x_{24}^{\nu }y_{24}^{\nu }+x_{34}^{\nu } y_{34}^{\nu }\right]{}^2+2
    \lambda _4 \left[(y_{14}^{\nu })^2+(y_{24}^{\nu })^2+(y_{34}^{\nu })^2\right] \\
    \beta^{(1)}(\lambda_{5}) &= \lambda _5 \Big\{
    -\frac{9}{20} g_1^2-\frac{9}{4} g_2^2+ \lambda _3-\lambda _4+\lambda _6+2 \lambda _7+2 \lambda _8+6 (x_{34}^{Q})^2+3 \left[(x_{42}^{u})^2+(x_{43}^{u})^2\right] \nonumber\\
    &+\frac{3}{2} (y_{43}^{u})^2+\frac{3}{2} \left[(y_{24}^{u})^2+(y_{34}^{u})^2\right]
    + 3 \left[(x_{42}^{d})^2+(x_{43}^{d})^2\right]+\frac{3}{2} (y_{43}^{d})^2+\frac{3}{2} \left[(y_{14}^{d})^2+(y_{24}^{d})^2+(y_{34}^{ d})^2\right]\nonumber\\
    &+2 (x_{34}^{L})^2+\left[(x_{42}^{e})^2+(x_{43}^{e})^2\right]+\frac{1}{2} (y_{43}^{ e})^2+\frac{1}{2} \left[(y_{24}^{e})^2+(y_{34}^{
    e})^2\right]+\frac{1}{2}
    \left[(x_{14}^{\nu })^2+(x_{24}^{\nu })^2+(x_{34}^{\nu})^2\right]\nonumber\\
    &+\frac{1}{2} \left[(y_{14}^{\nu })^2+(y_{24}^{ \nu })^2+(y_{34}^{\nu })^2\right]\Big\} \\
    \beta^{(1)}(\lambda_{6}) &= 2\lambda _5^2+10 \lambda _6^2+4\lambda _7^2+4 \lambda _8^2+24 \lambda _6 (x_{34}^{Q})^2-24 (x_{34}^{Q})^4+12
    \lambda _6 \left[(x_{42}^{u})^2+(x_{43}^{ u})^2\right]\nonumber\\
    &-12\left[(x_{42}^{u})^2+(x_{43}^{ u})^2\right]{}^2+12 \lambda _6 \left[(x_{42}^{d})^2+(x_{43}^{d})^2\right]-12\left[(x_{42}^{ d})^2+(x_{43}^{d})^2\right]{}^2\nonumber \\
    &+8 \lambda _6 (x_{34}^{L})^2-8(x_{34}^{L})^4+4 \lambda _6\left[(x_{42}^{e})^2+(x_{43}^{ e})^2\right]-4 \left[(x_{42}^{e})^2+(x_{43}^{ e})^2\right]{}^2\label{eq:betal6}\\
    \beta^{(1)}(\lambda_{7}) &= -\frac{9}{10}
    g_1^2 \lambda_7-\frac{9}{2} g_2^2 \lambda _7+2 \lambda_5^2+4 \lambda_7^2+6 \lambda _1 \lambda _7+4 \lambda_6 \lambda_7+4 \lambda _3 \lambda _8+2 \lambda_4 \lambda_8 \nonumber\\
    &-12 (x_{34}^{Q})^2 (y_{34}^{u})^2+12 \lambda _7 (x_{34}^{Q})^2+6\lambda _7 \left[(x_{42}^{u})^2+(x_{43}^{u})^2\right]-12(x_{43}^{u})^2 (y_{43}^{u})^2+6 \lambda _7 (y_{43}^{u})^2\nonumber\\\
    &+6\lambda _7 \left[(y_{24}^{u})^2+(y_{34}^{ u})^2\right]+6 \lambda _7 \left[(x_{42}^{ d})^2+(x_{43}^{d})^2\right]+4 \lambda _7 (x_{34}^{ L})^2-4 (x_{34}^{L})^2 (y_{34}^{\nu})^2\nonumber\\
    &+2 \lambda_7 \left[(x_{42}^{e})^2+(x_{43}^{ e})^2\right]+2 \lambda_7 \left[(y_{14}^{\nu })^2+(y_{24}^{\nu })^2+(y_{34}^{\nu})^2\right] \label{eq:betal7}\\
    \beta^{(1)} (\lambda_{8}) &= 
    -\frac{9}{10} g_1^2 \lambda _8-\frac{9}{2} g_2^2 \lambda _8+2 \lambda _5^2+4 \lambda _8^2 +4 \lambda _3 \lambda _7+2 \lambda _4 \lambda _7+6 \lambda _2 \lambda_8+4 \lambda _6 \lambda _8  -12 (y_{34}^{d})^2 (x_{34}^{Q})^2\nonumber\\
   &+12 \lambda _8 (x_{34}^{Q})^2+6 \lambda _8 \left[(x_{42}^{u})^2+(x_{43}^{u})^2\right]+6 \lambda _8 \left[(x_{42}^{d})^2+(x_{43}^{d})^2\right]-12 (x_{43}^{d})^2 (y_{43}^{d})^2+6\lambda _8 (y_{43}^{d})^2\nonumber\\
   &+6 \lambda _8 \left[(y_{14}^{d})^2+(y_{24}^{d})^2+(y_{34}^{ d})^2\right]+4 \lambda _8(x_{34}^{L})^2-4 (x_{34}^{L})^2(y_{34}^{e})^2-4
    (x_{34}^{L})^2(x_{34}^{\nu })^2\nonumber \\
    &+2 \lambda _8 \left[(x_{42}^{e})^2+(x_{43}^{
    e})^2\right]-4 (x_{43}^{e})^2 (y_{43}^{e})^2+2 \lambda _8 (y_{43}^{e})^2+2 \lambda _8 \left[(y_{24}^{ e})^2+(y_{34}^{e})^2\right]\nonumber \\
    &+2 \lambda _8\left[(x_{14}^{\nu })^2+(x_{24}^{ \nu })^2+(x_{34}^{\nu})^2\right] \\
    \beta^{(1)} (y_{43}^{d}) &=-\frac{1}{4} g_1^2 y_{43}^{d}-\frac{9}{4} g_2^2 y_{43}^{d}-8 g_3^2 y_{43}^{d}+y_{43}^{d} \left[(x_{14}^{\nu})^2+(x_{24}^{\nu})^2+(x_{34}^{\nu})^2\right]+\frac{1}{2} (x_{43}^{d})^2 y_{43}^{d} +\frac{1}{2} y_{43}^{d} (y_{43}^{u})^2\nonumber\\\
    &+3 y_{43}^{d} \left[(y_{14}^{d})^2+(y_{24}^{d})^2+(y_{34}^{d})^2\right]+y_{43}^{d} \left[(y_{24}^{e})^2+(y_{34}^{e})^2\right]+y_{43}^{d} (y_{43}^{e})^2+\frac{9}{2} (y_{43}^{d})^3 \\
    \beta^{(1)}(y_{14}^{d}) &= -\frac{1}{4} g_1^2 y_{14}^{d}-\frac{9}{4} g_2^2 y_{14}^{d}-8 g_3^2 y_{14}^{d}+y_{14}^{d} \left[(x_{14}^{\nu})^2+(x_{24}^{\nu})^2+(x_{34}^{\nu})^2\right] \nonumber\\\
    &+\frac{9}{2} y_{14}^{d} \left[(y_{14}^{d})^2+(y_{24}^{d})^2+(y_{34}^{d})^2\right] +y_{14}^{d} \left[(y_{24}^{e})^2+(y_{34}^{e})^2\right] +3 y_{14}^{d} (y_{43}^{d})^2+y_{14}^{d} (y_{43}^{e})^2 \\
   \beta^{(1)}(y_{24}^{d}) &= -\frac{1}{4} g_1^2 y_{24}^{d} -\frac{9}{4} g_2^2 y_{24}^{d}-8 g_3^2 y_{24}^{d}+y_{24}^{d} \left[(x_{14}^{\nu})^2+(x_{24}^{\nu})^2+(x_{34}^{\nu})^2\right]\nonumber\\\
   &+\frac{9}{2} y_{24}^{d} \left[(y_{14}^{d})^2+(y_{24}^{d})^2+(y_{34}^{d})^2\right]+y_{24}^{d} \left[(y_{24}^{e})^2+(y_{34}^{e})^2\right]+\frac{1}{2} y_{24}^{d} (y_{24}^{u})^2\nonumber\\\
   &+3 y_{24}^{d} (y_{43}^{d})^2+ y_{24}^{d} (y_{43}^{e})^2+\frac{1}{2} y_{24}^{u} y_{34}^{d} y_{34}^{u}\\
   \beta^{(1)}(y_{34}^{d}) &= -\frac{1}{4} g_1^2 y_{34}^{d}-\frac{9}{4} g_2^2 y_{34}^{d}-8 g_3^2 y_{34}^{d}+y_{34}^{d} \left[(x_{14}^{\nu})^2+(x_{24}^{\nu})^2+(x_{34}^{\nu})^2\right]+\frac{1}{2} (x_{34}^{Q})^2 y_{34}^{d} \nonumber\\\
   &+\frac{9}{2} y_{34}^{d} \left[(y_{14}^{d})^2+(y_{24}^{d})^2+(y_{34}^{d})^2\right] +\frac{1}{2} y_{24}^{d} y_{24}^{u} y_{34}^{u}+y_{34}^{d} \left[(y_{24}^{e})^2+(y_{34}^{e})^2\right]\nonumber\\\
   &+\frac{1}{2} y_{34}^{d} (y_{34}^{u})^2+3 y_{34}^{d} (y_{43}^{d})^2+y_{34}^{d} (y_{43}^{e})^2\\
   \beta^{(1)}(y_{43}^{u}) &= -\frac{17}{20} g_1^2 y_{43}^{u}-\frac{9}{4} g_2^2 y_{43}^{u}-8 g_3^2 y_{43}^{u}+\frac{1}{2} (x_{43}^{u})^2 y_{43}^{u}+y_{43}^{u} \left[(y_{14}^{\nu})^2+(y_{24}^{\nu})^2+(y_{34}^{\nu})^2\right]\nonumber\\\
   &+3 y_{43}^{u} \left[(y_{24}^{u})^2+(y_{34}^{u})^2\right]+\frac{1}{2} (y_{43}^{d})^2 y_{43}^{u}+\frac{9}{2} (y_{43}^{u})^3 \\
   \beta^{(1)}(y_{24}^{u}) &= -\frac{17}{20} g_1^2 y_{24}^{u}-\frac{9}{4} g_2^2 y_{24}^{u}-8 g_3^2 y_{24}^{u}+y_{24}^{u} \left[(y_{14}^{\nu})^2+(y_{24}^{\nu})^2+(y_{34}^{\nu})^2\right]+\frac{1}{2} (y_{24}^{d})^2 y_{24}^{u}\nonumber\\\
   &+\frac{1}{2} y_{24}^{d} y_{34}^{d} y_{34}^{u}+\frac{9}{2} y_{24}^{u} \left[(y_{24}^{u})^2+(y_{34}^{u})^2\right]+3 y_{24}^{u} (y_{43}^{u})^2 \\
   \beta^{(1)}(y_{34}^{u}) &= -\frac{17}{20} g_1^2 y_{34}^{u}-\frac{9}{4} g_2^2 y_{34}^{u}-8 g_3^2 y_{34}^{u}+\frac{1}{2} (x_{34}^{Q})^2 y_{34}^{u}+y_{34}^{u} \left[(y_{14}^{\nu})^2+(y_{24}^{\nu})^2+(y_{34}^{\nu})^2\right]\nonumber\\\
   &+\frac{1}{2} y_{24}^{d} y_{24}^{u} y_{34}^{d}+\frac{9}{2} y_{34}^{u} \left[(y_{24}^{u})^2+(y_{34}^{u})^2\right]+\frac{1}{2} (y_{34}^{d})^2 y_{34}^{u}+3 y_{34}^{u} (y_{43}^{u})^2 \\
   \beta^{(1)}(x_{42}^{d}) &= -\frac{2}{5} g_1^2 x_{42}^{d}-8 g_3^2 x_{42}^{d}+2 (x_{34}^{L})^2 x_{42}^{d}+6 (x_{34}^{Q})^2 x_{42}^{d}+x_{42}^{d} (x_{42}^{e})^2+3 x_{42}^{d} \left[(x_{42}^{u})^2+(x_{43}^{u})^2\right]\nonumber\\\
   &+4 x_{42}^{d} \left[(x_{42}^{d})^2+(x_{43}^{d})^2\right]+x_{42}^{d} (x_{43}^{e})^2\\
   \beta^{(1)}(x_{43}^{d}) &= -\frac{2}{5} g_1^2 x_{43}^{d}-8 g_3^2 x_{43}^{d}+2 (x_{34}^{L})^2 x_{43}^{d}+6 (x_{34}^{Q})^2 x_{43}^{d}+4 x_{43}^{d} \left[(x_{42}^{d})^2+(x_{43}^{d})^2\right]+(x_{42}^{e})^2 x_{43}^{d}\nonumber\\\
   &+3 x_{43}^{d} \left[(x_{42}^{u})^2+(x_{43}^{u})^2\right]+x_{43}^{d} (x_{43}^{e})^2+x_{43}^{d} (y_{43}^{d})^2 \\
   \beta^{(1)}(x_{34}^{Q}) &= -\frac{1}{10} g_1^2 x_{34}^{Q}-\frac{9}{2} g_2^2 x_{34}^{Q}-8 g_3^2 x_{34}^{Q}+2 (x_{34}^{L})^2 x_{34}^{Q}+3 x_{34}^{Q} \left[(x_{42}^{d})^2+ x_{43}^{d})^2\right]+x_{34}^{Q} (x_{42}^{e})^2\nonumber\\\
   &+3 x_{34}^{Q}\left[(x_{42}^{u})^2+(x_{43}^{u})^2\right]+x_{34}^{Q} (x_{43}^{e})^2+\frac{1}{2} x_{34}^{Q} (y_{34}^{d})^2+\frac{1}{2} x_{34}^{Q} (y_{34}^{u})^2+7 (x_{34}^{Q})^3 \\
   \beta^{(1)}(x_{42}^{u}) &= -\frac{8}{5} g_1^2 x_{42}^{u}-8 g_3^2 x_{42}^{u}+2 (x_{34}^{L})^2 x_{42}^{u}+6 (x_{34}^{Q})^2 x_{42}^{u}+3 x_{42}^{u} \left[(x_{42}^{d})^2+(x_{43}^{d})^2\right]\nonumber\\\
   &+x_{42}^{u} (x_{42}^{e})^2+x_{42}^{u} (x_{43}^{e})^2+4 x_{42}^{u}\left[(x_{42}^{u})^2+(x_{43}^{u})^2\right] \\
   \beta^{(1)}(x_{43}^{u}) &= -\frac{8}{5} g_1^2 x_{43}^{u}-8 g_3^2 x_{43}^{u}+2 (x_{34}^{L})^2 x_{43}^{u}+6 (x_{34}^{Q})^2 x_{43}^{u}+3 x_{43}^{u} \left[(x_{42}^{d})^2+(x_{43}^{d})^2\right]\nonumber\\\
   &+(x_{42}^{e})^2 x_{43}^{u}+4 x_{43}^{u} \left[(x_{42}^{u})^2+(x_{43}^{u})^2\right]+(x_{43}^{e})^2 x_{43}^{u}+x_{43}^{u} (y_{43}^{u})^2 \\
   \beta^{(1)}(y_{24}^{e}) &= -\frac{9}{4} g_1^2 y_{24}^{e}-\frac{9}{4} g_2^2 y_{24}^{e}+y_{24}^{e} \left[(x_{14}^{\nu})^2+(x_{24}^{\nu})^2+(x_{34}^{\nu})^2\right]-\frac{3}{2} x_{24}^{\nu} x_{34}^{\nu} y_{34}^{e}-\frac{3}{2} (x_{24}^{\nu})^2 y_{24}^{e}\nonumber\\\
   &+ \frac{1}{2} y_{24}^{e} (y_{24}^{\nu})^2 +3 y_{24}^{e} \left[(y_{14}^{d})^2+(y_{24}^{d})^2+(y_{34}^{d})^2\right]+\frac{5}{2} y_{24}^{e} \left[(y_{24}^{e})^2+(y_{34}^{e})^2\right]\nonumber\\\
   &+3 y_{24}^{e} (y_{43}^{d})^2+y_{24}^{e} (y_{43}^{e})^2+\frac{1}{2} y_{24}^{\nu} y_{34}^{e} y_{34}^{\nu} \\
   \beta^{(1)}(y_{34}^{e}) &= -\frac{9}{4} g_1^2 y_{34}^{e}-\frac{9}{4} g_2^2 y_{34}^{e}+y_{34}^{e} \left[(x_{14}^{\nu})^2+(x_{24}^{\nu})^2+(x_{34}^{\nu})^2\right]-\frac{3}{2} x_{24}^{\nu} x_{34}^{\nu} y_{24}^{e}-\frac{3}{2} (x_{34}^{\nu})^2 y_{34}^{e}\nonumber\\\
   &+\frac{1}{2} x_{34}^{L} x_{34}^{Q} y_{34}^{e} +3 y_{34}^{e} \left[(y_{14}^{d})^2+(y_{24}^{d})^2+(y_{34}^{d})^2\right]+\frac{1}{2} y_{24}^{e} y_{24}^{\nu} y_{34}^{\nu}+\frac{5}{2} y_{34}^{e} \left[(y_{24}^{e})^2+(y_{34}^{e})^2\right]\nonumber\\\
   &+\frac{1}{2} y_{34}^{e} (y_{34}^{\nu})^2+3 y_{34}^{e} (y_{43}^{d})^2+y_{34}^{e} (y_{43}^{e})^2 \\
   \beta^{(1)}(y_{43}^{e}) &= -\frac{9}{4} g_1^2 y_{43}^{e}-\frac{9}{4} g_2^2 y_{43}^{e}+y_{43}^{e} \left[(x_{14}^{\nu})^2+(x_{24}^{\nu})^2+(x_{34}^{\nu})^2\right]+\frac{1}{2} (x_{43}^{e})^2 y_{43}^{e}+y_{43}^{e} \left[(y_{24}^{e})^2+(y_{34}^{e})^2\right]\nonumber\\\
   &+3 y_{43}^{e} \left[(y_{14}^{d})^2+(y_{24}^{d})^2+(y_{34}^{d})^2\right]+3 (y_{43}^{d})^2 y_{43}^{e}+\frac{5}{2} (y_{43}^{e})^3 \\
   \beta^{(1)}(y_{14}^{\nu}) &= -\frac{9}{20} g_1^2 y_{14}^{\nu}-\frac{9}{4} g_2^2 y_{14}^{\nu}+\frac{1}{2} x_{14}^{\nu} x_{24}^{\nu} y_{24}^{\nu}+\frac{1}{2} x_{14}^{\nu} x_{34}^{\nu} y_{34}^{\nu}+\frac{1}{2} (x_{14}^{\nu})^2 y_{14}^{\nu}\nonumber\\\
   &+\frac{5}{2} y_{14}^{\nu} \left[(y_{14}^{\nu})^2+(y_{24}^{\nu})^2+(y_{34}^{\nu})^2\right]+3 y_{14}^{\nu} \left[(y_{24}^{u})^2+(y_{34}^{u})^2\right]+3 y_{14}^{\nu} (y_{43}^{u})^2 \\
   \beta^{(1)}(y_{24}^{\nu}) &= -\frac{9}{20} g_1^2 y_{24}^{\nu}-\frac{9}{4} g_2^2 y_{24}^{\nu}+\frac{1}{2} x_{14}^{\nu} x_{24}^{\nu} y_{14}^{\nu}+\frac{1}{2} x_{24}^{\nu} x_{34}^{\nu} y_{34}^{\nu}+\frac{1}{2} (x_{24}^{\nu})^2 y_{24}^{\nu}+3 y_{24}^{\nu} \left[(y_{24}^{u})^2+(y_{34}^{u})^2\right]\nonumber\\\
   &+\frac{5}{2} y_{24}^{\nu} \left[(y_{14}^{\nu})^2+(y_{24}^{\nu})^2+(y_{34}^{\nu})^2\right]+\frac{1}{2} (y_{24}^{e})^2 y_{24}^{\nu}+\frac{1}{2} y_{24}^{e} y_{34}^{e} y_{34}^{\nu}+3 y_{24}^{\nu} (y_{43}^{u})^2 \\
   \beta^{(1)}(y_{34}^{\nu}) &= -\frac{9}{20} g_1^2 y_{34}^{\nu}-\frac{9}{4} g_2^2 y_{34}^{\nu}+\frac{1}{2} x_{14}^{\nu} x_{34}^{\nu} y_{14}^{\nu}+\frac{1}{2} x_{24}^{\nu} x_{34}^{\nu} y_{24}^{\nu}+\frac{1}{2} (x_{34}^{L})^2 y_{34}^{\nu}+\frac{1}{2} (x_{34}^{\nu})^2 y_{34}^{\nu}+3 y_{34}^{\nu} (y_{43}^{u})^2\nonumber\\\
   &+\frac{5}{2} y_{34}^{\nu} \left[(y_{14}^{\nu})^2+(y_{24}^{\nu})^2+(y_{34}^{\nu})^2\right]+\frac{1}{2} y_{24}^{e} y_{24}^{\nu} y_{34}^{e}+3 y_{34}^{\nu} \left[(y_{24}^{u})^2+(y_{34}^{u})^2\right]+\frac{1}{2}
   (y_{34}^{e})^2 y_{34}^{\nu} \\
   \beta^{(1)}(x_{42}^{e}) &= -\frac{18}{5} g_1^2 x_{42}^{e}+2 (x_{34}^{L})^2 x_{42}^{e}+6 (x_{34}^{Q})^2 x_{42}^{e}+2 x_{42}^{e} (x_{43}^{e})^2+2 (x_{42}^{e})^3\nonumber\\\
   &+3 x_{42}^{e} \left[(x_{42}^{d})^2+(x_{43}^{d})^2\right]+3 x_{42}^{e} \left[(x_{42}^{u})^2+(x_{43}^{u})^2\right] \\
   \beta^{(1)}(x_{43}^{e}) &= -\frac{18}{5} g_1^2 x_{43}^{e}+2 (x_{34}^{L})^2 x_{43}^{e}+6 (x_{34}^{Q})^2 x_{43}^{e}+3 x_{43}^{e} \left[(x_{42}^{d})^2+(x_{43}^{d})^2\right]+2 (x_{42}^{e})^2 x_{43}^{e}\nonumber\\\
   &+3 x_{43}^{e} \left[(x_{42}^{u})^2+(x_{43}^{u})^2\right] +x_{43}^{e} (y_{43}^{e})^2+2 (x_{43}^{e})^3 \\
   \beta^{(1)}(x_{34}^{L}) &= -\frac{9}{10} g_1^2 x_{34}^{L}-\frac{9}{2} g_2^2 x_{34}^{L}+\frac{1}{2} x_{34}^{L} (x_{34}^{\nu})^2+6 x_{34}^{L} (x_{34}^{Q})^2+3 x_{34}^{L} \left[(x_{42}^{d})^2+(x_{43}^{d})^2\right]+x_{34}^{L} (x_{42}^{e})^2\nonumber\\\
   &+3 x_{34}^{L} \left[(x_{42}^{u})^2+(x_{43}^{u})^2\right]+x_{34}^{L} (x_{43}^{e})^2+\frac{1}{2} x_{34}^{L} (y_{34}^{e})^2+\frac{1}{2} x_{34}^{L} (y_{34}^{\nu})^2+3 (x_{34}^{L})^3 \\
   \beta^{(1)}(x_{14}^{\nu}) &= -\frac{9}{20} g1^2 x_{14}^{\nu}-\frac{9}{4} g2^2 x_{14}^{\nu}+\frac{5}{2} x_{14}^{\nu} \left[(x_{14}^{\nu})^2+(x_{24}^{\nu})^2+(x_{34}^{\nu})^2\right]+3 x_{14}^{\nu} \left[(y_{14}^{d})^2+(y_{24}^{d})^2+(y_{34}^{d})^2\right]\nonumber\\\
   &+\frac{1}{2} x_{14}^{\nu} (y_{14}^{\nu})^2+x_{14}^{\nu} \left[(y_{24}^{e})^2+(y_{34}^{e})^2\right]+3 x_{14}^{\nu} (y_{43}^{d})^2+x_{14}^{\nu} (y_{43}^{e})^2+\frac{1}{2} x_{24}^{\nu} y_{14}^{\nu} y_{24}^{\nu}+\frac{1}{2} x_{34}^{\nu} y_{14}^{\nu} y_{34}^{\nu} \\
   \beta^{(1)}(x_{24}^{\nu}) &= -\frac{9}{20} g1^2 x_{24}^{\nu}-\frac{9}{4} g2^2 x_{24}^{\nu}+\frac{5}{2} x_{24}^{\nu} \left[(x_{14}^{\nu})^2+(x_{24}^{\nu})^2+(x_{34}^{\nu})^2\right] +3 x_{24}^{\nu} \left[(y_{14}^{d})^2+(y_{24}^{d})^2+(y_{34}^{d})^2\right]\nonumber\\\
   &+\frac{1}{2} x_{14}^{\nu} y_{14}^{\nu} y_{24}^{\nu}+x_{24}^{\nu} \left[(y_{24}^{e})^2+(y_{34}^{e})^2\right]-\frac{3}{2} x_{24}^{\nu} (y_{24}^{e})^2+\frac{1}{2} x_{24}^{\nu} (y_{24}^{\nu})^2+3 x_{24}^{\nu} (y_{43}^{d})^2+\nonumber\\\
   &x_{24}^{\nu} (y_{43}^{e})^2-\frac{3}{2} x_{34}^{\nu} y_{24}^{e} y_{34}^{e}+\frac{1}{2} x_{34}^{\nu} y_{24}^{\nu} y_{34}^{\nu} \\
   \beta^{(1)}(x_{34}^{\nu}) &= -\frac{9}{20} g1^2 x_{34}^{\nu}-\frac{9}{4} g2^2 x_{34}^{\nu}+\frac{5}{2} x_{34}^{\nu} \left[(x_{14}^{\nu})^2+(x_{24}^{\nu})^2+(x_{34}^{\nu})^2\right]+\frac{1}{2} x_{14}^{\nu} y_{14}^{\nu} y_{34}^{\nu}-\frac{3}{2} x_{24}^{\nu} y_{24}^{e} y_{34}^{e}\nonumber\\\
   &+\frac{1}{2} x_{24}^{\nu} y_{24}^{\nu} y_{34}^{\nu}+\frac{1}{2} (x_{34}^{L})^2 x_{34}^{\nu}+3 x_{34}^{\nu} \left[(y_{14}^{d})^2+(y_{24}^{d})^2+(y_{34}^{d})^2\right]-\frac{3}{2} x_{34}^{\nu} (y_{34}^{e})^2\nonumber\\\
   &+x_{34}^{\nu} \left[(y_{24}^{e})^2+(y_{34}^{e})^2\right]+\frac{1}{2} x_{34}^{\nu} (y_{34}^{\nu})^2+3 x_{34}^{\nu} (y_{43}^{d})^2+x_{34}^{\nu} (y_{43}^{e})^2
\end{align}
}

\bibliography{bibliography}

\begin{thebibliography}{93}
\expandafter\ifx\csname natexlab\endcsname\relax\def\natexlab#1{#1}\fi
\expandafter\ifx\csname bibnamefont\endcsname\relax
  \def\bibnamefont#1{#1}\fi
\expandafter\ifx\csname bibfnamefont\endcsname\relax
  \def\bibfnamefont#1{#1}\fi
\expandafter\ifx\csname citenamefont\endcsname\relax
  \def\citenamefont#1{#1}\fi
\expandafter\ifx\csname url\endcsname\relax
  \def\url#1{\texttt{#1}}\fi
\expandafter\ifx\csname urlprefix\endcsname\relax\def\urlprefix{URL }\fi
\providecommand{\bibinfo}[2]{#2}
\providecommand{\eprint}[2][]{\url{#2}}

\bibitem[{\citenamefont{Froggatt and Nielsen}(1979)}]{Froggatt:1978nt}
\bibinfo{author}{\bibfnamefont{C.~D.} \bibnamefont{Froggatt}} \bibnamefont{and}
  \bibinfo{author}{\bibfnamefont{H.~B.} \bibnamefont{Nielsen}},
  \bibinfo{journal}{Nucl. Phys. B} \textbf{\bibinfo{volume}{147}},
  \bibinfo{pages}{277} (\bibinfo{year}{1979}).

\bibitem[{\citenamefont{Arkani-Hamed and
  Schmaltz}(2000)}]{Arkani-Hamed:1999ylh}
\bibinfo{author}{\bibfnamefont{N.}~\bibnamefont{Arkani-Hamed}}
  \bibnamefont{and} \bibinfo{author}{\bibfnamefont{M.}~\bibnamefont{Schmaltz}},
  \bibinfo{journal}{Phys. Rev. D} \textbf{\bibinfo{volume}{61}},
  \bibinfo{pages}{033005} (\bibinfo{year}{2000}), \eprint{hep-ph/9903417}.

\bibitem[{\citenamefont{Gherghetta and Pomarol}(2000)}]{Gherghetta:2000qt}
\bibinfo{author}{\bibfnamefont{T.}~\bibnamefont{Gherghetta}} \bibnamefont{and}
  \bibinfo{author}{\bibfnamefont{A.}~\bibnamefont{Pomarol}},
  \bibinfo{journal}{Nucl. Phys. B} \textbf{\bibinfo{volume}{586}},
  \bibinfo{pages}{141} (\bibinfo{year}{2000}), \eprint{hep-ph/0003129}.

\bibitem[{\citenamefont{Huber and Shafi}(2001)}]{Huber:2000ie}
\bibinfo{author}{\bibfnamefont{S.~J.} \bibnamefont{Huber}} \bibnamefont{and}
  \bibinfo{author}{\bibfnamefont{Q.}~\bibnamefont{Shafi}},
  \bibinfo{journal}{Phys. Lett. B} \textbf{\bibinfo{volume}{498}},
  \bibinfo{pages}{256} (\bibinfo{year}{2001}), \eprint{hep-ph/0010195}.

\bibitem[{\citenamefont{Minkowski}(1977)}]{Minkowski:1977sc}
\bibinfo{author}{\bibfnamefont{P.}~\bibnamefont{Minkowski}},
  \bibinfo{journal}{Phys. Lett. B} \textbf{\bibinfo{volume}{67}},
  \bibinfo{pages}{421} (\bibinfo{year}{1977}).

\bibitem[{\citenamefont{Gell-Mann et~al.}(1979)\citenamefont{Gell-Mann, Ramond,
  and Slansky}}]{Gell-Mann:1979vob}
\bibinfo{author}{\bibfnamefont{M.}~\bibnamefont{Gell-Mann}},
  \bibinfo{author}{\bibfnamefont{P.}~\bibnamefont{Ramond}}, \bibnamefont{and}
  \bibinfo{author}{\bibfnamefont{R.}~\bibnamefont{Slansky}},
  \bibinfo{journal}{Conf. Proc. C} \textbf{\bibinfo{volume}{790927}},
  \bibinfo{pages}{315} (\bibinfo{year}{1979}), \eprint{1306.4669}.

\bibitem[{\citenamefont{Yanagida}(1979)}]{Yanagida:1979as}
\bibinfo{author}{\bibfnamefont{T.}~\bibnamefont{Yanagida}},
  \bibinfo{journal}{Conf. Proc. C} \textbf{\bibinfo{volume}{7902131}},
  \bibinfo{pages}{95} (\bibinfo{year}{1979}).

\bibitem[{\citenamefont{Glashow}(1980)}]{Glashow:1979nm}
\bibinfo{author}{\bibfnamefont{S.~L.} \bibnamefont{Glashow}},
  \bibinfo{journal}{NATO Sci. Ser. B} \textbf{\bibinfo{volume}{61}},
  \bibinfo{pages}{687} (\bibinfo{year}{1980}).

\bibitem[{\citenamefont{Mohapatra and Senjanovic}(1981)}]{Mohapatra:1980yp}
\bibinfo{author}{\bibfnamefont{R.~N.} \bibnamefont{Mohapatra}}
  \bibnamefont{and}
  \bibinfo{author}{\bibfnamefont{G.}~\bibnamefont{Senjanovic}},
  \bibinfo{journal}{Phys. Rev. D} \textbf{\bibinfo{volume}{23}},
  \bibinfo{pages}{165} (\bibinfo{year}{1981}).

\bibitem[{\citenamefont{Schechter and Valle}(1982)}]{Schechter:1981cv}
\bibinfo{author}{\bibfnamefont{J.}~\bibnamefont{Schechter}} \bibnamefont{and}
  \bibinfo{author}{\bibfnamefont{J.~W.~F.} \bibnamefont{Valle}},
  \bibinfo{journal}{Phys. Rev. D} \textbf{\bibinfo{volume}{25}},
  \bibinfo{pages}{774} (\bibinfo{year}{1982}).

\bibitem[{\citenamefont{Schechter and Valle}(1980)}]{Schechter:1980gr}
\bibinfo{author}{\bibfnamefont{J.}~\bibnamefont{Schechter}} \bibnamefont{and}
  \bibinfo{author}{\bibfnamefont{J.~W.~F.} \bibnamefont{Valle}},
  \bibinfo{journal}{Phys. Rev. D} \textbf{\bibinfo{volume}{22}},
  \bibinfo{pages}{2227} (\bibinfo{year}{1980}).

\bibitem[{\citenamefont{King}(2018)}]{King:2018fcg}
\bibinfo{author}{\bibfnamefont{S.~F.} \bibnamefont{King}},
  \bibinfo{journal}{JHEP} \textbf{\bibinfo{volume}{09}}, \bibinfo{pages}{069}
  (\bibinfo{year}{2018}), \eprint{1806.06780}.

\bibitem[{\citenamefont{Lee and C\'arcamo~Hern\'andez}(2022)}]{Lee:2022sic}
\bibinfo{author}{\bibfnamefont{H.}~\bibnamefont{Lee}} \bibnamefont{and}
  \bibinfo{author}{\bibfnamefont{A.~E.} \bibnamefont{C\'arcamo~Hern\'andez}}
  (\bibinfo{year}{2022}), \eprint{2207.01710}.

\bibitem[{\citenamefont{C\'arcamo~Hern\'andez
  et~al.}(2022)\citenamefont{C\'arcamo~Hern\'andez, King, and
  Lee}}]{CarcamoHernandez:2021yev}
\bibinfo{author}{\bibfnamefont{A.~E.} \bibnamefont{C\'arcamo~Hern\'andez}},
  \bibinfo{author}{\bibfnamefont{S.~F.} \bibnamefont{King}}, \bibnamefont{and}
  \bibinfo{author}{\bibfnamefont{H.}~\bibnamefont{Lee}},
  \bibinfo{journal}{Phys. Rev. D} \textbf{\bibinfo{volume}{105}},
  \bibinfo{pages}{015021} (\bibinfo{year}{2022}), \eprint{2110.07630}.

\bibitem[{\citenamefont{C\'arcamo~Hern\'andez
  et~al.}(2020)\citenamefont{C\'arcamo~Hern\'andez, King, Lee, and
  Rowley}}]{CarcamoHernandez:2019ydc}
\bibinfo{author}{\bibfnamefont{A.~E.} \bibnamefont{C\'arcamo~Hern\'andez}},
  \bibinfo{author}{\bibfnamefont{S.~F.} \bibnamefont{King}},
  \bibinfo{author}{\bibfnamefont{H.}~\bibnamefont{Lee}}, \bibnamefont{and}
  \bibinfo{author}{\bibfnamefont{S.~J.} \bibnamefont{Rowley}},
  \bibinfo{journal}{Phys. Rev. D} \textbf{\bibinfo{volume}{101}},
  \bibinfo{pages}{115016} (\bibinfo{year}{2020}), \eprint{1910.10734}.

\bibitem[{\citenamefont{Hern\'andez et~al.}(2021)\citenamefont{Hern\'andez,
  King, and Lee}}]{Hernandez:2021tii}
\bibinfo{author}{\bibfnamefont{A.~E.~C.} \bibnamefont{Hern\'andez}},
  \bibinfo{author}{\bibfnamefont{S.~F.} \bibnamefont{King}}, \bibnamefont{and}
  \bibinfo{author}{\bibfnamefont{H.}~\bibnamefont{Lee}},
  \bibinfo{journal}{Phys. Rev. D} \textbf{\bibinfo{volume}{103}},
  \bibinfo{pages}{115024} (\bibinfo{year}{2021}), \eprint{2101.05819}.

\bibitem[{\citenamefont{Wolfenstein}(1983)}]{Wolfenstein:1983yz}
\bibinfo{author}{\bibfnamefont{L.}~\bibnamefont{Wolfenstein}},
  \bibinfo{journal}{Phys. Rev. Lett.} \textbf{\bibinfo{volume}{51}},
  \bibinfo{pages}{1945} (\bibinfo{year}{1983}).

\bibitem[{\citenamefont{Workman et~al.}(2022)}]{ParticleDataGroup:2022pth}
\bibinfo{author}{\bibfnamefont{R.~L.} \bibnamefont{Workman}}
  \bibnamefont{et~al.} (\bibinfo{collaboration}{Particle Data Group}),
  \bibinfo{journal}{PTEP} \textbf{\bibinfo{volume}{2022}},
  \bibinfo{pages}{083C01} (\bibinfo{year}{2022}).

\bibitem[{\citenamefont{Muhlleitner et~al.}(2017)\citenamefont{Muhlleitner,
  Sampaio, Santos, and Wittbrodt}}]{Muhlleitner:2016mzt}
\bibinfo{author}{\bibfnamefont{M.}~\bibnamefont{Muhlleitner}},
  \bibinfo{author}{\bibfnamefont{M.~O.~P.} \bibnamefont{Sampaio}},
  \bibinfo{author}{\bibfnamefont{R.}~\bibnamefont{Santos}}, \bibnamefont{and}
  \bibinfo{author}{\bibfnamefont{J.}~\bibnamefont{Wittbrodt}},
  \bibinfo{journal}{JHEP} \textbf{\bibinfo{volume}{03}}, \bibinfo{pages}{094}
  (\bibinfo{year}{2017}), \eprint{1612.01309}.

\bibitem[{\citenamefont{Camargo-Molina and
  O'Leary}(2014)}]{Camargo-Molina-OLeary:2014}
\bibinfo{author}{\bibfnamefont{J.~E.} \bibnamefont{Camargo-Molina}}
  \bibnamefont{and} \bibinfo{author}{\bibfnamefont{B.}~\bibnamefont{O'Leary}}
  (\bibinfo{year}{2014}),
  \eprint{https://github.com/JoseEliel/VevaciousPlusPlus}.

\bibitem[{\citenamefont{Camargo-Molina
  et~al.}(2013)\citenamefont{Camargo-Molina, O'Leary, Porod, and
  Staub}}]{Camargo-Molina:2013qva}
\bibinfo{author}{\bibfnamefont{J.~E.} \bibnamefont{Camargo-Molina}},
  \bibinfo{author}{\bibfnamefont{B.}~\bibnamefont{O'Leary}},
  \bibinfo{author}{\bibfnamefont{W.}~\bibnamefont{Porod}}, \bibnamefont{and}
  \bibinfo{author}{\bibfnamefont{F.}~\bibnamefont{Staub}},
  \bibinfo{journal}{Eur. Phys. J. C} \textbf{\bibinfo{volume}{73}},
  \bibinfo{pages}{2588} (\bibinfo{year}{2013}), \eprint{1307.1477}.

\bibitem[{\citenamefont{Czarnecki et~al.}(2003)\citenamefont{Czarnecki,
  Marciano, and Vainshtein}}]{Czarnecki:2002nt}
\bibinfo{author}{\bibfnamefont{A.}~\bibnamefont{Czarnecki}},
  \bibinfo{author}{\bibfnamefont{W.~J.} \bibnamefont{Marciano}},
  \bibnamefont{and}
  \bibinfo{author}{\bibfnamefont{A.}~\bibnamefont{Vainshtein}},
  \bibinfo{journal}{Phys. Rev. D} \textbf{\bibinfo{volume}{67}},
  \bibinfo{pages}{073006} (\bibinfo{year}{2003}), \bibinfo{note}{[Erratum:
  Phys.Rev.D 73, 119901 (2006)]}, \eprint{hep-ph/0212229}.

\bibitem[{\citenamefont{Melnikov and Vainshtein}(2004)}]{Melnikov:2003xd}
\bibinfo{author}{\bibfnamefont{K.}~\bibnamefont{Melnikov}} \bibnamefont{and}
  \bibinfo{author}{\bibfnamefont{A.}~\bibnamefont{Vainshtein}},
  \bibinfo{journal}{Phys. Rev. D} \textbf{\bibinfo{volume}{70}},
  \bibinfo{pages}{113006} (\bibinfo{year}{2004}), \eprint{hep-ph/0312226}.

\bibitem[{\citenamefont{Aoyama et~al.}(2012)\citenamefont{Aoyama, Hayakawa,
  Kinoshita, and Nio}}]{Aoyama:2012wk}
\bibinfo{author}{\bibfnamefont{T.}~\bibnamefont{Aoyama}},
  \bibinfo{author}{\bibfnamefont{M.}~\bibnamefont{Hayakawa}},
  \bibinfo{author}{\bibfnamefont{T.}~\bibnamefont{Kinoshita}},
  \bibnamefont{and} \bibinfo{author}{\bibfnamefont{M.}~\bibnamefont{Nio}},
  \bibinfo{journal}{Phys. Rev. Lett.} \textbf{\bibinfo{volume}{109}},
  \bibinfo{pages}{111808} (\bibinfo{year}{2012}), \eprint{1205.5370}.

\bibitem[{\citenamefont{Kurz et~al.}(2014)\citenamefont{Kurz, Liu, Marquard,
  and Steinhauser}}]{Kurz:2014wya}
\bibinfo{author}{\bibfnamefont{A.}~\bibnamefont{Kurz}},
  \bibinfo{author}{\bibfnamefont{T.}~\bibnamefont{Liu}},
  \bibinfo{author}{\bibfnamefont{P.}~\bibnamefont{Marquard}}, \bibnamefont{and}
  \bibinfo{author}{\bibfnamefont{M.}~\bibnamefont{Steinhauser}},
  \bibinfo{journal}{Phys. Lett. B} \textbf{\bibinfo{volume}{734}},
  \bibinfo{pages}{144} (\bibinfo{year}{2014}), \eprint{1403.6400}.

\bibitem[{\citenamefont{Davier et~al.}(2011)\citenamefont{Davier, Hoecker,
  Malaescu, and Zhang}}]{Davier:2010nc}
\bibinfo{author}{\bibfnamefont{M.}~\bibnamefont{Davier}},
  \bibinfo{author}{\bibfnamefont{A.}~\bibnamefont{Hoecker}},
  \bibinfo{author}{\bibfnamefont{B.}~\bibnamefont{Malaescu}}, \bibnamefont{and}
  \bibinfo{author}{\bibfnamefont{Z.}~\bibnamefont{Zhang}},
  \bibinfo{journal}{Eur. Phys. J. C} \textbf{\bibinfo{volume}{71}},
  \bibinfo{pages}{1515} (\bibinfo{year}{2011}), \bibinfo{note}{[Erratum:
  Eur.Phys.J.C 72, 1874 (2012)]}, \eprint{1010.4180}.

\bibitem[{\citenamefont{Gnendiger et~al.}(2013)\citenamefont{Gnendiger,
  St\"ockinger, and St\"ockinger-Kim}}]{Gnendiger:2013pva}
\bibinfo{author}{\bibfnamefont{C.}~\bibnamefont{Gnendiger}},
  \bibinfo{author}{\bibfnamefont{D.}~\bibnamefont{St\"ockinger}},
  \bibnamefont{and}
  \bibinfo{author}{\bibfnamefont{H.}~\bibnamefont{St\"ockinger-Kim}},
  \bibinfo{journal}{Phys. Rev. D} \textbf{\bibinfo{volume}{88}},
  \bibinfo{pages}{053005} (\bibinfo{year}{2013}), \eprint{1306.5546}.

\bibitem[{\citenamefont{Colangelo et~al.}(2014)\citenamefont{Colangelo,
  Hoferichter, Nyffeler, Passera, and Stoffer}}]{Colangelo:2014qya}
\bibinfo{author}{\bibfnamefont{G.}~\bibnamefont{Colangelo}},
  \bibinfo{author}{\bibfnamefont{M.}~\bibnamefont{Hoferichter}},
  \bibinfo{author}{\bibfnamefont{A.}~\bibnamefont{Nyffeler}},
  \bibinfo{author}{\bibfnamefont{M.}~\bibnamefont{Passera}}, \bibnamefont{and}
  \bibinfo{author}{\bibfnamefont{P.}~\bibnamefont{Stoffer}},
  \bibinfo{journal}{Phys. Lett. B} \textbf{\bibinfo{volume}{735}},
  \bibinfo{pages}{90} (\bibinfo{year}{2014}), \eprint{1403.7512}.

\bibitem[{\citenamefont{Davier et~al.}(2017)\citenamefont{Davier, Hoecker,
  Malaescu, and Zhang}}]{Davier:2017zfy}
\bibinfo{author}{\bibfnamefont{M.}~\bibnamefont{Davier}},
  \bibinfo{author}{\bibfnamefont{A.}~\bibnamefont{Hoecker}},
  \bibinfo{author}{\bibfnamefont{B.}~\bibnamefont{Malaescu}}, \bibnamefont{and}
  \bibinfo{author}{\bibfnamefont{Z.}~\bibnamefont{Zhang}},
  \bibinfo{journal}{Eur. Phys. J. C} \textbf{\bibinfo{volume}{77}},
  \bibinfo{pages}{827} (\bibinfo{year}{2017}), \eprint{1706.09436}.

\bibitem[{\citenamefont{Masjuan and Sanchez-Puertas}(2017)}]{Masjuan:2017tvw}
\bibinfo{author}{\bibfnamefont{P.}~\bibnamefont{Masjuan}} \bibnamefont{and}
  \bibinfo{author}{\bibfnamefont{P.}~\bibnamefont{Sanchez-Puertas}},
  \bibinfo{journal}{Phys. Rev. D} \textbf{\bibinfo{volume}{95}},
  \bibinfo{pages}{054026} (\bibinfo{year}{2017}), \eprint{1701.05829}.

\bibitem[{\citenamefont{Colangelo et~al.}(2017)\citenamefont{Colangelo,
  Hoferichter, Procura, and Stoffer}}]{Colangelo:2017fiz}
\bibinfo{author}{\bibfnamefont{G.}~\bibnamefont{Colangelo}},
  \bibinfo{author}{\bibfnamefont{M.}~\bibnamefont{Hoferichter}},
  \bibinfo{author}{\bibfnamefont{M.}~\bibnamefont{Procura}}, \bibnamefont{and}
  \bibinfo{author}{\bibfnamefont{P.}~\bibnamefont{Stoffer}},
  \bibinfo{journal}{JHEP} \textbf{\bibinfo{volume}{04}}, \bibinfo{pages}{161}
  (\bibinfo{year}{2017}), \eprint{1702.07347}.

\bibitem[{\citenamefont{Hoferichter et~al.}(2018)\citenamefont{Hoferichter,
  Hoid, Kubis, Leupold, and Schneider}}]{Hoferichter:2018kwz}
\bibinfo{author}{\bibfnamefont{M.}~\bibnamefont{Hoferichter}},
  \bibinfo{author}{\bibfnamefont{B.-L.} \bibnamefont{Hoid}},
  \bibinfo{author}{\bibfnamefont{B.}~\bibnamefont{Kubis}},
  \bibinfo{author}{\bibfnamefont{S.}~\bibnamefont{Leupold}}, \bibnamefont{and}
  \bibinfo{author}{\bibfnamefont{S.~P.} \bibnamefont{Schneider}},
  \bibinfo{journal}{JHEP} \textbf{\bibinfo{volume}{10}}, \bibinfo{pages}{141}
  (\bibinfo{year}{2018}), \eprint{1808.04823}.

\bibitem[{\citenamefont{Keshavarzi et~al.}(2018)\citenamefont{Keshavarzi,
  Nomura, and Teubner}}]{Keshavarzi:2018mgv}
\bibinfo{author}{\bibfnamefont{A.}~\bibnamefont{Keshavarzi}},
  \bibinfo{author}{\bibfnamefont{D.}~\bibnamefont{Nomura}}, \bibnamefont{and}
  \bibinfo{author}{\bibfnamefont{T.}~\bibnamefont{Teubner}},
  \bibinfo{journal}{Phys. Rev. D} \textbf{\bibinfo{volume}{97}},
  \bibinfo{pages}{114025} (\bibinfo{year}{2018}), \eprint{1802.02995}.

\bibitem[{\citenamefont{Colangelo et~al.}(2019)\citenamefont{Colangelo,
  Hoferichter, and Stoffer}}]{Colangelo:2018mtw}
\bibinfo{author}{\bibfnamefont{G.}~\bibnamefont{Colangelo}},
  \bibinfo{author}{\bibfnamefont{M.}~\bibnamefont{Hoferichter}},
  \bibnamefont{and} \bibinfo{author}{\bibfnamefont{P.}~\bibnamefont{Stoffer}},
  \bibinfo{journal}{JHEP} \textbf{\bibinfo{volume}{02}}, \bibinfo{pages}{006}
  (\bibinfo{year}{2019}), \eprint{1810.00007}.

\bibitem[{\citenamefont{Hoferichter et~al.}(2019)\citenamefont{Hoferichter,
  Hoid, and Kubis}}]{Hoferichter:2019mqg}
\bibinfo{author}{\bibfnamefont{M.}~\bibnamefont{Hoferichter}},
  \bibinfo{author}{\bibfnamefont{B.-L.} \bibnamefont{Hoid}}, \bibnamefont{and}
  \bibinfo{author}{\bibfnamefont{B.}~\bibnamefont{Kubis}},
  \bibinfo{journal}{JHEP} \textbf{\bibinfo{volume}{08}}, \bibinfo{pages}{137}
  (\bibinfo{year}{2019}), \eprint{1907.01556}.

\bibitem[{\citenamefont{Davier et~al.}(2020)\citenamefont{Davier, Hoecker,
  Malaescu, and Zhang}}]{Davier:2019can}
\bibinfo{author}{\bibfnamefont{M.}~\bibnamefont{Davier}},
  \bibinfo{author}{\bibfnamefont{A.}~\bibnamefont{Hoecker}},
  \bibinfo{author}{\bibfnamefont{B.}~\bibnamefont{Malaescu}}, \bibnamefont{and}
  \bibinfo{author}{\bibfnamefont{Z.}~\bibnamefont{Zhang}},
  \bibinfo{journal}{Eur. Phys. J. C} \textbf{\bibinfo{volume}{80}},
  \bibinfo{pages}{241} (\bibinfo{year}{2020}), \bibinfo{note}{[Erratum:
  Eur.Phys.J.C 80, 410 (2020)]}, \eprint{1908.00921}.

\bibitem[{\citenamefont{Keshavarzi et~al.}(2020)\citenamefont{Keshavarzi,
  Nomura, and Teubner}}]{Keshavarzi:2019abf}
\bibinfo{author}{\bibfnamefont{A.}~\bibnamefont{Keshavarzi}},
  \bibinfo{author}{\bibfnamefont{D.}~\bibnamefont{Nomura}}, \bibnamefont{and}
  \bibinfo{author}{\bibfnamefont{T.}~\bibnamefont{Teubner}},
  \bibinfo{journal}{Phys. Rev. D} \textbf{\bibinfo{volume}{101}},
  \bibinfo{pages}{014029} (\bibinfo{year}{2020}), \eprint{1911.00367}.

\bibitem[{\citenamefont{G\'erardin et~al.}(2019)\citenamefont{G\'erardin,
  Meyer, and Nyffeler}}]{Gerardin:2019vio}
\bibinfo{author}{\bibfnamefont{A.}~\bibnamefont{G\'erardin}},
  \bibinfo{author}{\bibfnamefont{H.~B.} \bibnamefont{Meyer}}, \bibnamefont{and}
  \bibinfo{author}{\bibfnamefont{A.}~\bibnamefont{Nyffeler}},
  \bibinfo{journal}{Phys. Rev. D} \textbf{\bibinfo{volume}{100}},
  \bibinfo{pages}{034520} (\bibinfo{year}{2019}), \eprint{1903.09471}.

\bibitem[{\citenamefont{Bijnens et~al.}(2019)\citenamefont{Bijnens,
  Hermansson-Truedsson, and Rodr\'\i{}guez-S\'anchez}}]{Bijnens:2019ghy}
\bibinfo{author}{\bibfnamefont{J.}~\bibnamefont{Bijnens}},
  \bibinfo{author}{\bibfnamefont{N.}~\bibnamefont{Hermansson-Truedsson}},
  \bibnamefont{and}
  \bibinfo{author}{\bibfnamefont{A.}~\bibnamefont{Rodr\'\i{}guez-S\'anchez}},
  \bibinfo{journal}{Phys. Lett. B} \textbf{\bibinfo{volume}{798}},
  \bibinfo{pages}{134994} (\bibinfo{year}{2019}), \eprint{1908.03331}.

\bibitem[{\citenamefont{Colangelo et~al.}(2020)\citenamefont{Colangelo,
  Hagelstein, Hoferichter, Laub, and Stoffer}}]{Colangelo:2019uex}
\bibinfo{author}{\bibfnamefont{G.}~\bibnamefont{Colangelo}},
  \bibinfo{author}{\bibfnamefont{F.}~\bibnamefont{Hagelstein}},
  \bibinfo{author}{\bibfnamefont{M.}~\bibnamefont{Hoferichter}},
  \bibinfo{author}{\bibfnamefont{L.}~\bibnamefont{Laub}}, \bibnamefont{and}
  \bibinfo{author}{\bibfnamefont{P.}~\bibnamefont{Stoffer}},
  \bibinfo{journal}{JHEP} \textbf{\bibinfo{volume}{03}}, \bibinfo{pages}{101}
  (\bibinfo{year}{2020}), \eprint{1910.13432}.

\bibitem[{\citenamefont{Blum et~al.}(2020)\citenamefont{Blum, Christ, Hayakawa,
  Izubuchi, Jin, Jung, and Lehner}}]{Blum:2019ugy}
\bibinfo{author}{\bibfnamefont{T.}~\bibnamefont{Blum}},
  \bibinfo{author}{\bibfnamefont{N.}~\bibnamefont{Christ}},
  \bibinfo{author}{\bibfnamefont{M.}~\bibnamefont{Hayakawa}},
  \bibinfo{author}{\bibfnamefont{T.}~\bibnamefont{Izubuchi}},
  \bibinfo{author}{\bibfnamefont{L.}~\bibnamefont{Jin}},
  \bibinfo{author}{\bibfnamefont{C.}~\bibnamefont{Jung}}, \bibnamefont{and}
  \bibinfo{author}{\bibfnamefont{C.}~\bibnamefont{Lehner}},
  \bibinfo{journal}{Phys. Rev. Lett.} \textbf{\bibinfo{volume}{124}},
  \bibinfo{pages}{132002} (\bibinfo{year}{2020}), \eprint{1911.08123}.

\bibitem[{\citenamefont{Aoyama et~al.}(2019)\citenamefont{Aoyama, Kinoshita,
  and Nio}}]{Aoyama:2019ryr}
\bibinfo{author}{\bibfnamefont{T.}~\bibnamefont{Aoyama}},
  \bibinfo{author}{\bibfnamefont{T.}~\bibnamefont{Kinoshita}},
  \bibnamefont{and} \bibinfo{author}{\bibfnamefont{M.}~\bibnamefont{Nio}},
  \bibinfo{journal}{Atoms} \textbf{\bibinfo{volume}{7}}, \bibinfo{pages}{28}
  (\bibinfo{year}{2019}).

\bibitem[{\citenamefont{Aoyama et~al.}(2020)}]{Aoyama:2020ynm}
\bibinfo{author}{\bibfnamefont{T.}~\bibnamefont{Aoyama}} \bibnamefont{et~al.},
  \bibinfo{journal}{Phys. Rept.} \textbf{\bibinfo{volume}{887}},
  \bibinfo{pages}{1} (\bibinfo{year}{2020}), \eprint{2006.04822}.

\bibitem[{\citenamefont{Bennett et~al.}(2006)}]{Muong-2:2006rrc}
\bibinfo{author}{\bibfnamefont{G.~W.} \bibnamefont{Bennett}}
  \bibnamefont{et~al.} (\bibinfo{collaboration}{Muon g-2}),
  \bibinfo{journal}{Phys. Rev. D} \textbf{\bibinfo{volume}{73}},
  \bibinfo{pages}{072003} (\bibinfo{year}{2006}), \eprint{hep-ex/0602035}.

\bibitem[{\citenamefont{Abi et~al.}(2021)}]{Muong-2:2021ojo}
\bibinfo{author}{\bibfnamefont{B.}~\bibnamefont{Abi}} \bibnamefont{et~al.}
  (\bibinfo{collaboration}{Muon g-2}), \bibinfo{journal}{Phys. Rev. Lett.}
  \textbf{\bibinfo{volume}{126}}, \bibinfo{pages}{141801}
  (\bibinfo{year}{2021}), \eprint{2104.03281}.

\bibitem[{\citenamefont{Aguillard et~al.}(2023)}]{Muong-2:2023cdq}
\bibinfo{author}{\bibfnamefont{D.~P.} \bibnamefont{Aguillard}}
  \bibnamefont{et~al.} (\bibinfo{collaboration}{Muon g-2})
  (\bibinfo{year}{2023}), \eprint{2308.06230}.

\bibitem[{\citenamefont{Kowalska and Sessolo}(2017)}]{Kowalska:2017iqv}
\bibinfo{author}{\bibfnamefont{K.}~\bibnamefont{Kowalska}} \bibnamefont{and}
  \bibinfo{author}{\bibfnamefont{E.~M.} \bibnamefont{Sessolo}},
  \bibinfo{journal}{JHEP} \textbf{\bibinfo{volume}{09}}, \bibinfo{pages}{112}
  (\bibinfo{year}{2017}), \eprint{1707.00753}.

\bibitem[{\citenamefont{Athron et~al.}(2021)\citenamefont{Athron, Bal\'azs,
  Jacob, Kotlarski, St\"ockinger, and St\"ockinger-Kim}}]{Athron:2021iuf}
\bibinfo{author}{\bibfnamefont{P.}~\bibnamefont{Athron}},
  \bibinfo{author}{\bibfnamefont{C.}~\bibnamefont{Bal\'azs}},
  \bibinfo{author}{\bibfnamefont{D.~H.~J.} \bibnamefont{Jacob}},
  \bibinfo{author}{\bibfnamefont{W.}~\bibnamefont{Kotlarski}},
  \bibinfo{author}{\bibfnamefont{D.}~\bibnamefont{St\"ockinger}},
  \bibnamefont{and}
  \bibinfo{author}{\bibfnamefont{H.}~\bibnamefont{St\"ockinger-Kim}},
  \bibinfo{journal}{JHEP} \textbf{\bibinfo{volume}{09}}, \bibinfo{pages}{080}
  (\bibinfo{year}{2021}), \eprint{2104.03691}.

\bibitem[{\citenamefont{Porod}(2003)}]{Porod:2003um}
\bibinfo{author}{\bibfnamefont{W.}~\bibnamefont{Porod}},
  \bibinfo{journal}{Comput. Phys. Commun.} \textbf{\bibinfo{volume}{153}},
  \bibinfo{pages}{275} (\bibinfo{year}{2003}), \eprint{hep-ph/0301101}.

\bibitem[{\citenamefont{Porod and Staub}(2012)}]{Porod:2011nf}
\bibinfo{author}{\bibfnamefont{W.}~\bibnamefont{Porod}} \bibnamefont{and}
  \bibinfo{author}{\bibfnamefont{F.}~\bibnamefont{Staub}},
  \bibinfo{journal}{Comput. Phys. Commun.} \textbf{\bibinfo{volume}{183}},
  \bibinfo{pages}{2458} (\bibinfo{year}{2012}), \eprint{1104.1573}.

\bibitem[{\citenamefont{Mandal and Pich}(2019)}]{Mandal:2019gff}
\bibinfo{author}{\bibfnamefont{R.}~\bibnamefont{Mandal}} \bibnamefont{and}
  \bibinfo{author}{\bibfnamefont{A.}~\bibnamefont{Pich}},
  \bibinfo{journal}{JHEP} \textbf{\bibinfo{volume}{12}}, \bibinfo{pages}{089}
  (\bibinfo{year}{2019}), \eprint{1908.11155}.

\bibitem[{\citenamefont{Abdesselam et~al.}(2021)}]{Belle:2021ysv}
\bibinfo{author}{\bibfnamefont{A.}~\bibnamefont{Abdesselam}}
  \bibnamefont{et~al.} (\bibinfo{collaboration}{Belle}),
  \bibinfo{journal}{JHEP} \textbf{\bibinfo{volume}{10}}, \bibinfo{pages}{19}
  (\bibinfo{year}{2021}), \eprint{2103.12994}.

\bibitem[{\citenamefont{Hayasaka et~al.}(2010)}]{Hayasaka:2010np}
\bibinfo{author}{\bibfnamefont{K.}~\bibnamefont{Hayasaka}}
  \bibnamefont{et~al.}, \bibinfo{journal}{Phys. Lett. B}
  \textbf{\bibinfo{volume}{687}}, \bibinfo{pages}{139} (\bibinfo{year}{2010}),
  \eprint{1001.3221}.

\bibitem[{\citenamefont{Grossman et~al.}(2020)\citenamefont{Grossman, Passemar,
  and Schacht}}]{Grossman:2019bzp}
\bibinfo{author}{\bibfnamefont{Y.}~\bibnamefont{Grossman}},
  \bibinfo{author}{\bibfnamefont{E.}~\bibnamefont{Passemar}}, \bibnamefont{and}
  \bibinfo{author}{\bibfnamefont{S.}~\bibnamefont{Schacht}},
  \bibinfo{journal}{JHEP} \textbf{\bibinfo{volume}{07}}, \bibinfo{pages}{068}
  (\bibinfo{year}{2020}), \eprint{1911.07821}.

\bibitem[{\citenamefont{Belfatto and Trifinopoulos}(2023)}]{Belfatto:2023tbv}
\bibinfo{author}{\bibfnamefont{B.}~\bibnamefont{Belfatto}} \bibnamefont{and}
  \bibinfo{author}{\bibfnamefont{S.}~\bibnamefont{Trifinopoulos}},
  \bibinfo{journal}{Phys. Rev. D} \textbf{\bibinfo{volume}{108}},
  \bibinfo{pages}{035022} (\bibinfo{year}{2023}), \eprint{2302.14097}.

\bibitem[{\citenamefont{Crivellin and Hoferichter}(2020)}]{Crivellin:2020lzu}
\bibinfo{author}{\bibfnamefont{A.}~\bibnamefont{Crivellin}} \bibnamefont{and}
  \bibinfo{author}{\bibfnamefont{M.}~\bibnamefont{Hoferichter}},
  \bibinfo{journal}{Phys. Rev. Lett.} \textbf{\bibinfo{volume}{125}},
  \bibinfo{pages}{111801} (\bibinfo{year}{2020}), \eprint{2002.07184}.

\bibitem[{\citenamefont{Czarnecki et~al.}(2020)\citenamefont{Czarnecki,
  Marciano, and Sirlin}}]{Czarnecki:2019iwz}
\bibinfo{author}{\bibfnamefont{A.}~\bibnamefont{Czarnecki}},
  \bibinfo{author}{\bibfnamefont{W.~J.} \bibnamefont{Marciano}},
  \bibnamefont{and} \bibinfo{author}{\bibfnamefont{A.}~\bibnamefont{Sirlin}},
  \bibinfo{journal}{Phys. Rev. D} \textbf{\bibinfo{volume}{101}},
  \bibinfo{pages}{091301} (\bibinfo{year}{2020}), \eprint{1911.04685}.

\bibitem[{\citenamefont{Cheung et~al.}(2020)\citenamefont{Cheung, Keung, Lu,
  and Tseng}}]{Cheung:2020vqm}
\bibinfo{author}{\bibfnamefont{K.}~\bibnamefont{Cheung}},
  \bibinfo{author}{\bibfnamefont{W.-Y.} \bibnamefont{Keung}},
  \bibinfo{author}{\bibfnamefont{C.-T.} \bibnamefont{Lu}}, \bibnamefont{and}
  \bibinfo{author}{\bibfnamefont{P.-Y.} \bibnamefont{Tseng}},
  \bibinfo{journal}{JHEP} \textbf{\bibinfo{volume}{05}}, \bibinfo{pages}{117}
  (\bibinfo{year}{2020}), \eprint{2001.02853}.

\bibitem[{\citenamefont{Branco et~al.}(2021)\citenamefont{Branco, Penedo,
  Pereira, Rebelo, and Silva-Marcos}}]{Branco:2021vhs}
\bibinfo{author}{\bibfnamefont{G.~C.} \bibnamefont{Branco}},
  \bibinfo{author}{\bibfnamefont{J.~T.} \bibnamefont{Penedo}},
  \bibinfo{author}{\bibfnamefont{P.~M.~F.} \bibnamefont{Pereira}},
  \bibinfo{author}{\bibfnamefont{M.~N.} \bibnamefont{Rebelo}},
  \bibnamefont{and} \bibinfo{author}{\bibfnamefont{J.~I.}
  \bibnamefont{Silva-Marcos}}, \bibinfo{journal}{JHEP}
  \textbf{\bibinfo{volume}{07}}, \bibinfo{pages}{099} (\bibinfo{year}{2021}),
  \eprint{2103.13409}.

\bibitem[{\citenamefont{Albergaria and Branco}(2023)}]{Albergaria:2023vls}
\bibinfo{author}{\bibfnamefont{F.}~\bibnamefont{Albergaria}} \bibnamefont{and}
  \bibinfo{author}{\bibfnamefont{G.~C.} \bibnamefont{Branco}}
  (\bibinfo{year}{2023}), \eprint{2307.13073}.

\bibitem[{\citenamefont{Staub}(2014)}]{Staub:2013tta}
\bibinfo{author}{\bibfnamefont{F.}~\bibnamefont{Staub}},
  \bibinfo{journal}{Comput. Phys. Commun.} \textbf{\bibinfo{volume}{185}},
  \bibinfo{pages}{1773} (\bibinfo{year}{2014}), \eprint{1309.7223}.

\bibitem[{\citenamefont{Staub}(2015)}]{Staub:2015kfa}
\bibinfo{author}{\bibfnamefont{F.}~\bibnamefont{Staub}}, \bibinfo{journal}{Adv.
  High Energy Phys.} \textbf{\bibinfo{volume}{2015}}, \bibinfo{pages}{840780}
  (\bibinfo{year}{2015}), \eprint{1503.04200}.

\bibitem[{\citenamefont{Darm\'e et~al.}(2018)\citenamefont{Darm\'e, Kowalska,
  Roszkowski, and Sessolo}}]{Darme:2018hqg}
\bibinfo{author}{\bibfnamefont{L.}~\bibnamefont{Darm\'e}},
  \bibinfo{author}{\bibfnamefont{K.}~\bibnamefont{Kowalska}},
  \bibinfo{author}{\bibfnamefont{L.}~\bibnamefont{Roszkowski}},
  \bibnamefont{and} \bibinfo{author}{\bibfnamefont{E.~M.}
  \bibnamefont{Sessolo}}, \bibinfo{journal}{JHEP}
  \textbf{\bibinfo{volume}{10}}, \bibinfo{pages}{052} (\bibinfo{year}{2018}),
  \eprint{1806.06036}.

\bibitem[{\citenamefont{Kowalska and Sessolo}(2021)}]{Kowalska:2020zve}
\bibinfo{author}{\bibfnamefont{K.}~\bibnamefont{Kowalska}} \bibnamefont{and}
  \bibinfo{author}{\bibfnamefont{E.~M.} \bibnamefont{Sessolo}},
  \bibinfo{journal}{Phys. Rev. D} \textbf{\bibinfo{volume}{103}},
  \bibinfo{pages}{115032} (\bibinfo{year}{2021}), \eprint{2012.15200}.

\bibitem[{\citenamefont{Kou et~al.}(2019)\citenamefont{Kou, Urquijo,
  Altmannshofer, Beaujean, Bell, Beneke, Bigi, Bishara, Blanke, Bobeth
  et~al.}}]{10.1093/ptep/ptz106}
\bibinfo{author}{\bibfnamefont{E.}~\bibnamefont{Kou}},
  \bibinfo{author}{\bibfnamefont{P.}~\bibnamefont{Urquijo}},
  \bibinfo{author}{\bibfnamefont{W.}~\bibnamefont{Altmannshofer}},
  \bibinfo{author}{\bibfnamefont{F.}~\bibnamefont{Beaujean}},
  \bibinfo{author}{\bibfnamefont{G.}~\bibnamefont{Bell}},
  \bibinfo{author}{\bibfnamefont{M.}~\bibnamefont{Beneke}},
  \bibinfo{author}{\bibfnamefont{I.~I.} \bibnamefont{Bigi}},
  \bibinfo{author}{\bibfnamefont{F.}~\bibnamefont{Bishara}},
  \bibinfo{author}{\bibfnamefont{M.}~\bibnamefont{Blanke}},
  \bibinfo{author}{\bibfnamefont{C.}~\bibnamefont{Bobeth}},
  \bibnamefont{et~al.}, \bibinfo{journal}{Progress of Theoretical and
  Experimental Physics} \textbf{\bibinfo{volume}{2019}},
  \bibinfo{pages}{123C01} (\bibinfo{year}{2019}), ISSN
  \bibinfo{issn}{2050-3911},
  \eprint{https://academic.oup.com/ptep/article-pdf/2019/12/123C01/32693980/ptz106.pdf},
  \urlprefix\url{https://doi.org/10.1093/ptep/ptz106}.

\bibitem[{\citenamefont{Aggarwal et~al.}(2022)}]{Belle-II:2022cgf}
\bibinfo{author}{\bibfnamefont{L.}~\bibnamefont{Aggarwal}} \bibnamefont{et~al.}
  (\bibinfo{collaboration}{Belle-II}) (\bibinfo{year}{2022}),
  \eprint{2207.06307}.

\bibitem[{\citenamefont{Kowalska and Kumar}(2019)}]{Kowalska:2019qxm}
\bibinfo{author}{\bibfnamefont{K.}~\bibnamefont{Kowalska}} \bibnamefont{and}
  \bibinfo{author}{\bibfnamefont{D.}~\bibnamefont{Kumar}},
  \bibinfo{journal}{JHEP} \textbf{\bibinfo{volume}{12}}, \bibinfo{pages}{094}
  (\bibinfo{year}{2019}), \eprint{1910.00847}.

\bibitem[{\citenamefont{Olivas et~al.}(2022)\citenamefont{Olivas, Kowalska, and
  Kumar}}]{Olivas:2021nft}
\bibinfo{author}{\bibfnamefont{U.~C.} \bibnamefont{Olivas}},
  \bibinfo{author}{\bibfnamefont{K.}~\bibnamefont{Kowalska}}, \bibnamefont{and}
  \bibinfo{author}{\bibfnamefont{D.}~\bibnamefont{Kumar}},
  \bibinfo{journal}{JHEP} \textbf{\bibinfo{volume}{03}}, \bibinfo{pages}{132}
  (\bibinfo{year}{2022}), \eprint{2112.11742}.

\bibitem[{\citenamefont{Aad et~al.}(2023{\natexlab{a}})}]{ATLAS:2022tla}
\bibinfo{author}{\bibfnamefont{G.}~\bibnamefont{Aad}} \bibnamefont{et~al.}
  (\bibinfo{collaboration}{ATLAS}), \bibinfo{journal}{Eur. Phys. J. C}
  \textbf{\bibinfo{volume}{83}}, \bibinfo{pages}{719}
  (\bibinfo{year}{2023}{\natexlab{a}}), \eprint{2212.05263}.

\bibitem[{\citenamefont{Tumasyan et~al.}(2023)}]{CMS:2022fck}
\bibinfo{author}{\bibfnamefont{A.}~\bibnamefont{Tumasyan}} \bibnamefont{et~al.}
  (\bibinfo{collaboration}{CMS}), \bibinfo{journal}{JHEP}
  \textbf{\bibinfo{volume}{07}}, \bibinfo{pages}{020} (\bibinfo{year}{2023}),
  \eprint{2209.07327}.

\bibitem[{\citenamefont{Aad et~al.}(2023{\natexlab{b}})}]{ATLAS:2022hnn}
\bibinfo{author}{\bibfnamefont{G.}~\bibnamefont{Aad}} \bibnamefont{et~al.}
  (\bibinfo{collaboration}{ATLAS}), \bibinfo{journal}{Phys. Lett. B}
  \textbf{\bibinfo{volume}{843}}, \bibinfo{pages}{138019}
  (\bibinfo{year}{2023}{\natexlab{b}}), \eprint{2210.15413}.

\bibitem[{\citenamefont{Alwall et~al.}(2014)\citenamefont{Alwall, Frederix,
  Frixione, Hirschi, Maltoni, Mattelaer, Shao, Stelzer, Torrielli, and
  Zaro}}]{Alwall:2014hca}
\bibinfo{author}{\bibfnamefont{J.}~\bibnamefont{Alwall}},
  \bibinfo{author}{\bibfnamefont{R.}~\bibnamefont{Frederix}},
  \bibinfo{author}{\bibfnamefont{S.}~\bibnamefont{Frixione}},
  \bibinfo{author}{\bibfnamefont{V.}~\bibnamefont{Hirschi}},
  \bibinfo{author}{\bibfnamefont{F.}~\bibnamefont{Maltoni}},
  \bibinfo{author}{\bibfnamefont{O.}~\bibnamefont{Mattelaer}},
  \bibinfo{author}{\bibfnamefont{H.~S.} \bibnamefont{Shao}},
  \bibinfo{author}{\bibfnamefont{T.}~\bibnamefont{Stelzer}},
  \bibinfo{author}{\bibfnamefont{P.}~\bibnamefont{Torrielli}},
  \bibnamefont{and} \bibinfo{author}{\bibfnamefont{M.}~\bibnamefont{Zaro}},
  \bibinfo{journal}{JHEP} \textbf{\bibinfo{volume}{07}}, \bibinfo{pages}{079}
  (\bibinfo{year}{2014}), \eprint{1405.0301}.

\bibitem[{\citenamefont{Aad et~al.}(2023{\natexlab{c}})}]{ATLAS:2023bfh}
\bibinfo{author}{\bibfnamefont{G.}~\bibnamefont{Aad}} \bibnamefont{et~al.}
  (\bibinfo{collaboration}{ATLAS}) (\bibinfo{year}{2023}{\natexlab{c}}),
  \eprint{2307.07584}.

\bibitem[{\citenamefont{Aad et~al.}(2023{\natexlab{d}})}]{ATLAS:2023pja}
\bibinfo{author}{\bibfnamefont{G.}~\bibnamefont{Aad}} \bibnamefont{et~al.}
  (\bibinfo{collaboration}{ATLAS}), \bibinfo{journal}{JHEP}
  \textbf{\bibinfo{volume}{08}}, \bibinfo{pages}{153}
  (\bibinfo{year}{2023}{\natexlab{d}}), \eprint{2305.03401}.

\bibitem[{\citenamefont{Aad et~al.}(2022)}]{ATLAS:2022ozf}
\bibinfo{author}{\bibfnamefont{G.}~\bibnamefont{Aad}} \bibnamefont{et~al.}
  (\bibinfo{collaboration}{ATLAS}), \bibinfo{journal}{Phys. Rev. D}
  \textbf{\bibinfo{volume}{105}}, \bibinfo{pages}{092012}
  (\bibinfo{year}{2022}), \eprint{2201.07045}.

\bibitem[{\citenamefont{Aad et~al.}(2023{\natexlab{e}})}]{ATLAS:2023qqf}
\bibinfo{author}{\bibfnamefont{G.}~\bibnamefont{Aad}} \bibnamefont{et~al.}
  (\bibinfo{collaboration}{ATLAS}) (\bibinfo{year}{2023}{\natexlab{e}}),
  \eprint{2308.02595}.

\bibitem[{\citenamefont{Faroughy et~al.}(2017)\citenamefont{Faroughy, Greljo,
  and Kamenik}}]{Faroughy:2016osc}
\bibinfo{author}{\bibfnamefont{D.~A.} \bibnamefont{Faroughy}},
  \bibinfo{author}{\bibfnamefont{A.}~\bibnamefont{Greljo}}, \bibnamefont{and}
  \bibinfo{author}{\bibfnamefont{J.~F.} \bibnamefont{Kamenik}},
  \bibinfo{journal}{Phys. Lett. B} \textbf{\bibinfo{volume}{764}},
  \bibinfo{pages}{126} (\bibinfo{year}{2017}), \eprint{1609.07138}.

\bibitem[{\citenamefont{Kowalska et~al.}(2019)\citenamefont{Kowalska, Sessolo,
  and Yamamoto}}]{Kowalska:2018ulj}
\bibinfo{author}{\bibfnamefont{K.}~\bibnamefont{Kowalska}},
  \bibinfo{author}{\bibfnamefont{E.~M.} \bibnamefont{Sessolo}},
  \bibnamefont{and} \bibinfo{author}{\bibfnamefont{Y.}~\bibnamefont{Yamamoto}},
  \bibinfo{journal}{Phys. Rev. D} \textbf{\bibinfo{volume}{99}},
  \bibinfo{pages}{055007} (\bibinfo{year}{2019}), \eprint{1812.06851}.

\bibitem[{\citenamefont{Aad et~al.}(2023{\natexlab{f}})}]{ATLAS:2023sbu}
\bibinfo{author}{\bibfnamefont{G.}~\bibnamefont{Aad}} \bibnamefont{et~al.}
  (\bibinfo{collaboration}{ATLAS}), \bibinfo{journal}{JHEP}
  \textbf{\bibinfo{volume}{07}}, \bibinfo{pages}{118}
  (\bibinfo{year}{2023}{\natexlab{f}}), \eprint{2303.05441}.

\bibitem[{\citenamefont{Sirunyan
  et~al.}(2019{\natexlab{a}})}]{PhysRevD.100.052003}
\bibinfo{author}{\bibfnamefont{A.~M.} \bibnamefont{Sirunyan}}
  \bibnamefont{et~al.} (\bibinfo{collaboration}{CMS Collaboration}),
  \bibinfo{journal}{Phys. Rev. D} \textbf{\bibinfo{volume}{100}},
  \bibinfo{pages}{052003} (\bibinfo{year}{2019}{\natexlab{a}}),
  \urlprefix\url{https://link.aps.org/doi/10.1103/PhysRevD.100.052003}.

\bibitem[{\citenamefont{Aad et~al.}(2023{\natexlab{g}})}]{ATLAS:2023lfr}
\bibinfo{author}{\bibfnamefont{G.}~\bibnamefont{Aad}} \bibnamefont{et~al.}
  (\bibinfo{collaboration}{ATLAS}) (\bibinfo{year}{2023}{\natexlab{g}}),
  \eprint{2305.09322}.

\bibitem[{\citenamefont{Kling et~al.}(2020)\citenamefont{Kling, Su, and
  Su}}]{Kling:2020hmi}
\bibinfo{author}{\bibfnamefont{F.}~\bibnamefont{Kling}},
  \bibinfo{author}{\bibfnamefont{S.}~\bibnamefont{Su}}, \bibnamefont{and}
  \bibinfo{author}{\bibfnamefont{W.}~\bibnamefont{Su}}, \bibinfo{journal}{JHEP}
  \textbf{\bibinfo{volume}{06}}, \bibinfo{pages}{163} (\bibinfo{year}{2020}),
  \eprint{2004.04172}.

\bibitem[{\citenamefont{Bahl et~al.}(2023)\citenamefont{Bahl, Biek\"otter,
  Heinemeyer, Li, Paasch, Weiglein, and Wittbrodt}}]{Bahl:2022igd}
\bibinfo{author}{\bibfnamefont{H.}~\bibnamefont{Bahl}},
  \bibinfo{author}{\bibfnamefont{T.}~\bibnamefont{Biek\"otter}},
  \bibinfo{author}{\bibfnamefont{S.}~\bibnamefont{Heinemeyer}},
  \bibinfo{author}{\bibfnamefont{C.}~\bibnamefont{Li}},
  \bibinfo{author}{\bibfnamefont{S.}~\bibnamefont{Paasch}},
  \bibinfo{author}{\bibfnamefont{G.}~\bibnamefont{Weiglein}}, \bibnamefont{and}
  \bibinfo{author}{\bibfnamefont{J.}~\bibnamefont{Wittbrodt}},
  \bibinfo{journal}{Comput. Phys. Commun.} \textbf{\bibinfo{volume}{291}},
  \bibinfo{pages}{108803} (\bibinfo{year}{2023}), \eprint{2210.09332}.

\bibitem[{\citenamefont{Aad et~al.}(2020{\natexlab{a}})}]{ATLAS:2020zms}
\bibinfo{author}{\bibfnamefont{G.}~\bibnamefont{Aad}} \bibnamefont{et~al.}
  (\bibinfo{collaboration}{ATLAS}), \bibinfo{journal}{Phys. Rev. Lett.}
  \textbf{\bibinfo{volume}{125}}, \bibinfo{pages}{051801}
  (\bibinfo{year}{2020}{\natexlab{a}}), \eprint{2002.12223}.

\bibitem[{\citenamefont{Sirunyan et~al.}(2020{\natexlab{a}})}]{CMS:2019pzc}
\bibinfo{author}{\bibfnamefont{A.~M.} \bibnamefont{Sirunyan}}
  \bibnamefont{et~al.} (\bibinfo{collaboration}{CMS}), \bibinfo{journal}{JHEP}
  \textbf{\bibinfo{volume}{04}}, \bibinfo{pages}{171}
  (\bibinfo{year}{2020}{\natexlab{a}}), \bibinfo{note}{[Erratum: JHEP 03, 187
  (2022)]}, \eprint{1908.01115}.

\bibitem[{\citenamefont{Sirunyan et~al.}(2018{\natexlab{a}})}]{CMS:2018rmh}
\bibinfo{author}{\bibfnamefont{A.~M.} \bibnamefont{Sirunyan}}
  \bibnamefont{et~al.} (\bibinfo{collaboration}{CMS}), \bibinfo{journal}{JHEP}
  \textbf{\bibinfo{volume}{09}}, \bibinfo{pages}{007}
  (\bibinfo{year}{2018}{\natexlab{a}}), \eprint{1803.06553}.

\bibitem[{\citenamefont{Sirunyan et~al.}(2019{\natexlab{b}})}]{CMS:2019mij}
\bibinfo{author}{\bibfnamefont{A.~M.} \bibnamefont{Sirunyan}}
  \bibnamefont{et~al.} (\bibinfo{collaboration}{CMS}), \bibinfo{journal}{Phys.
  Lett. B} \textbf{\bibinfo{volume}{798}}, \bibinfo{pages}{134992}
  (\bibinfo{year}{2019}{\natexlab{b}}), \eprint{1907.03152}.

\bibitem[{\citenamefont{Aaboud et~al.}(2019)}]{ATLAS:2019odt}
\bibinfo{author}{\bibfnamefont{M.}~\bibnamefont{Aaboud}} \bibnamefont{et~al.}
  (\bibinfo{collaboration}{ATLAS}), \bibinfo{journal}{JHEP}
  \textbf{\bibinfo{volume}{07}}, \bibinfo{pages}{117} (\bibinfo{year}{2019}),
  \eprint{1901.08144}.

\bibitem[{\citenamefont{Sirunyan et~al.}(2018{\natexlab{b}})}]{CMS:2018hir}
\bibinfo{author}{\bibfnamefont{A.~M.} \bibnamefont{Sirunyan}}
  \bibnamefont{et~al.} (\bibinfo{collaboration}{CMS}), \bibinfo{journal}{JHEP}
  \textbf{\bibinfo{volume}{08}}, \bibinfo{pages}{113}
  (\bibinfo{year}{2018}{\natexlab{b}}), \eprint{1805.12191}.

\bibitem[{\citenamefont{Aad et~al.}(2020{\natexlab{b}})}]{ATLAS:2019tpq}
\bibinfo{author}{\bibfnamefont{G.}~\bibnamefont{Aad}} \bibnamefont{et~al.}
  (\bibinfo{collaboration}{ATLAS}), \bibinfo{journal}{Phys. Rev. D}
  \textbf{\bibinfo{volume}{102}}, \bibinfo{pages}{032004}
  (\bibinfo{year}{2020}{\natexlab{b}}), \eprint{1907.02749}.

\bibitem[{\citenamefont{Sirunyan et~al.}(2020{\natexlab{b}})}]{CMS:2019rlz}
\bibinfo{author}{\bibfnamefont{A.~M.} \bibnamefont{Sirunyan}}
  \bibnamefont{et~al.} (\bibinfo{collaboration}{CMS}), \bibinfo{journal}{JHEP}
  \textbf{\bibinfo{volume}{01}}, \bibinfo{pages}{096}
  (\bibinfo{year}{2020}{\natexlab{b}}), \eprint{1908.09206}.

\bibitem[{\citenamefont{Aad et~al.}(2021)}]{ATLAS:2021upq}
\bibinfo{author}{\bibfnamefont{G.}~\bibnamefont{Aad}} \bibnamefont{et~al.}
  (\bibinfo{collaboration}{ATLAS}), \bibinfo{journal}{JHEP}
  \textbf{\bibinfo{volume}{06}}, \bibinfo{pages}{145} (\bibinfo{year}{2021}),
  \eprint{2102.10076}.

\bibitem[{\citenamefont{Bhattacharyya and Das}(2016)}]{Bhattacharyya:2015nca}
\bibinfo{author}{\bibfnamefont{G.}~\bibnamefont{Bhattacharyya}}
  \bibnamefont{and} \bibinfo{author}{\bibfnamefont{D.}~\bibnamefont{Das}},
  \bibinfo{journal}{Pramana} \textbf{\bibinfo{volume}{87}}, \bibinfo{pages}{40}
  (\bibinfo{year}{2016}), \eprint{1507.06424}.

\end{thebibliography}
\end{document}